# THE EUROPEAN FAR-INFRARED SPACE ROADMAP

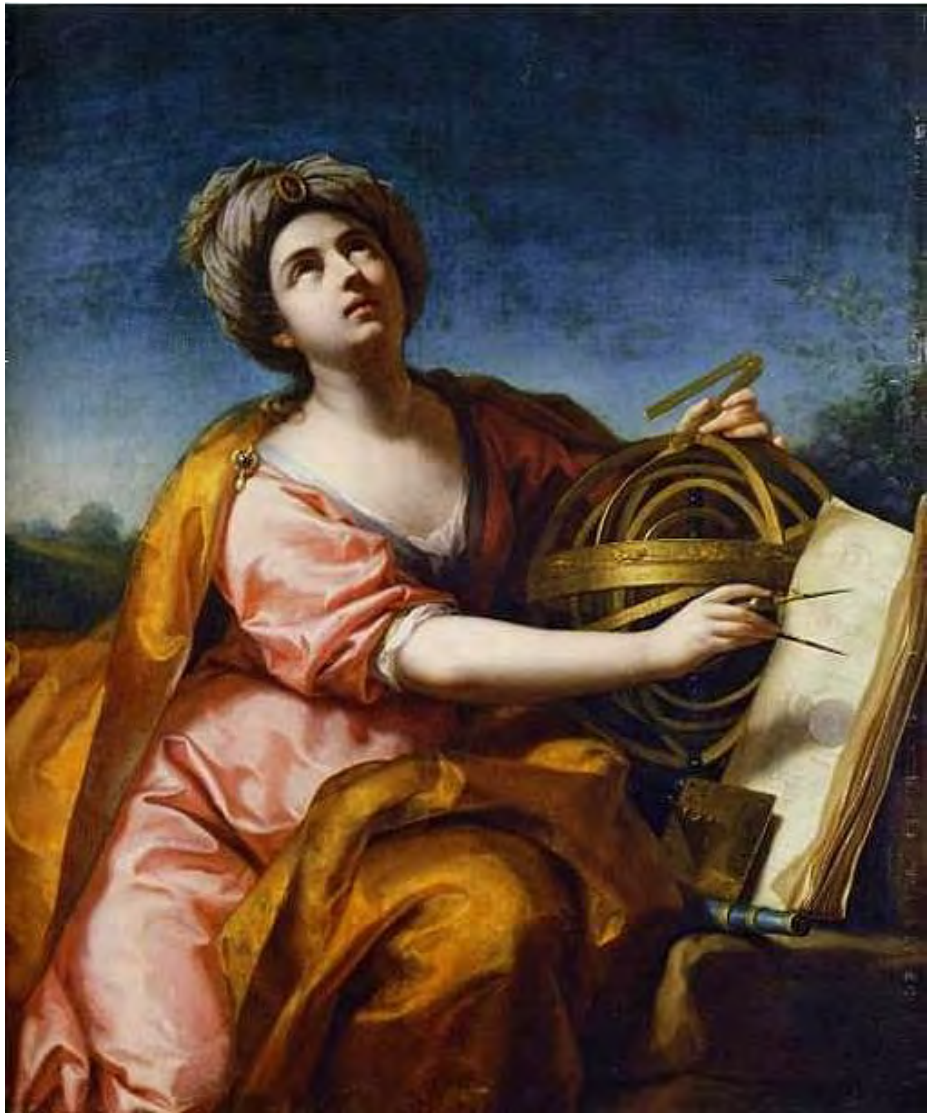

*Front page:*
*Urania, the muse of Astronomy (Jean Roux – 1730)*
*In Greek mythology, Urania (Greek : Ourania "Heavenly" of Ouranos, "Heaven") was the Muse who presided over Astronomy and Geometry.*
*Urania is holding a globe with the constellations and a compass for her measurements.*

# 1 Executive summary

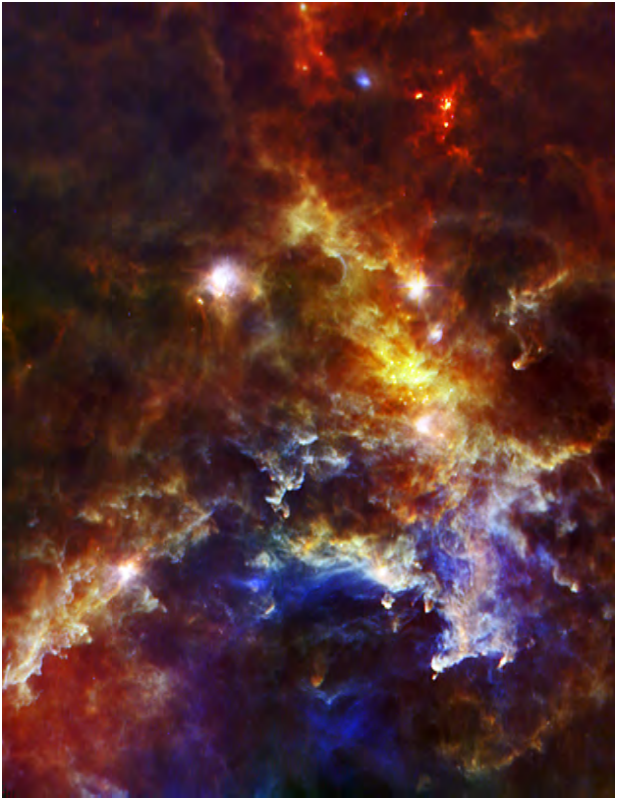

(Rosetta Nebula: ESA/PACS&SPIRE/HOBYS Key Program)

We live in a highly dynamic Universe. Small density fluctuations after the Big Bang grew into larger gravitational potential wells. Within the wells dark matter decoupled from gaseous matter, the latter condensed and formed the first stars and galaxies. These first galaxies transformed rapidly accreting gas and matter from the cosmic web. As time went on the primordial pristine gas was polluted with heavy metals (elements heavier than hydrogen, helium and lithium) originating in stellar nucleosynthesis and ejected into the interstellar gas at the end of the stellar cycle. The role of the interstellar gas is pivotal as it provides both the material to form stars and maintains the fossil record of star formation activity in the form of metal enrichment. These metals became locked up in dust grains, created through condensations in the winds of evolved stars and supernovae (SNe) ejecta.

Today we can see the effects of dust in our Galaxy through dark patches obscuring the starlight in the night sky. Indeed, ultraviolet (UV) and optical light are absorbed and scattered by dust grains, while the mid-infrared (MIR) and far-infrared (FIR) spectral regimes capture dust emission because of the relatively low grain temperature, 20-60 K. Such emission is important within our Solar System, in proto-planetary disks where we believe planetary systems like our own are forming, and throughout the disk of the Milky Way. Because of the copious amounts of dust present, *it is virtually impossible to study the details of how stars and planets form at UV or optical wavelengths; it must be done in the FIR.*

Space-borne observatories are necessary to probe the complete FIR regime. Previous missions (IRAS, ISO, Spitzer, AKARI, Herschel, Planck) have convincingly shown that most of the star formation in the universe is enshrouded in dust. Over cosmic time, since the Big Bang, half of the integrated energy and most of the photons emerge in the FIR. Thus, *to study processes ranging from the formation of planets to star formation in our Galaxy and in distant galaxies, access to the FIR wavelength regime is crucial.* The process of star formation and the build-up of galaxies are intimately linked to the exchange of energy within the ISM; gas is heated by diffuse background radiation, stars, cosmic rays and shocks which subsequently cools down through radiative processes. This "ISM energy cycle" maintains or disrupts the stellar environment, thus enhancing or quenching star formation through "feedback" mechanisms. *The main coolants of the ISM are visible only in the FIR.*

This document describes fundamental, yet still unresolved, astrophysical questions that can only be answered through a FIR space mission, and gives an overview of the technology required to answer them: essentially a "roadmap". These queries regard:

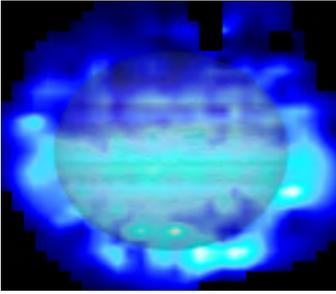
Jupiter: ESA/Herschel/T.Cavalié et al.

*Our Solar System:*
- origin of the water in the Earth's oceans and on Mars and the giant planets
- origin and composition of small bodies in the Solar System (that would be called "debris disks" around more distant stars)

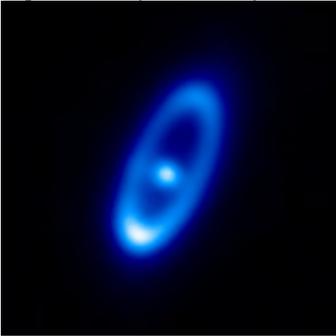
Formalhaut: ESA/Herschel/PACS/ Bram Acke

*Planet formation:*
- mechanisms of exoplanet formation
- origin of water (and other volatiles) on exoplanets

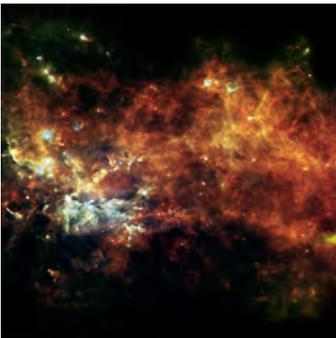
Vulpecula: ESA/PACS/SPIRE/Hi-GAL

*Our Galaxy, the Milky Way:*
- origin and confinement of the ubiquitous ISM filaments discovered by Herschel
- the "CO-dark" gas and the physical processes governing the different gas phases in the ISM energy cycle
- origin of dust grains from evolved stars and supernovae (literally "star dust")

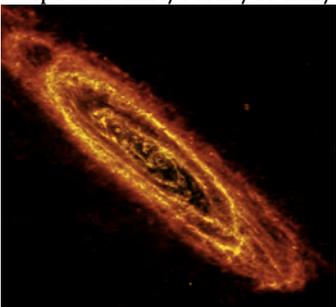
Andromeda: ESA/Herschel/PACS/ SPIRE/O. Krause, HSC, H.Linz

*Nearby galaxies:*
- probing of star-formation activity through FIR cooling lines
- effects of feedback on the dust and gas in galaxies' ISM
- regulation of dust content in galaxies

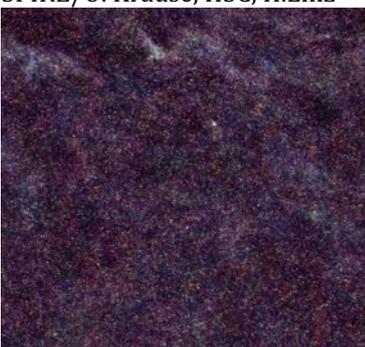
H-ATLAS: ESA/SPIRE KP: S. Eales

*Galaxy evolution in the early universe:*
- quantifying the star-formation history of the Universe
- assessing the physical conditions for galaxy assembly
- origin of the co-evolution of galaxies and supermassive black holes

The questions are intimately related to three of the four given in ESA's Cosmic Vision 2015-2025:

1. what are the conditions for planet formation and the emergence of life?
2. how does the Solar System work?
3. what are the fundamental physical laws of the Universe?
4. how did the Universe originate and what is it made of?

For two of them (1. conditions for planet formation and 4. the origin of the Universe) a Far-Infrared Observatory is listed explicitly as a Candidate Project. We believe that even for the Solar System (for which a FIR Observatory is not considered as a Candidate Project), a FIR space mission would provide unique answers to the questions posed, because they are directly applicable to the workings of the Solar System especially in the context of exoplanets and planetary systems.

Also described in this document are various options for a FIR Observatory, including a consideration of the importance of angular resolution and spectroscopic capabilities. Large, but light, single mirrors are contrasted against interferometric concepts. Spectroscopy and imaging provide complementary views of the astrophysical processes under scrutiny, and we discuss the sometimes mutually exclusive approaches required for their implementation.

Previous FIR missions have already demonstrated an impressive track record and opened a new era of astronomy. The discovery with IRAS of a new class of galaxies, the IR luminous and ultra-luminous galaxies; ISO's detailed characterization of the emission of small interstellar dust grains (Polycyclic Aromatic Hydrocarbons); the resolution into individual galaxies of the cosmic IR background by Spitzer and Herschel; the discovery with Herschel of tenuous pc-wide filaments within which clouds condense to form stars, the discovery of copious amounts of water in the Universe have revolutionized our understanding of how galaxies form and evolve and how planetary systems like our own may originate. However, even Herschel, the most advanced FIR observatory so far, had insufficient spatial resolution and sensitivity to probe the dust-enshrouded galaxies in the early universe without the aid of cosmic lenses. In addition, spectroscopic observations with Herschel covered less than 1% of the sky!

We believe that the time is right to remedy this situation, and to define, develop and support a new advanced FIR Observatory. In subsequent chapters, we describe what has been achieved so far within the themes of the science questions posed here, but most importantly we discuss the necessity to pursue the answers with FIR space missions. Chapter 2 discusses the goals of a FIR observatory for Solar System science; Chapter 3 for protoplanetary disks and planet formation; Chapter 4 for our Galaxy; Chapter 5 for nearby galaxies, and Chapter 6 for distant galaxies in the early Universe. Chapter 7 describes the technology development already underway to achieve these goals and examples of FIR missions. Concepts like SPICA and FIRSPEX would come earlier than large projects like TALC, CALISTO or interferometric concepts like SPIRIT or ESPRIT.

## 2 Solar System Science

Herschel, with its unprecedented sensitivity in the far infrared wavelengths range, has provided exciting new insights into solar system science addressing topics such as the origin and formation of the solar system, the water cycle of Mars, the source of water in the stratospheres of the outer planets, the isotopic ratios in cometary and planetary atmospheres and a number of new detections (possibly related to cryo-volcanic activity) including the Enceladus water torus, water atmospheres and emissions of the Galilean satellites and Ceres and the ocean like water in a Jupiter family comet. Furthermore together with Spitzer a large number of albedos of Kuiper belt objects have been determined, helping to constrain formation processing from protoplanetary/debris disks to solar system bodies. Most of these observations can only be performed in the FIR and from space. The following chapters will describe the findings in more detail and address, the open science questions

### 2.1 Mars

The water cycle is a key aspect of the Martian atmosphere/ surface system. Temporal and spatial variations of the column-integrated amount of water have been characterized by a number of space missions including Mars Global Viking, Mars Global Surveyor and Mars Express. Only the latter provided vertical profiles of water by solar occultation observations in the middle atmosphere of Mars (Fedorova, 2009). Complementary HIFI vertical profiles of water from ground into the middle atmosphere were scheduled to exactly constrain the variable hygropause level, which is supposed to change in altitude between 10 and 50 km over a martian season, however, the HIFI prime instrument failure short after launch resulted in only a small seasonal coverage around Northern summer with the redundant HIFI instrument (Figure 2.1) showing a generally low hygropause level. Full seasonal coverage and monitoring observations are required and can this can only be achieved with a future FIR telescope.

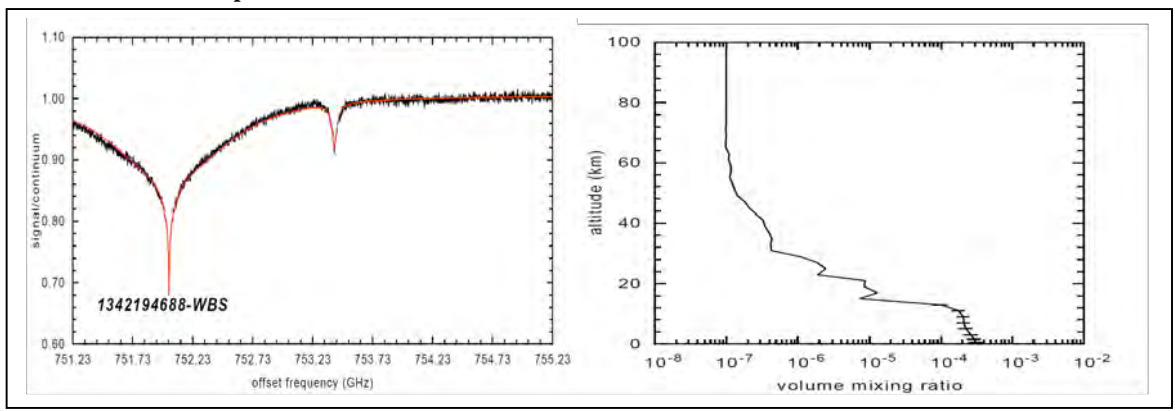

Figure 2.1: HIFI spectra of $H_2O$ and HDO and retrieved vertical profile of water vapor during solar longitude LS=78 ° (Hartogh et al., A&A 2017, in prep.)

Hydroxyl chemistry is tightly related to the water abundance and is essential for understanding the stability of the Martian atmosphere. HIFI observed $H_2O_2$, but the detected volume mixing ratio was considerably smaller than predicted by photochemical models and infrared observations. The SNR of these observations was very low so that vertical profiles could not be derived. Recent observations by SOFIA and the GREAT instrument only provided upper limits. Higher sensitivity observations are required to understand the discrepancy between models, and FIR observations. Future heterodyne receivers will take advantage of the recent progress in Hot Electron Bolometer (HEB) technology, providing higher sensitivity. Together with a larger collecting area compared to Herschel observations of dedicated solar longitudes over a full martian year would provide for the first time vertical $H_2O_2$ profiles and in a unique way constrain photochemical models. Other important species are $HO_2$ and OH not observed by HIFI, since Mars' apparent diameter was too small during the Herschel observation windows so that the planet was not resolved. With larger collection area and higher sensitivity a new advanced FIR telescope will either detect these molecules or provide > 1 order of magnitude lower upper limits. This is the case also for line surveys. HIFI performed line surveys in bands 1-5, however due to the limited SNR no new species were detected and only some new upper limits were derived. HIFI detected molecular oxygen (Figure 2.2) for the first time in the submm and for the first time since 1972 (Hartogh et al. 2010). While the column density is (within the error bars) compatible with the 1972 observations and shows only small deviations to recent ground-based in situ measurements of Curiosity, HIFI derived deviations from the constant vertical altitude profile not predicted by photochemical models and not yet fully understood. Observations of $O_2$ with SNR > 100 and, preferably at different locations on the Martian disk are required. High SNR observations of hydrogen and oxygen isotopes (in $H_2O$ and CO) will constrain atmospheric escape processes and isotopic fractionation with altitude. During the Mars opposition in October 2020, for instance, Mars' apparent diameter will be > 22 arcsec. Resolved FIR observations of temperature (via CO transitions) and the species mentioned above would constrain chemistry and dynamics (especially the meridional transport) of the martian atmosphere.

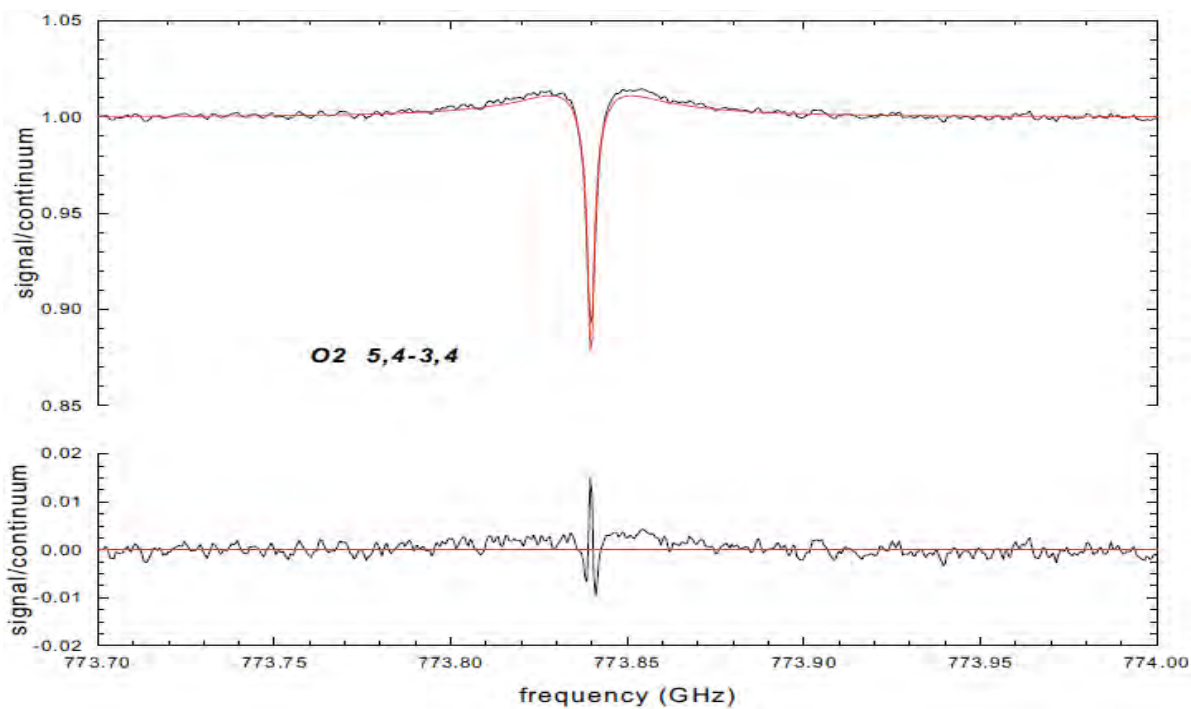

Figure 2.2 Observed and fitted O2 line at 774 GHz and residuals indicating deviations from constant vertical profile (Hartogh et al. 2010)

## 2.2 Outer Planets and Titan

ISO discovered water vapor in the stratospheres of the four giant planets (Feuchtgruber et al., 1997), implying the existence of external sources of water. These sources may be interplanetary dust particle fluxes (IDP), local sources (rings, satellites) or cometary impacts. Disentangling the various sources is a key objective. It bears implications on the variety of poorly understood phenomena such as the production of dust at large heliocentric distances, the transport and ionization of solid and gaseous material from satellites and rings in planetary magnetospheres and the frequency of cometary impacts in the outer Solar System. Horizontally and vertically resolved observations of Herschel resulted in the conclusion that the SL9 impact in 1994 delivered most of the water in Jupiter's stratosphere (Cavalie et al., 2013, Figure 2.4), while the Enceladus water torus (Hartogh et al, 2011) for the first time directly detected by Herschel (Figure 2.3) is the main source of water in Saturn's stratosphere. However, open questions remain and the relative contributions of the other sources have to be quantified. The submm wave instrument (SWI) on JUICE (Hartogh et al, 2014) will complement remote FIR observations by providing very highly resolved 3-d observations of temperature, winds, a number of important atmospheric gases, isotopic ratios, the OPR of water and atmospheric waves.

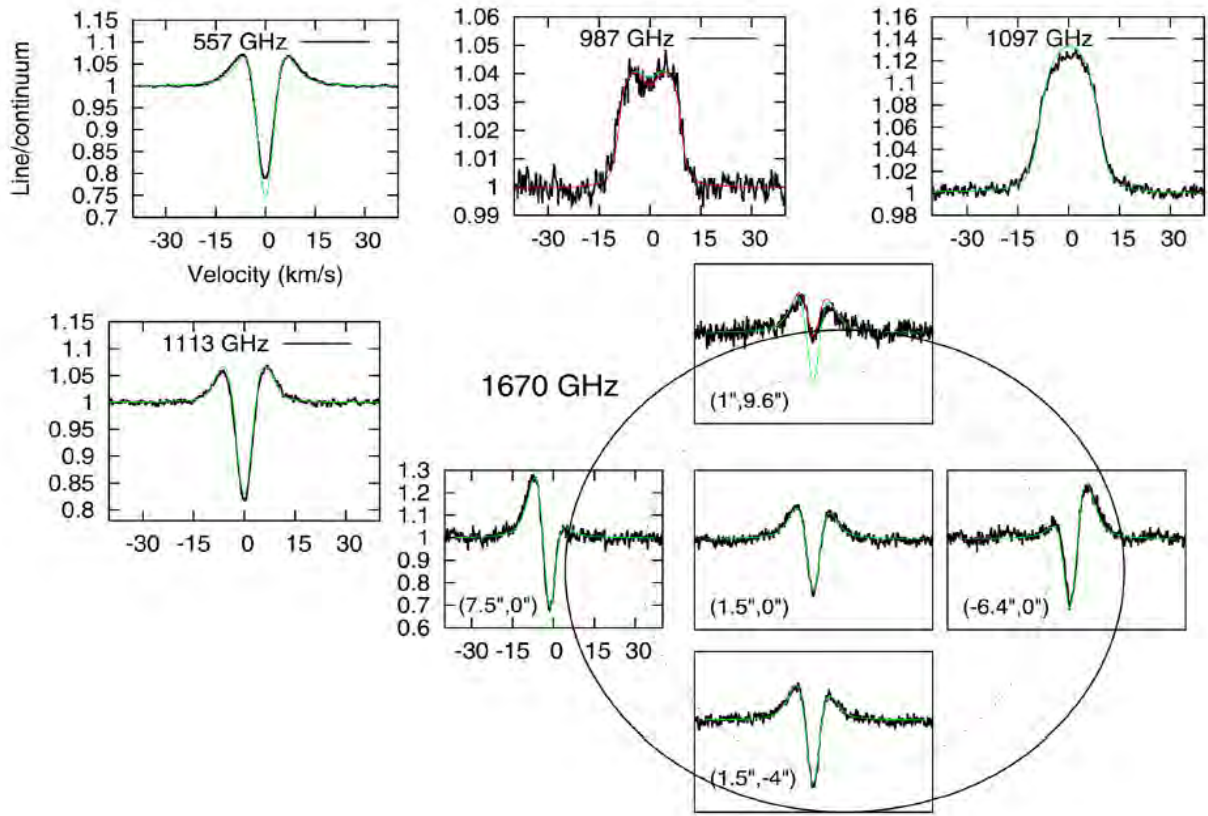

Figure 2.3 HIFI observations of the Enceladus torus at ground-state and higher transitions. At 987 GHz and 1097 GHz the torus is almost transparent, the lines show the stratospheric emission of Saturn (Hartogh et al. 2011a)

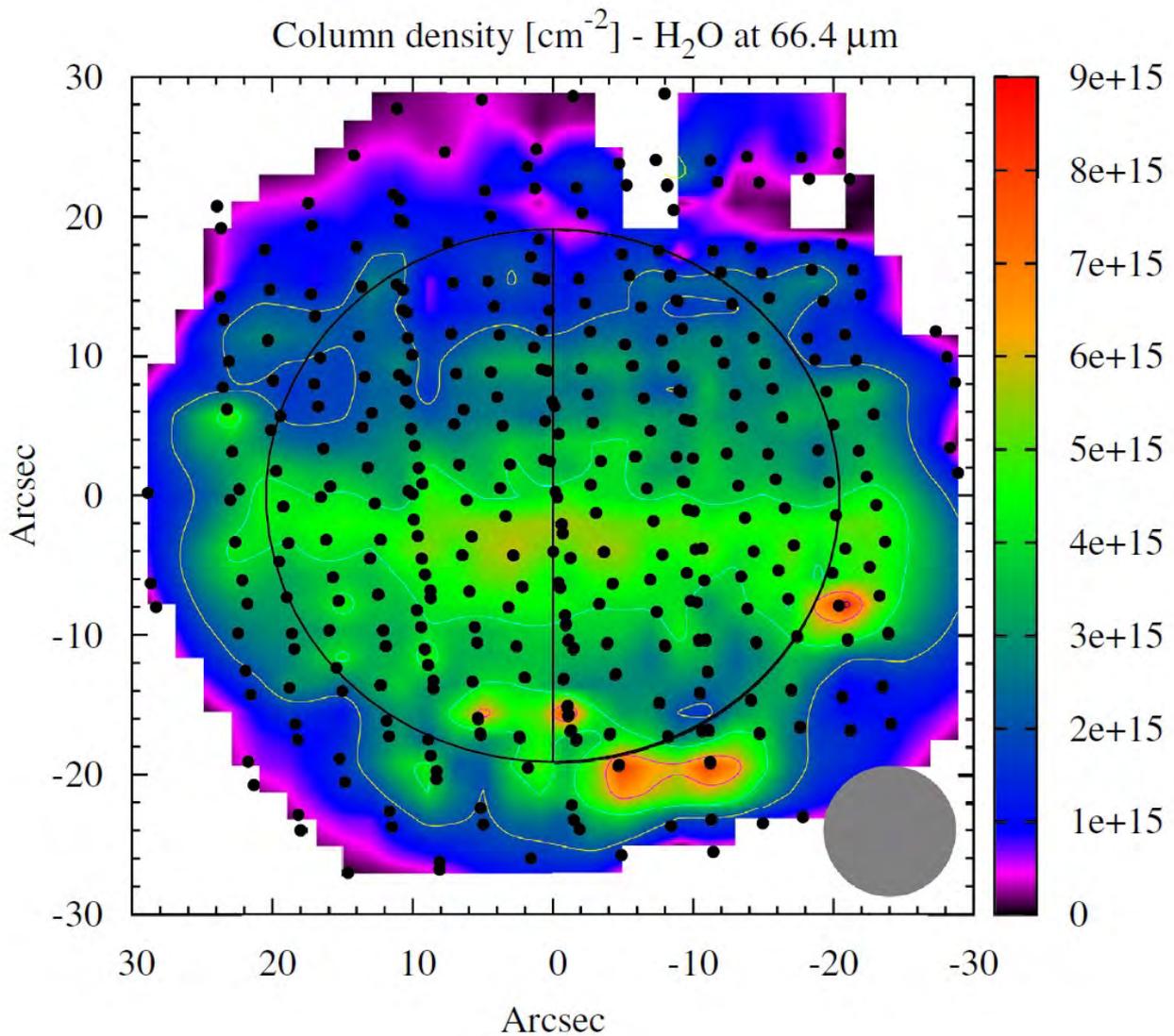

Figure 2.4 PACS map of water vapor column density of Jupiter (Cavalie et al. 2013)

High SNR observations with a new FIR telescope with higher spatial resolution will not only better constrain the sources, but also enable a deeper search for emission lines of new molecules in the Enceladus torus. The higher spatial resolution will also reduce the rotational smearing effect of the atmospheric spectral lines due to the rapid rotation of Jupiter and Saturn. As a result the retrieval of vertical profiles of all species will extend to higher altitudes. Besides water at least $PH_3$, $NH_3$ and $CH_4$ in Jupiter and Saturn and HCN and CO in Neptune are of high interest. Higher spatial resolution than that afforded by Herschel is needed to derive high SNR observation of Titan's water transitions. Although HIFI observed Titan water transitions for more than 15 h (Moreno at al., 2012), the SNR was not sufficient (Figure 2.5) to directly retrieve a vertical profile. Instead a vertical profile based on chemistry model calculation was proposed to fit HIFI and PACS data. A new heterodyne instrument has the potential to provide water spectra with an SNR of > 100 required for a direct retrieval of the vertical profile of water from the measured line shape. Herschel SPIRE and PACS observations confirmed the composition of Titan's atmosphere as known from former Cassini-CIRS and ground-based observations (Courtin, 2011; Rengel, 2014). A particularly interesting result was the 16-O to 18-O isotopic ratio of only 380 ±60 in CO, about 24 % lower than the telluric value. Future line surveys with PACS or SPIRE-like instrument would provide more than one order of magnitude higher sensitivity compared to HIFI and will likely discover a number of new molecules in Titan's atmosphere.

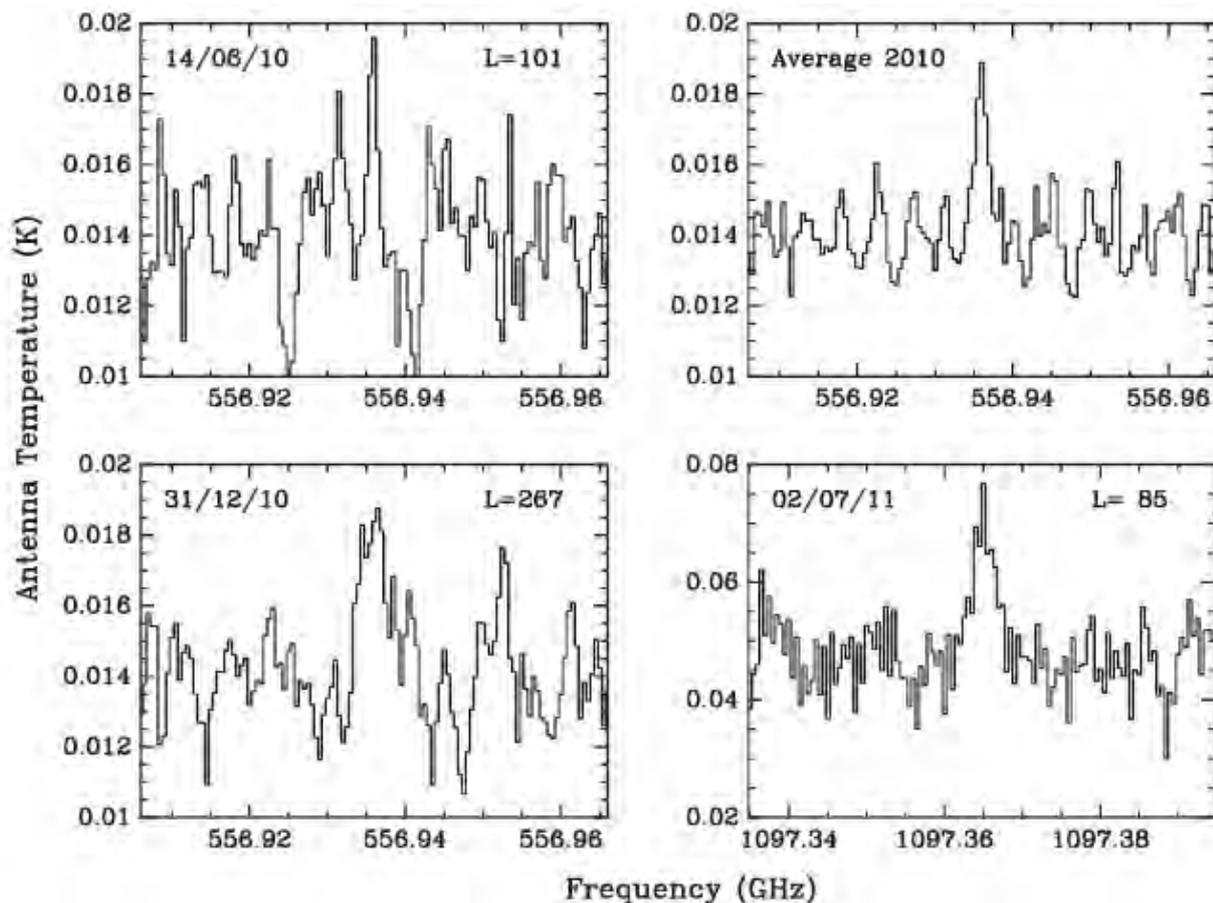

Figure 2.5 HIFI observations of the 557 GHz and 1097 GHz water transitions on Titan (Moreno et al. 2012)

## 2.3   Enceladus and the Galilean Satellites

Cryo-volcanic activity was discovered for the first time in Enceladus by the Cassini spacecraft in 2006 (e.g. Porco et al, 2006). Volcanic plumes feed the Enceladus torus with about 300 kg of water per second (Cassidy and Johnson, 2010). HST observations found transient water vapor at Europa's south pole (Roth et al, 2013) pointing to cryovolcanic activity or changing surface stresses based on Europa's orbital phases. Herschel HIFI observations found asymmetric water vapor atmospheres in Ganymede and Callisto (Hartogh et al, in preparation). Their potential sources may include sputtering processes, sublimation or unknown surface processes. A new FIR heterodyne instrument has the capability to monitor the spatiotemporal evolution of these water atmospheres/plumes, relate them in more detail to orbital phases and possibly determine their sources and sinks. Broadband surface observations will add valuable information about the thermophysical properties of the ice/regolith layer and the composition of ices.

## 2.4   Ceres and the asteroid belt, main belt comets

HIFI observations of water plumes on Ceres with a production rate of about 6 kg/s pointed to cryo-volcanism (Kueppers et al, 2014). Recent observations of hazes by the DAWN camera however suggest sublimation of water ice over the Occator crater (Figure 2.6) that has been probably collected from areas beyond the snow line (Nathues et al, 2015).

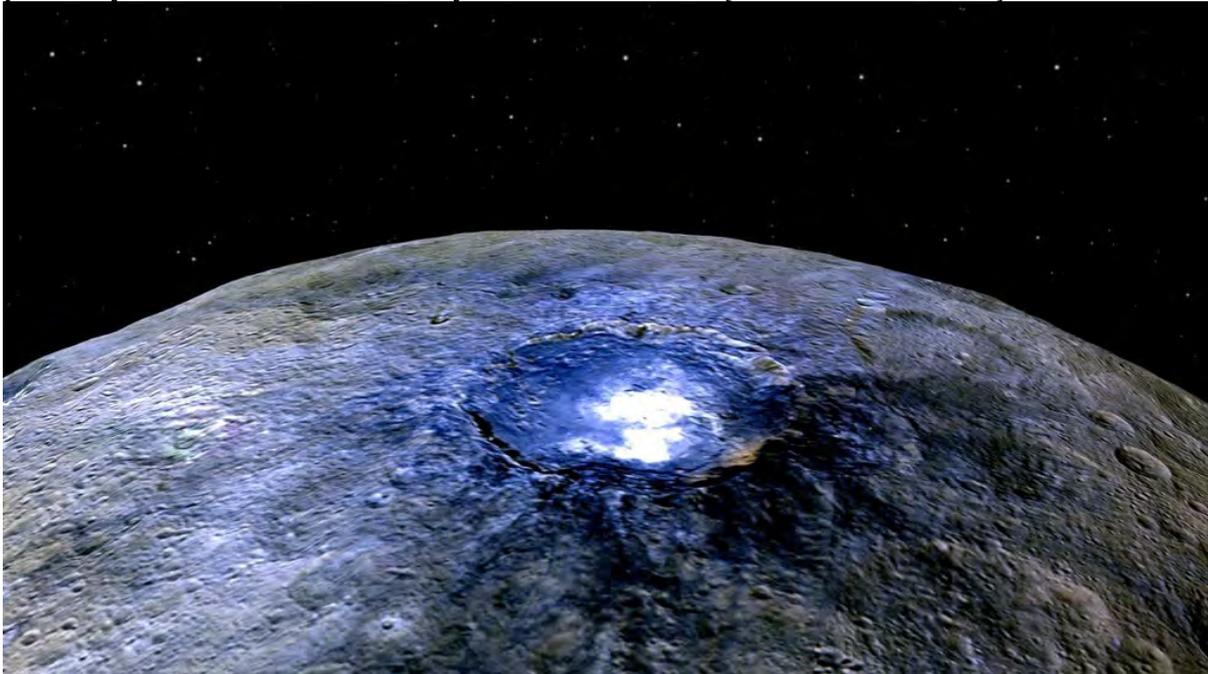

Figure 2.6 DAWN Framing Camera image of the Occator crater on Ceres from 4425 km distance. Under certain circumstances, hazes can be observed over the crater (Image: NASA/JPL-Caltech/UCLA/MPS/DLR/IDA)

The presence and abundance of water in asteroids are relevant to many areas of research on the Solar System, ranging from the origin of water and life on Earth to the large-scale migration of giant planets such as Jupiter. The initial HIFI observations of Ceres provided ambiguous results with 4 sigma detections in only one polarization and no detection in the other. Their repetitions lead to similar results, indicating that the water emission was related to a local source. Finally 10 hour observations, covering the "light curve" of Ceres lead to the crucial detection. The sensitivity of the observation was around 1 kg/s of water production rate. A future FIR heterodyne instrument may reduce the sensitivity to about 100 g/s or less, depending on the observed transition. This high sensitivity opens a new field of asteroids research. Water emissions may be found in other asteroids, for instance in carbonaceous chondrites or main belt comets (MBCs). Water sublimation could be one cause explaining the observed dust comas in MBCs, however all attempts to detect water were unsuccessful thus far (de Val-Borro et al. 2012, O'Rourke et al. 2013) due to the limited sensitivity of former observations including Herschel/HIFI.

### 2.4.1 Comets and the D/H rations in the solar system

Having retained and preserved pristine material from the Solar Nebula at the moment of their accretion, comets contain unique clues to the history and evolution of the Solar System. Their study assesses the natural link between interstellar matter and Solar System bodies and their formation. Ironically, although being the most abundant cometary volatile, water is one of the most difficult species to observe. Since cometary gas is cold (10–100 K) and water is rotationally relaxed at fluorescence equilibrium, the rotational transitions between the lowest energy states are the most intense. The water line at 557GHz is expected to be among the strongest lines of the radio spectrum of comets. Water plays an important role in the thermal balance of cometary atmospheres, as a cooling agent via emission in its rotational lines. This role is crucial in determining the expansion velocity and temperature of the atmosphere, which are two fundamental parameters for the physical description of this medium. Indeed, cooling becomes effective only in the outer coma where the transitions become optically thin. The observations of several water lines with Herschel provided insights into the excitation of this molecule and optical depth effects (Hartogh et al, 2010, de Val-Borro et al, 2010). Observations of $H_3O+$ constrain the excitation by ionic collisions. This will lead us to more realistic models of the thermodynamics of the atmosphere. A FIR Space telescope is necessary to continue research on water excitation.

Studying water and its isotopologues in Solar System bodies (Hartogh et al, 2009), provides key information about their formation and evolution. A crucial parameter, in particular, is the deuterium/hydrogen (D/H) ratio measured in water. It is known from laboratory experiments, and confirmed by observations in the interstellar medium (ISM), that deuterium is enriched in ices, due to ion-molecule and grain-surface reactions at low temperature. The D/H ratio in Solar System objects (see Figure 2.7) provides information about the physico-chemical conditions under which the water formed and about mixing processes of equilibrated water with cometary ices with increasing D/H as function of heliocentric distance. It may also provide information about formation processes of the outer planets and Jeans-escape of the terrestrial planets.

Prior to the first Herschel observations, the D/H-ratio in 6 Oort cloud comets (OCCs) was determined to be in average twice as high as VSMOW (Vienna Standard Mean Ocean Water~ 156 ppm), which ruled out comets as an external supply of Earth water.  Herschel/HIFI determined for the first time the D/H-ratio in a Jupiter family comet (JFC) (103P/Hartley 2) to agree with VSMOW (Hartogh et al, 2011). Observations by Lis et al. 2013 of the Jupiter family comets 45P Honda-Mrkos-Pajdusakova confirmed the low D/H, however HIFI observations of the Oort cloud comet C/2009 P1 Garradd with about 200 ppm (Bockelée-Morvan, et al, 2012) relativizes the seemingly D/H dichotomy between OCCs and JFCs. Gibb et al. 2012 provided another constraint for D/H in C/2007 N3 Lulin. The in-situ measurement of the highest-ever measured cometary D/H ratio in the JFC 67P/Churyumov-Gerasimenko point into the same direction (Altwegg et al. 2012). For some time JFCs were the comets with the largest scatter in D/H ratios until this feature went back to OCCs. (Biver et al., 2016) determined D/H ratios of ~ 140 and 650 ppm in 2 further OCCs by submm observations (Odin and IRAM).  D/H-ratios from a larger sample of comets (together with the 14N/15N ratio which currently favours asteroids as the agents of water) are essential for a better understanding of solar system formation processes.

While the D/H-radio in the gas giants, measured in hydrogen corresponds to the protosolar value (Figure 2.7), it should be increased by isotopic exchange reactions between water and hydrogen in the ice giants. Recent Herschel/PACS observations of HD confirmed a value about twice as high as the protosolar value (Lellouch et al, 2010; Feuchtgruber et al, 2013). Based on formation models and measured D/H ratios in comets they conclude that the ice mass fraction in both planets may be smaller than predicted by previous models. Future FIR heterodyne observations may determine not only the D/H-radio in hydrogen with higher sensitivity and resolution but for the first time also provide D/H-ratios in water in the giant planets' atmospheres, providing an additional constraint to external sources of water.

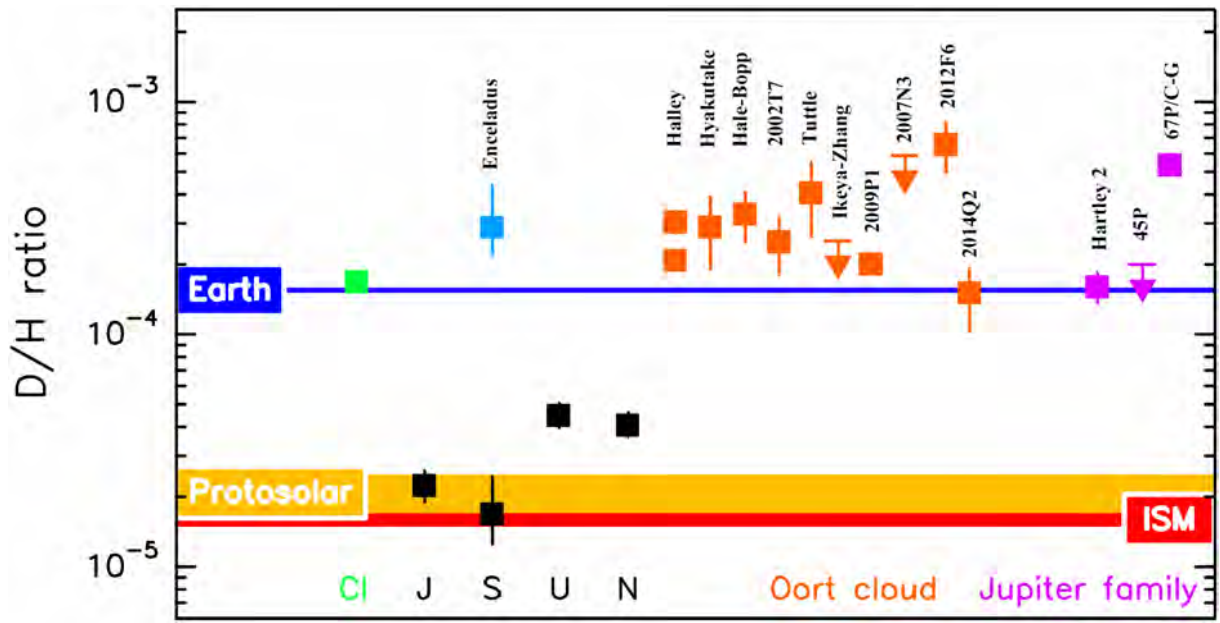

Figure 2.7: Measured D/H distribution in the solar system (taken and adapted from Hartogh et al. 2011b)

Recently, the ROSINA mass spectrometer suite on board Rosetta discovered an abundant amount of molecular oxygen, $O_2$, in the coma of 67P of nearly 4 % relative to water (Bieler et al, 2015), which is in contradiction to current solar system formation models. It could be shown that $O_2$ is indeed a parent species and that the derived abundances point to a primordial origin. Crucial questions are whether the $O_2$ abundance is peculiar to comet 67P/Churyumov-Gerasimenko or Jupiter family comets in general, and also whether Oort cloud comets such as comet 1P/Halley contain similar amounts of molecular oxygen. Re-analysis of Giotto data showed indeed a similar amount of molecular oxygen (Rubin et al, 2015). These unexpected results raise the question whether this high $O_2$ to $H_2O$ ratio is a general feature of comets and whether planetesimals commonly contain $O_2$ at a level of a few percent. If so, a substantial part of $O_2$ may have been delivered to Earth by impacts rather a sign of biological activity. The close abundance despite very different dynamical histories and erosion rates of both comets indicates that the observed $O_2$ has already been formed in the ices of the pre- and protosolar nebula, before the comet ultimately formed. This would turn require that ice grains did not, or only partially, sublimate and reform during the collapse of the protosolar nebula, perhaps due to formation of planetesimals at a very early stage. Because the large amount of molecular oxygen found in 67P and 1 P was unexpected, Herschel/HIFI observations of $O_2$ in comets were never scheduled. Future FIR heterodyne observations of $O_2$ for a large set of JFCs and OCCs are required in order to better constrain the recent findings.

## 2.5 Kuiper Belt Objects

Kuiper belt objects (KBOs) are the best-preserved remnants of the formation of our solar system. They retain information on the chemistry of the protoplanetary/debris disc and the physical processes that led to the formation of the planets. A fundamental property of KBOs, their size, is hard to measure from Earth. Due to their large distances, most KBOs are unresolved and their apparent brightness is degenerate with the uncertainties in size and albedo. The KBO size distribution can tell us how these objects formed and on how collisions have eroded the initial. In solar system astronomy, the most fundamental application of thermal observations is precisely the calculation of sizes and albedos of unresolved small bodies. But much more can be achieved in this uncharted wavelength range.

A future FIR telescope may detect the Rayleigh-Jeans tail of the thermal emission of hundreds of KBOs. In that regime, the flux is nearly proportional to the instantaneous cross-section of the object, with a very weak (1/4th power) dependence on poorly constrained parameters such as the IR emissivity and the temperature distribution across the surface. New observations will hence offer a reliable, unbiased estimate of the size distribution of KBOs.

Albedos, on the other hand, provide an important constraint on the surface chemistry of KBOs. Of the ~120 KBOs with reliable Spitzer and Herschel albedo estimates appear very dark (albedos 0.04 - 0.25), the largest bodies (Pluto-scale) show highly reflective surfaces, probably indicating fresh, ice-rich surfaces. A family of objects associated to the KBO and dwarf planet Haumea show water-ice rich (Barkume et al 2006), high albedo surfaces, and is suspected to have formed through a massive collision onto the proto-Haumea. Accurate albedos are also crucial for spectral modelling of KBO surfaces (De Bergh et al 2013).Cooled down to 5 K, a future FIR telescope will improve on the sensitivity of Herschel by two orders of magnitude. This is a very promising scientific enterprise given the success of Herschel in the study of KBOs (e.g. Müller et al 2010).

In addition, sensitive spectroscopic capabilities will yield KBO SEDs. These can be fit with very detailed thermo-physical models in which the effects of spin state, thermal inertia and infrared beaming are all taken into account. For example, high surface thermal inertia generally implies a porous, regolith-covered surface. High beaming parameter (FIR flux enhancement) indicates a cratered surface (Lellouch et al 2013). It is interesting to see how these properties vary with size.

Thermal light curves are also particularly interesting. The phase difference between the optical and thermal light curves of an object can be used to decide if the variability is mainly caused by surface patches or by a shape cross-section effect. In addition, the angle between spin axis and line-of-sight can be constrained from the average thermal flux (a pole-on object will be warmer than one closer to equinox). This is important as some models of planetesimal formation predict that the largest KBOs should have aligned spins (Johansen & Lacerda 2010). Herschel thermal light curve data on the fast spinning (P=4 hr) dwarf planet Haumea confirms its triaxial shape and high density ~2.5 g/cc, and supports the presence of a dark surface spot of unknown nature (Lacerda et al 2008, Lacerda 2009).

Future spectroscopic capabilities will let us search the surfaces of KBOs for long wavelength ice and mineral features that are stronger and less ambiguous than those at shorter wavelengths. Currently, only the brightest few KBOs can be studied in a useful way using visible and NIR spectroscopy. The little we know about the surface composition of KBOs is based on broadband photometry. With the future telescope we will be able to map out the composition of small body surfaces throughout the outer solar system. The relative unexplored wavelength regime has also the potential to lead to new discoveries.

## 2.6  Spectroscopic and telescope requirements

Achieving the science goals mentioned above requires a considerable progress in the telescope and instrument performance. This includes i) extension of the heterodyne bands from 400 GHz to 5 THz (access to new species like HD, OI and to stronger transitions), ii) increase of the sensitivity in this range to < 3 hv/k (HIFI achieved this sensitivity only in band 1. Many HIFI observations could have been done more efficient at higher frequencies, provided the sensitivity was as good as in band 1), iii) increase of sensitivity of incoherent instruments to $1\times10^{-20}$ Wm$^{-2}$ (5σ 1hr/FOV), iv) spectral resolution of 100 kHz for heterodyne and > 5000 for incoherent instruments, v) increase of telescope diameter to 10 m or more (spatial resolution of planets, wind measurements, point source observations).

# 3 Protoplanetary and debris disks

Two decades after the detection of the first exo-planet by Mayor & Queloz (1995, 51 Peg B) the number of known exoplanets is over 3000. Interestingly, none of the planetary systems found to date is similar to our own Solar System, instead a wide variety of planetary architectures are found in terms of planetary mass and orbital properties. As an example, Figure 3.1 shows the planetary mass vs semi-major axis diagram. In spite of this large number of discoveries, our understanding of the planet formation process is still rather limited: how do planets form? what is the origin of the large heterogeneity of planetary systems? how are volatiles delivered to planets? Where do the asteroids and comets in these systems reside?

Our poor understanding of the planet formation process is largely due to our lack of knowledge of the physical and chemical conditions during the formation phase within the optically thick gas-rich protoplanetary disks that surround all young stars. Similarly, outside the Solar System we have little knowledge of processes that occur beyond this gas-rich phase, such as the final growth of Earth-like planets and dynamical instabilities that cause Late Heavy Bombardment-like events. Learning about such processes, and placing the architecture of the Solar System's planets, asteroids, and comets in context will require detailed characterisation of the optically thin gas-poor debris disks that almost certainly surround all main-sequence stars.

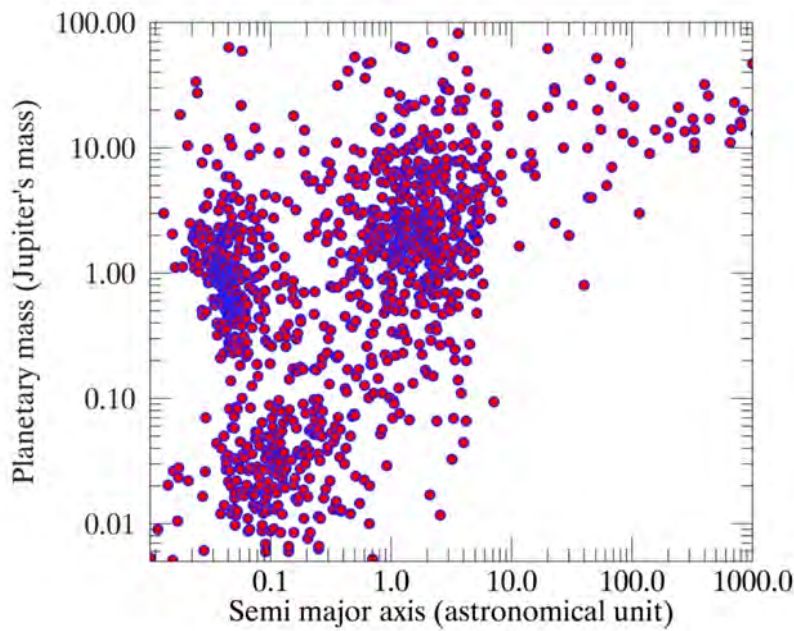

Figure 3.1 Exoplanets demography (from exoplanet.eu)

## 3.1 Observations of protoplanetary disks: disk molecular layers

The interior of a protoplanetary disk is characterized by strong temperature and density gradients both in the radial and vertical direction (Figure 3-2). Along the radial axis the temperature drops from a few $10^3$ K in the inner part (close to the star) to a few 10 K at larger disk radii. A similar temperature gradient occurs in the vertical direction with the outermost layers hotter than the disk interior. Simultaneously, the density increases from $n_H$ ~ $10^5$ cm$^{-3}$ in the outermost layers to $n_H$ ~$10^{15}$ cm$^{-3}$ in the disk midplane. These strong gradients lead to different physical and chemical processes taking place in different part of the disk. We can identify three major chemical layers: 1) the *photon-dominated layer*; 2) the *warm molecular layer* and 3) the *cold midplane*. The three layers are sketched in Figure 3-2. In the photon-dominated layer the chemistry is regulated by photo-processes where the gas interacts with the stellar and interstellar radiation. Here the gas is expected to be mostly in atomic form due to photo-dissociation of molecules by means of ultraviolet photons. The warm molecular layer is partly shielded from the photo-dissociating radiation and thanks to the warm temperature and increasing density, several gas-phase chemical reactions can occur including formation of simple molecules through ion-neutral and neutral-neutral reactions. The timescale of these gas-phase reactions is fast ($10^4$ yr) leading to a rich molecular zone. Finally, in the cold midplane, many species condense on the surface of dust grains, and the chemistry is controlled by grain-surface reactions.

Non-static physical processes occur in disks leading to mixing between the three chemical regions outlined above. Vertical mixing will occur in the inner region of the disk (close to the star) where turbulence is likely to be present (e.g., because of viscous accretion). Radial mixing is also expected because of the inward drift of small dust particles and migration of planetesimals. Chemical-dynamical models of disks show that the continuous exchange of material between the cold midplane and the the warm molecular layer accelerate the formation of complex molecules allowing the chemical enrichment in the disk interior (e.g., Semenov & Diebe 2011; Furuya & Aikawa 2014).

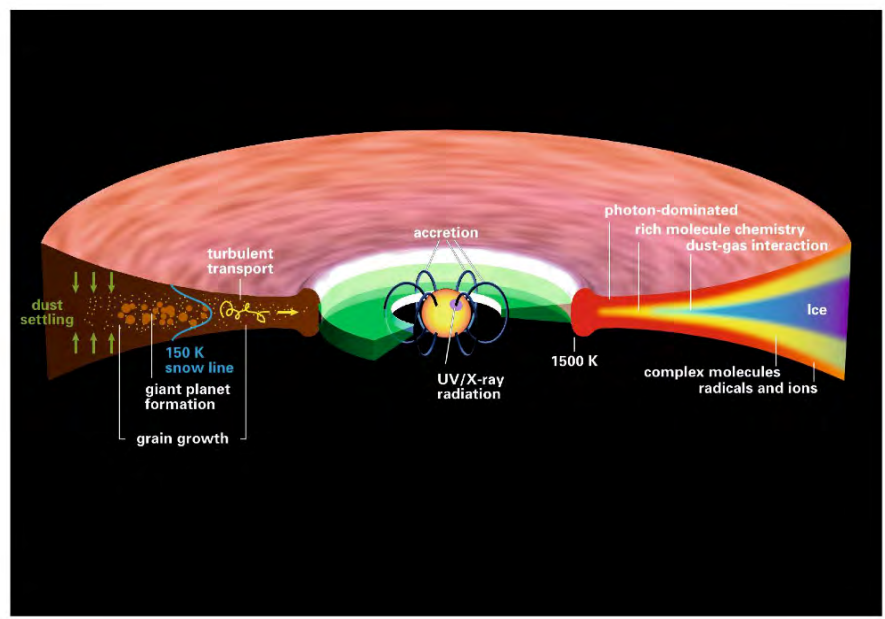

Figure 3-2 Schematic of the interior of a protoplanetary disk (from Henning & Semenov 2013)

Observations at infrared wavelengths ($\lambda$ ~ 40 – 600 µm) are best suited to study the physical (e.g., density and temperature) and chemical (e.g., molecular composition, relative abundance of different species) conditions at the time of planet formation. The disk emission peaks in the far-infrared (~ 100 µm) and this spectral window is unique as it gives us access to several gas- and solid-state features. Only at these wavelengths is instead possible to study:
- the gas mass and its temporal evolution
- the water reservoirs
- the origin of gas in debris disks
- the disk thermal structure
- the dust composition
- detect the true Kuiper belt analogues

All of these issues are tightly connected to the formation of planets and their atmosphere.

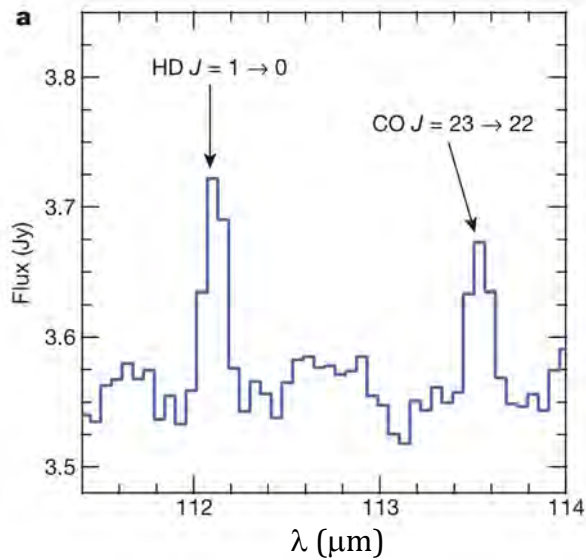

## 3.2  The gas mass and disk evolution

Missions like Kepler have started to characterize the mass distribution of exo-planets, and distinguish smaller rocky planets and larger ice and gas giants. In the standard *core-accretion* planet formation model, planets grow from slow coagulation of dust grains into larger and larger entities and ultimately planets. Some grow out to 10 Earth masses or more, enough to start capturing gas directly from the surrounding disk and evolve into gas giants. In addition to the dust mass, the gas mass of the disk is therefore of great importance to understand the formation of planets.

Figure 3.3 Detection of HD in TW Hya (Bergin et al. 2013)

Measuring the gas mass is challenging. Its main constituent, $H_2$, is very difficult to detect and, if detected, only probes a very small section of the disk that is warm or strongly irradiated by ultraviolet radiation. Measurements of the CO molecule and its isotopologues are affected by freeze out in the cold and dense disk midplane as well as isotope-selective photodissociation and chemical processing of carbon into longer carbon chains (Best & Williams 2014; Miotello et al. in 2014; Kama et al. 2016).

Often, gas masses inferred from CO are lower by factors to 10-1000 compared to the expected values extrapolated from the dust mass. While this may be due to gas-to-dust mass ratios that differ from the standard ISM values, at least in one case this does not seem to be the case. In TW Hya, the Herschel satellite detected emission from deuterated molecular hydrogen, HD (Figure 3.3; Bergin et al. 2013). Using simple chemical reasoning and the extensive knowledge about the structure of this particularly well studied disk, a high gas mass was inferred, consistent with the value extrapolated from the dust mass and suggesting CO freeze out, photodissociation, and chemical processing.

The HD lines from disks are not strong, and Herschel only managed to detect it toward a few single disks (Bergin et al. 2013; McClure et al. 2016). Its interpretation is affected by the fact that the Herschel observations do not resolve the emission spatially or spectrally. This means that the association of the emission with the disk cannot be directly proven: any contribution from, for example, a warm disk wind cannot be separated from that from the disk itself. The HD lines at 112 and 56 micrometer can only be detected from space. Future space instrumentation can explore these lines with greater sensitivity and spectral resolution. Accessing both lines is important to constrain the excitation and remove dependencies on modeled temperature structures.

## 3.3 Water reservoirs

Water is a key species for the formation and chemical composition of planets. Water on Earth allowed the early biological evolution of the planet and the origin of life. The origin of $H_2O$ (and volatiles in general) on Earth is still unclear: the planet could have been formed in a "wet" environment accreting $H_2O$ from hydrated silicates or in a "dry" environment with $H_2O$ being delivered to the Planet through the collisions with comets and asteroids. Watery atmospheres are being detected in a growing number of exo-planets (e.g., Kreidberg et al. 2015).

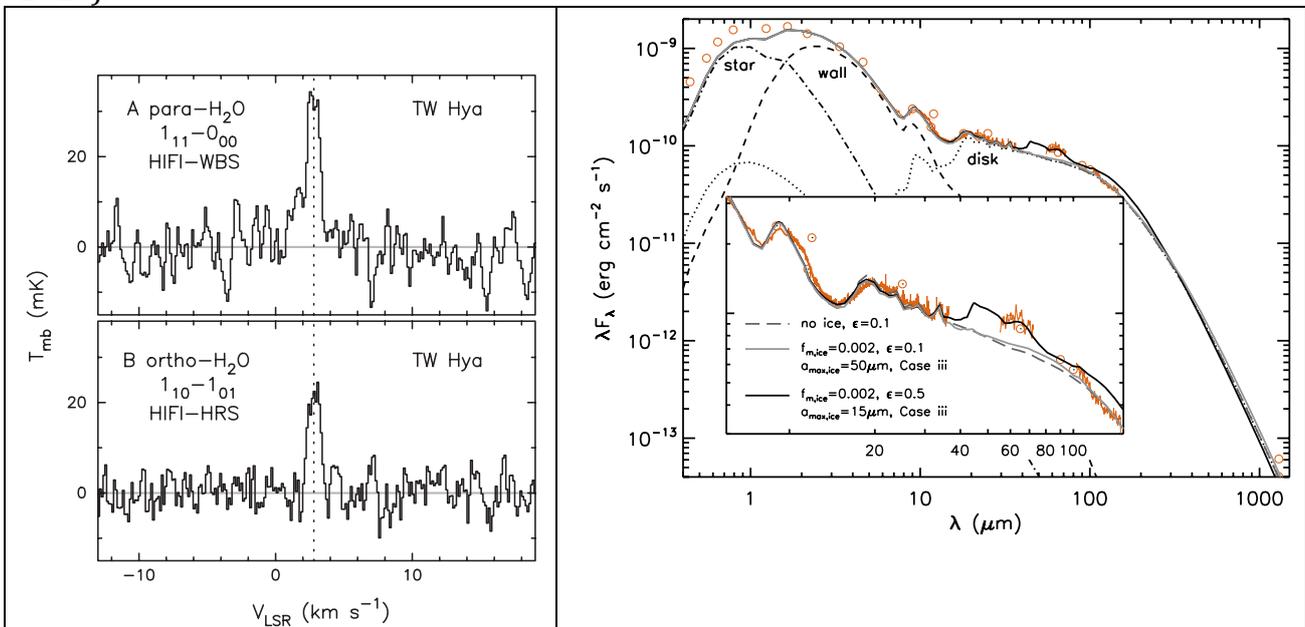

Figure 3.4 (left) Detection of the ground state $H_2O$ lines in TW Hya with Herchel/HIFI (Hogerheijde et al. 2011). (right) Water ice in GQ Lup with Herscel/PACS (McClure et al. 2012)

Spectroscopic observations of protoplanetary disks in the far-infrared are mandatory to determine the $H_2O$ abundance distribution at the epoch of planet formation. This information is very valuable to understand the delivery of $H_2O$ to planets and the origin of planetary atmospheres. While observations from the ground in the near-infrared (1–5 μm and 8–13 μm) give access to the ro-vibrational spectrum of $H_2O$, tracing only the hot component in the disk outer layers, the far-infrared range covers hundreds of $H_2O$ pure rotational transitions (including the ground-state ones) spanning a wide range in upper energy level ($E_u \sim 50 - 1000$ K). Given the temperature and density gradients in the disk interior, the different far-infrared rotational transitions are sensitive to different regions in the disk, allowing us to measure the $H_2O$ reservoirs in the entire disk. The high-J (J > 3) rotational transitions are sensitive to the warm molecular layer of disks (e.g., Fedele et al. 2013). The ground-state $H_2O$ lines, (o- $H_2O$ $1_{10} - 1_{01}$ at 538.3 µm and p- $H_2O$ $1_{11} - 0_{00}$ at 269.3 µm, give us access to the cold $H_2O$ reservoirs in the disk interior. Deep integration with the Herschel Space Observatory have detected these transitions in 3 disks only, TW Hya, HD 100546 and DG Tau (Figure 3-4, Hogerheijde et al. 2011 and in prep, Podio et al 2013) suggesting an overall low abundance of cold $H_2O$ in disks. The low abundance of cold $H_2O$ may be due with most of the $H_2O$ being in solid phase. A wide-band far infrared spectrograph is needed to determine the abundance of $H_2O$ ice in disks. The two most prominent ice bands peaks at λ ∼ 40 µm and 60 µm. The intensity, the shape and the flux ratio of the two ice bands provides us with information about the ice abundance, its structure (amorphous or crystalline) and formation temperature. Both ice bands have been previously detected in an handful of disks with ISO and Herschel and the $H_2O$ ice properties remain still unknown for the vast majority of protoplanetary disks.

In main-sequence debris disk systems, the water content of extrasolar comets can also be probed using far-IR lines. As described below, gas released in collisions between comets is detected in a handful of debris disk systems, which provides a way to probe the composition. Indeed, using the CII (157 microns) and OI (63, 145 microns) fine structure lines, in addition to CO observations, one can compute the $H_2O$/CO ratio. This calculation has been done for beta Pic (Kral et al. 2016), so far the only mature planetary system with these 3 key observations. A far-IR instrument would be key to increase our sample of known atomic carbon and oxygen gas disks and on-going ALMA observations are already increasing the number of systems with CO detected every year. This would provide for the first time a taxonomy of exocomets around main sequence stars. This work is complementary to the above studies of water during the protoplanetary disk phase, allowing us to "follow the water" throughout the lifetime of a planetary system.

The contribution of instruments like JWST (MIRI) or the E-ELT (METIS) will be very limited for this science case for several reasons. Both instruments can only observe high- energy level transitions (ro-vibrational transitions and high-J rotational transitions) with upper energy levels $E_u$ ∼ 1000K. These $H_2O$ lines come from the outermost layer of the inner disk and are not a probe of the global $H_2O$ budget. The 3 µm ice bands observable with JWST (NIRSPEC) can only be observed in absorption (only possible for a few systems where the disk is seen almost edge-on) or in emission through the scattering of small dust grains. Thus the 3 µm band is only sensitive to the ice present in the outermost layers.

## 3.4 Origin of gas in debris disks

The detection of gas in debris disks is a relatively new, but highly promising, phenomenon. These observations detect molecular gas (CO) and atomic species (such as CI, CII, OI and some metals). In most cases the lifetime of CO is short compared to the system age, so it is interpreted as being produced in cometary collisions (i.e. is secondary) rather than primordial (e.g. Marino et al. 2016). However, a few systems appear more likely to host a long-lived remnant of the primordial disk (Moor et al. 2013), and detailed study of these will provide new insight into the final dispersal of protoplanetary disks. Detailed study of secondary gas composition will yield insight into the parent bodies, for example constraining the composition of extrasolar comets and providing complementary information on the protoplanetary disk conditions in which they formed. The lifetime of CO in an optically thin environment is short, and therefore the bulk of the gas mass in such systems is expected to lie in the atomic species. These spread towards the star by as-yet unclear viscous mechanisms, thus pervading the entire system (Kral et al. 2016). Imaging and spectroscopy could map out the atomic gas distribution, and yield insight into fundamental disk processes such as the magneto-rotational instabililty (Kral & Latter). With sufficient spectral and/or spatial resolution and S/N, differences between the observed and expected atomic gas distribution may reveal the gravitational influence of planets.

An understanding of the origin of secondary gas and its evolution will be critical to further our understanding of planetary systems. While detections are restricted to just a few young stars, these are limited by sensitivity and the known sample could be increased by 1-2 orders of magnitude with a sensitive far-IR mission.

## 3.5 Dust composition

Dust plays a key role in the disk evolution and planet formation process. Because of its large opacity, the dust grains: control the temperature and density structure; shield the disk interior from the energetic radiation; regulate the ionization structure and helps the formation of ices and complex molecular species.

Thus, knowing the dust properties, and in particular its composition, is of fundamental importance to understand 1) planet formation and disk evolution processes and 2) the formation material of stars and planets.

Far-infrared spectroscopy is a powerful tool to characterize the dust composition in disks. As shown in Figure 3-5, the 35 – 90 μm spectral range provides a unique window to study several solid-state features of astrophysical and astrobiological relevance, including water, forsterite ($Mg_2SiO_4$) and enstatite ($MgSiO_3$) as well as carbon biomarkers as calcite ($CaCO_3$) and dolomite ($CaMg(CO_3)_2$). The compositional and structural properties of dust in disks can be precisely determined from these resonances. The (wavelength) position and width of the far-infrared resonances reflect the lattice structure and composition of the materials. In particular, far-infared observations allow us to accurately measure: the [Fe/Mg] ratio in olivine ($[Mg,Fe]_2SiO_4$) and pyroxene ($[Mg,Fe]SiO_3$); the lattice structure of pyroxene; the abundance and structure of water ice. The Fe/Mg ratio yields information about the dust alteration during the planet formation and planet evolution phases, therefore constrains the properties of the mineral's parent-body, for example its formation timescale, size and the efficiency of possible heating sources such as radioactive decay of unstable elements.

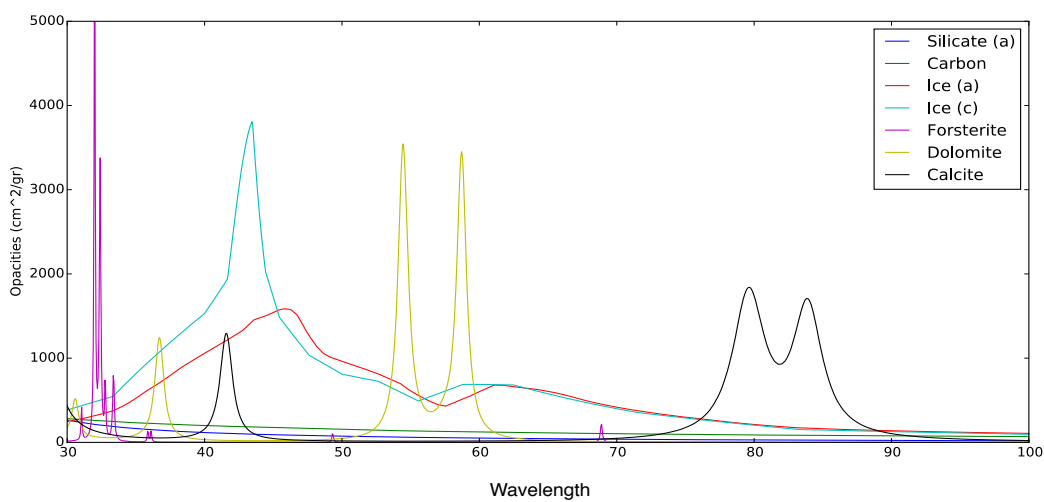

Figure 3.5 Opacity of dust minerals (© B. de Vries)

Far-infrared spectroscopic observations of protoplanetary and debris disks allow us to follow the dust evolution from the early stage of planet formation to the later stage of dynamical interactions, where a.o secondary gas-production occurs. These measurements will put our own Solar system with its asteroids and Kuiper Belt into a much broader context.

Previous observations with ISO (LWS), Spitzer (MIPS-SED) and Herschel (PACS) have detected only the strong 69μm band of crystalline olivine (e.g., Bowey et al. 2002, Su et al. 2015, De Vries et al 2012, Sturm et al. 2013, Blommaert et al 2014). This feature has only been detected towards two debris disks. The limited baseline stability of Herschel/PACS over a broad wavelength range has hampered the detection of other solid-state features.

Far-infrared spectroscopic observations of protoplanetary and debris disks would allow us to follow the dust evolution from the early stage of planet formation to the later stage of dynamical interactions.

## 3.6  The disk temperature structure

The knowledge of the initial density and temperature structure inside a protoplanetary disk is of major importance for the formation of planets. Sub-millimeter observations from the ground are well suited to determine the density structure because at these wavelengths the dust is optically thin allowing to study the cold disk midplane. The disk temperature structure instead is better investigated in the far-infrared and an ideal "thermometer" is the rotational ladder of CO and its rarer isotopologues ($^{13}CO$ and $C^{18}O$). CO is one of the most abundant species and its well understood chemistry makes it an ideal tracer of the gas properties in disks. The high-$J$ ($J_u > 10$) transitions of CO emerge from the warm molecular layer and the different rotational lines comes from slightly different orbital radii. The same is true for the other CO isotopologues with the main difference that, because of the lower optical depth (for the same $J$), $^{12}CO$, $^{13}CO$ and $C^{18}O$ forms a sequence in the vertical direction. Thus, far-infrared observations of multiple CO isotopologues will provide direct insights on the vertical (different $\tau$) and radial (different $J$) temperature structure of protoplanetary disks. As an example, Figure 3-6 shows the emitting region of 4 different CO and $^{13}CO$ transitions (Fedele et al. 2016): the low-$J$ transitions ($J = 3 - 2$ & $6 - 5$, observable from the ground) are mostly emerging from the outer, colder, part of the disk, while the high-J transitions provides information on the density and temperature structure down to the planet forming region.

## 3.7  The faintest debris disks – true Kuiper belt analogues

The locations and orbital structure of the Solar System's asteroids and comets provide strong constraints on the Solar System's history, a story that includes Neptune's outward migration, capture of Jupiter's Trojans, and the Late Heavy Bombardment of the terrestrial planets (e.g. Gomes et al. 2005). This story lacks context because true analogues of our Asteroid and Kuiper belts remain invisible. While all other stars must host debris disks at some level, only the brightest 20% are currently detectable, and we know neither our rank in the remaining 80%, nor how this rank is related to the Solar System's history. A decade from now we will have detected or set stringent limits on planets around most nearby stars, but the limits on small body populations will be as poor as they are now. A future far-IR mission can be designed to image true Kuiper belt analogues, completing the planetary

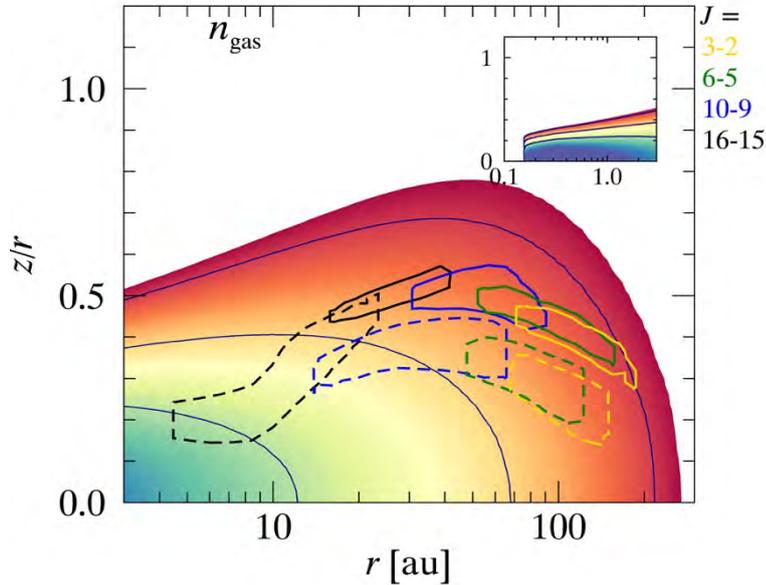

Figure 3.6 Line emitting region of a mix of low- and high-J $^{12}$CO and $^{13}$CO transitions for a disk around an Herbig Ae star (Fedele et al. 2016).

system inventory around nearby stars to the extent that we can place our own in context.

## 3.8 Telescope requirements

Each of the science cases outlined above has different telescope requirements. The primary requirement for the observations of gaseous transitions (CO, $H_2O$, HD,...) is spectral resolution: the gas lines are emitted on top of a bright infrared continuum and the line-to-continuum ratio is low (< 0.05); and from spectrally resolved lines it is possible to extract the velocity profile and measure the radial distribution of the gas.

To improve the line-to-continuum ratio a resolving power of > 5000 is needed. To resolve the line velocity profile a resolving power of > $3 \times 10^5$ is required.

On the other hand, observations of the water ice bands and of the solid-state dust features require a wide-band, low resolution ( $\sim 10^2$) spectrograph: the baseline stability over a wide wavelength range (35 – 100 μm) is very <u>critical</u> to detect the broad-band ice features, to measure their shape and relative flux.

Detection of true Kuiper belt analogues requires imaging capability. To survey O(100) nearby stars requires an estimated angular resolution of better than 2" at 60 microns, and an achievable contrast of better than $10^{-4}$ at 2λ/D.

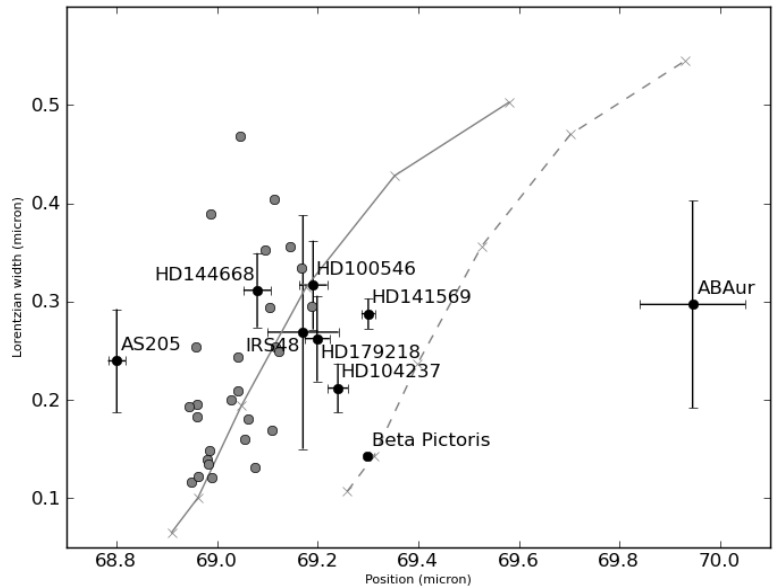

Figure 3.7 Diagram showing the dependence of far-infrared bands on grain temperature and composition. In this diagram the 69 μm feature is given as an example. It shows the width and central wavelength of the 69 μm band for six temperatures (from lowest to highest width: 50, 100, 150, 200, 300K) and for crystalline olivine ($Mg_{2-2x}Fe_{2x}SiO4$) with x = 0.0 (grey solid), x = 0.01 (grey dashed). The measured width and wavelength positions show how the band broadens and shifts as a function of temperature or iron content. The black dots are the positions of the 69 μm band in the Herschel spectra of protoplanetary disks and the debris disk of Beta Pictoris (de Vries et al 2012; Maaskant et al 2015).

# 4   The ISM in the Milky Way as a pathfinder

Stars form inside the molecular clouds of the interstellar medium (ISM) but we still don't fully understand what sets the stellar birth rate in a given galaxy. FIR photometric images of nearby giant molecular clouds (GMCs) of the Milky Way have shown that most of the star formation within GMCs takes place along filaments (Evans et al. 2009; André et al. 2010; Molinari et al. 2010). The dense filaments are embedded in a more diffuse medium in which turbulent dynamics seems to be driven on larger scales (Brunt et al. 2009). These extended regions constitute the bulk of the mass of GMCs, as much as 90% (McKee & Ostriker 2007), and thus must play a critical role in their evolution. The mechanical (winds and supernova explosions) and radiative (UV-radiation) feedback from stars themselves determines the properties of the ISM (energy balance, turbulence, chemical composition). By studying the structure, dynamics, and physical properties of the ISM as a whole we can begin to unravel the detailed processes that ultimately regulate the star formation in our Galaxy. Even before star formation occurs, cold and dense molecular clouds need to assemble from diffuse atomic clouds. Many questions still remain about this process. How do GMCs actually form? Which processes shape the diversity of observed structures? (halos, converging flows, filaments, cores, etc.). A new paradigm in which turbulence and magnetic fields are important actors emerges (Hennebelle & Falgarone 2012), but their specific role is largely unknown. Spectroscopic observations of large areas of the Milky Way at the appropriate angular and spectral resolution are challenging.

Observing in the FIR is essential, as most of the above radiative and mechanical processes produce strong FIR line and continuum emission. These include the thermal emission from dust grains heated by stellar radiation, the emission from the most relevant gas cooling lines that determine the energy balance, and a variety of atomic and molecular lines that arise from shocked material (e.g. van Dishoeck et al. 2011; Gerin et al. 2016). Because of their specific chemistry, some molecules emitting/absorbing in the FIR are also very sensitive to the turbulence dissipation (e.g. Godard et al. 2012) and to the ionization sources (cosmic rays, X-rays, etc.). Thus, they can be used to constrain the ionization rate (Neufeld et al. 2010; Indriolo et al. 2015). Velocity resolved observations can discriminate between the various heating processes. For instance, turbulent dissipation and shoks leqave their imprint on the line profiles. With the exception of the cold gas in pre-stellar cores (a very small volume of the ISM), most of the neutral interstellar gas (i.e. hydrogen in neutral form) in the Milky Way is found at low extinction ($A_V$<10) levels. Therefore, all of the neutral atomic and at least 90% of the molecular gas is permeated by stellar UV and visible photons and emits back in the FIR (Hollenbach & Tielens 1999). At the scales of an entire galaxy, the interaction between stellar radiation and interstellar matter results in strong FIR emission. Indeed, half of the luminosity of the Galaxy originates from these FIR photons from the ISM.

## 4.1   Big questions, FIR answers

The superb photometric capabilities of Herschel have provided panoramic views of large areas of the Galaxy by imaging the FIR dust emission and showing the projected structure of the ISM (e.g. André et al. 2010; Molinari et al. 2010; Schneider et al. 2012). The dominant gas coolants of UV-irradiated gas are emitted in the FIR. The [CII]158 μm line in particular is typically the brightest emission line of the ISM (Dalgarno & McCray 1972). These FIR lines are unique tracers of the physical conditions prevailing in diffuse clouds and in the extended component of denser GMCs (e.g. Hollenbach & Tielens 1999), the environment that sets the initial conditions for star formation. These lines can be used to measure star-forming rates and constrain feedback processes. At much smaller spatial scales, FIR lines can also be used to determine the cooling processes in individual protostars and protoclusters. FIR lines are especially sensitive to the properties of the shocked gas in protostellar outflows (e.g. Herczeg et al. 2012; Karska et al. 2013). However, only high-spectral resolution (velocity-resolved) line observations allow for decomposing the emission into its different phases and constituents (e.g. Langer et al. 2010; Pineda et al. 2013; Gerin et al. 2015), allowing us to characterize and reconstruct their properties and dynamics individually. Such observations are extremely limited (in particular for large-scale mapping). The power of velocity-resolved FIR imaging-spectroscopy, will help us to address crucial questions that are likely to bring a paradigm shift in our understanding of molecular cloud formation, the role of environment and feedback in star formation and its link to galaxy evolution; **(1)** Which processes transform diffuse atomic clouds into denser molecular clouds? **(2)** What fraction of giant molecular clouds is converted into stars during their lifetime and how does this depend on local conditions? (*Star formation efficiency and timescale*); **(3)** What internal sources of energy drive the dynamics of molecular clouds after their formation?

As we discuss below, only by mapping large areas of diverse environments (diffuse clouds, quiescent molecular clouds, triggered star-forming regions, galactic plane, high-latitude clouds, etc.) in key FIR diagnostic lines and in FIR dust polarization emission (i.e. magnetic field orientation) will allow us to start answering these questions. SOFIA is now making progress with questions 1 and 3 through heterodyne [CII] mapping, but it is limited by the total of observing hours per flight, thus leaving question 2 completely out of reach.

To complicate matters further, velocity-resolved images of cold molecular gas tracers in low-mass star forming regions show that some of the filaments observed in the (dust) photometric images are indeed formed by velocity-coherent sub-filaments or fibers (e.g. Hacar & Tafalla, 2013). This exacerbates the need for high-spectral resolution mapping of FIR tracers of the cloud environment ([CII], [CI], excited CO, etc.). Whether the same scenario scales to high-mass star forming regions is still uncertain. Together with the thermal instability, the magnetic field is another missing component in this paradigm. Despite the fact that magnetic energy is a significant fraction of the ISM energy budget, it remains too poorly constrained observationally (e.g. Crutcher 2012). Planck's low-angular resolution images of the magnetic field orientation in the ISM (from the dust polarized emission, see Figure 4-6) start to reveal the role of magnetic field at the large scales of the Milky Way (both in star-forming and starless clouds). *FIR dust polarization images at a few arcsec-resolution are clearly needed to move forward.*

While FIR photometric images of the dust thermal emission provide a static "snapshot" of the impact of the star formation process over entire molecular cloud complexes, it is only by pursuing large scale maps of key spectrally resolved lines that we can probe the extended cloud/filament/core dynamics and kinematics in detail. Each of these physical stages leaves a particular signature in the prevailing physical conditions and chemistry, and vice-versa, the particular physical conditions and chemistry influences the cloud evolution (through gas cooling, ionization fraction, coupling with the magnetic field, etc.). WISE MIR- and Herschel FIR-photometric cameras have imaged a significant fraction of the Milky Way with angular resolutions down to ~35'' (Herschel/SPIRE at 500μm). Dust SED analysis has allowed the construction of $H_2$ integrated column density maps in many star-forming clouds. Unfortunately, similar maps of the key gas tracers at similar angular resolution and size as those with Herschel/SPIRE, do not exist while Planck's angular resolution is arcminutes rather than arc seconds **(Figure 4-6)**. Velocity-resolved maps in the bright FIR [NII], [CII], [OI], and [CI] lines at comparable angular resolution to Herschel photometric maps will**:**
**1)** provide access to the kinematics and turbulence of the emitting gas in the Milky Way,
**2)** constrain the thermal properties of the gas (linking with the different energy sources),
**3)** enable the study of global dynamics of the gaseous disks.
**4)** allow detailed scaling-laws of these bright FIR line diagnostics with the FIR dust, PAH, and CO luminosities that can later be used to understand and characterize their emission from the distant universe with ALMA, once the excitation is understood.
*Velocity-resolved, large-scale maps of the Milky Way of the bright FIR fine structure lines, will add the gas kinematics and turbulence piece of information needed to progress in our understanding of ISM cloud formation and evolution.*

## 4.2 The physical processes governing the different gas phases in the ISM energy cycle

The ISM is a key component of galaxies and plays a pivotal role in their evolution. It is the reservoir of baryonic matter, and, as galaxies evolve, the constituents of the ISM are gradually converted into stars. A fraction of the enriched products of stellar evolution goes back to the ISM through stellar winds and supernova explosions. Thus, the life cycle of the interstellar matter itself (atoms, molecules and dust grains) is closely related with that of the stars. This is a *non-equilibrium* system dissipating all types of energy forms (kinetic, thermal, stellar radiation, magnetic, self-gravity and cosmic-ray energy) injected in the ISM. It is far from understood how these different energy sources contribute to the dynamical and thermal properties of the ISM, but eventually, all the injected energy is radiated as photons emitted by dust grains and by specific gas constituents. The FIR is the key band to characterize this radiative cooling (and indirectly the energy sources and feedback).

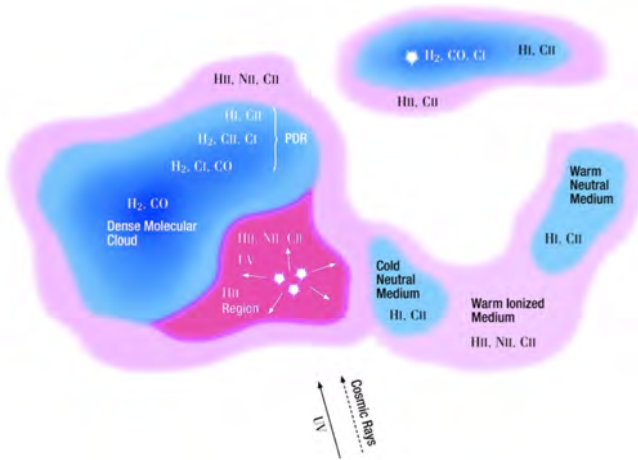

Figure 4-1 The various phase of the ISM together with the atomic, ionic and molecular lines that are produced from each phase (credit J. Pineda).

Most of the radiation emitted from a Milky-Way type galaxy is either in the visible (i.e., from stars) or in the FIR where dust in the ISM absorbs and re-radiates the starlight. Despite their important role, only ~1% of the ISM mass is in dust grains, 99% of the mass is made up of gas. The gas in the ISM in our Galaxy and in external galaxies can be found either in atomic, ionized or molecular state, and appears in a very wide range of densities and temperatures. Most of the ISM can be characterized by a few "*phases*" determined by their ionization and thermal state (how is the medium heated and cooled). Ionized gas can be found as "coronal" gas in the "hot ionized medium" (HIM, collisionally-ionized gas at $T\approx10^6$K and $n\approx0.04$cm$^{-3}$) and as "Warm ionized medium" (WIM, at $T\approx10^4$K gas in which most of the hydrogen is in H$^+$, and includes diffuse intercloud gas with $n \approx 0.3$cm$^{-3}$ and denser "HII regions" with $n\approx10^4$cm$^{-3}$ photoionized by nearby massive stars). In addition, two neutral atomic phases approximately coexist at roughly thermal pressure equilibrium (e.g. Wolfire et al. 2003); the "Warm neutral medium" (WNM, with $T\approx10^4$K and $n\approx0.3$cm$^{-3}$) and the "Cold neutral medium" (CNM, $T\approx 80$K and $n\approx 40$cm$^{-3}$, see Figure 4-1). As we show below, this is the domain where key FIR cooling lines dominate the thermal state of the gas. Indeed, by mechanisms that are not yet fully understood, the very extended cold atomic clouds form high-pressure filamentary molecular (most hydrogen is in H$_2$) clouds with $T\approx 10$-$50$K and $n\approx 10^{3-6}$cm$^{-3}$. Dense cores inside molecular clouds become gravitationally unstable, and they gradually collapse and form stars. Understanding the neutral phases of the ISM and their dependence with the various energy forms and cooling mechanisms is a major step in understanding the formation of clouds themselves, and the global star formation rates in galaxies. In recent years however, it has become apparent that the traditional distinction of the ISM into these well-separated, thermally and chemically stable phases does not reflect the dynamic nature of the ISM evolution. A large fraction of the gas can be found in transitional regions. To obtain a full inventory of the interstellar gas, observations of all phases and the transitions between them are needed.

The ionized gas has been traditionally studied through radio continuum, radio recombination lines and H$\alpha$ observations (Haffner et al. 2009), the neutral atomic gas with observations of the HI 21cm line (e.g. Kalberla & Kerp 2009), and the rotational transitions of the CO molecule have been used to study the cold molecular gas (e.g. Dame et al. 2001).
The volume occupied for the different phases scales inversely with the gas density. Thus, most of the baryonic mass is in the *neutral* phases (the relevant material for star formation), whereas most of the volume is in the *hot* and *warm* phases. The ISM is very dynamic, change of phases occur, and multiple interfaces (or boundaries between phases) appear along a given line-of-sight (e.g. the HI line at 21cm traces the total hydrogen atom column density but is nearly impossible to distinguish between the WNM and the CNM). Unfortunately, even for our galaxy, most of the relevant processes and parameters discussed above are still very controversial and debated. In particular, it is crucial to constrain the masses, filling factors, energy-balance, ionization sources, dynamics and topology of each of the phases that make up the ISM throughout the Milky Way. In addition, this will help us understanding how and at which rate the ISM is converted into stars.

### 4.2.1 Characterizing and studying the ISM phases and their common boundaries

The neutral carbon atom has an ionization potential of 11.3 eV (below that of hydrogen), so that the ion C$^+$ traces the H$^+$/H/H$_2$ transition, the critical conversion from atomic to molecular ISM. As a consequence, [CII] 158μm emission/absorption from different ISM phases is expected along a given sightline. The [CII] line is the most important gas coolant of the CNM (Dalgarno & McCray 1972). It is also a key tracer of the surfaces of the much denser (star-forming) molecular clouds illuminated by UV photons from nearby massive stars, the so-called photodissociation regions or PDRs (cf., Hollenbach and Tielens 1999). The [CII] line dominates the cooling of low density (<10$^4$ cm$^{-3}$) and low far-UV field PDRs (<10$^4$ times the far-UV radiation field in the diffuse ISM). For the higher densities and FUV fields found locally in PDRs surrounding HII regions around newly formed massive stars, the [OI]63μm line is more luminous (it has a higher critical density). In these HII/cloud interfaces, both [OI] and [CII] lines are the dominant gas coolants. At the larger spatial scales of a whole star-forming molecular cloud (a few parsec size), or even of an entire galaxy, the [CII] line is again the dominant interstellar emission, carrying up to 1% of their total FIR luminosity (Crawford et al. 1985). On a very large-scale (~7 degree angular resolution), FIRAS/COBE observations of the Milky Way have shown that the [CII] line is the strongest cooling line in the ISM at about 0.3% of the infrared emission (Bennett et al 1994). These are very important results, as bright atomic FIR fine structure lines are easily detectable and provide unique information on the stellar UV field, the gas density and temperature, and, if spectrally-resolved (i.e velocity-resolved), the ISM gas dynamics.

There is a long debate on which of the ISM components, including the WIM, the CNM and the externally illuminated surfaces of molecular clouds dominates the [CII] emission in any given galaxy. Only high spectral-resolution observations of [CII] can tackle this problem. Recent velocity-resolved line surveys in the Milky Way, carried out with HIFI, suggest that the average contribution to the [CII] emission from molecular gas illuminated by FUV photons is ∼55%–75%, the cold atomic gas ∼20%–25%, and the HII ionized gas ∼5%–20% (Pineda et al 2014, Figure 4-2). Unfortunately, Herschel surveys only sampled a very limited number of sightlines in the Milky Way, so a large statistical support is missing. In fact, these fractions likely change from galaxy to galaxy as metallicity and star formation histories change.

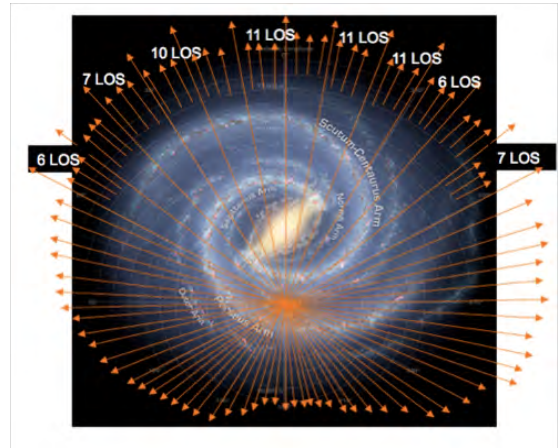

Figure 4-2 C+ observations along a number of lines of sight towards the Galactic Plane (figure taken from Pineda et al. 2014).

Nitrogen is the most abundant heavy element after oxygen and carbon. Nitrogen atoms have an ionization potential of 14.5 eV, and owing to the their low critical densities (a few tens of electrons $cm^{-3}$), the [NII] 122, 205μm fine structure lines trace the extended, low-density WIM and also the localized and denser ionized gas in HII regions surrounding massive stars. The [NII] 205μm fine structure line is an excellent probe to characterize the ionized gas and is also very bright in star-forming regions (typically a factor ∼10 fainter than [CII]). Used in tandem with the [CII] 158μm line, the properties and dynamics of the ionized and neutral interfaces can be constrained. In particular, a good estimation of the C+ fraction in ionized gas, and a more precise estimation of the stellar UV-field (thus indirectly the star-formation rate) can be achieved if both [NII] and [CII] lines are observed simultaneously (e.g. Persson et al. 2014, see Figure 4-3). Resolving their line profiles in the Doppler velocity space is needed to distinguish the different components and phases producing the C+ and N+ emission/absorption.

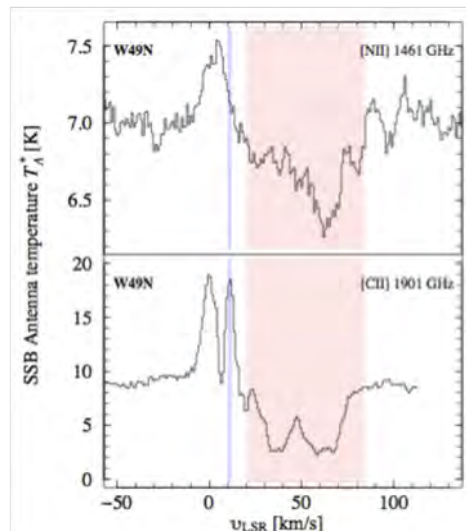

Figure 4-3 Spectra of N+ 205 μm and C+ 158 μm towards W31C. The $v_{LSR}$ of the HII region is marked in blue and the velocity of the LOS gas is marked in red (from Persson et al 2014).

*Although Herschel demonstrated the diagnostic power of velocity-resolved [CII] and [NII] line observations toward a few sightlines (Figure 4-3, e.g. Persson et al. 2014), Herschel was not designed to carry out large-scale spectral-mapping of these lines. Thus, the spatial and velocity distribution of the key FIR cooling lines (at the relevant angular resolution for Milky Way's studies) remains unknown.*

## 4.3 "CO-dark" molecular gas

Stars are born in clouds where most of the mass is in molecular form ($H_2$). Molecular clouds are the fuel for star formation and thus it is crucial to constrain the mass of a given galaxy contained in molecular gas. This has been traditionally done by mapping the lowest energy transitions of CO in the millimeter domain, and by assuming a (admittedly uncertain) conversion factor from CO luminosities to $H_2$ masses (the "X factor").

The detection of an excess γ-ray emission, produced by the interaction of cosmic rays and molecular clouds, with respect to predictions based on the ISM column density as traced by HI and CO, led Grenier et al. (2005) to the existence of so-called "CO-dark molecular gas" (see also Abdo et al. 2010), just as earlier [CII] measurements of IC10 by Madden et al. 1997. In this view, molecular clouds are surrounded by large envelopes of atomic gas and by a transition region where $H_2$ molecules start to form and survive as dust extinction and $H_2$ self-shielding attenuates the UV flux. This component where hydrogen is primarily molecular but most carbon is still in $C^+$ and not in CO is known as "CO-faint" or "CO-dark $H_2$ gas", and bridges the local dense star-forming clumps to the much more extended cold diffuse atomic clouds. Although it may contain a significant fraction of the baryonic matter in a galaxy, the "CO-dark" molecular gas component is difficult to trace observationally ($H_2$ is a symmetrical molecule and radiates inefficiently).

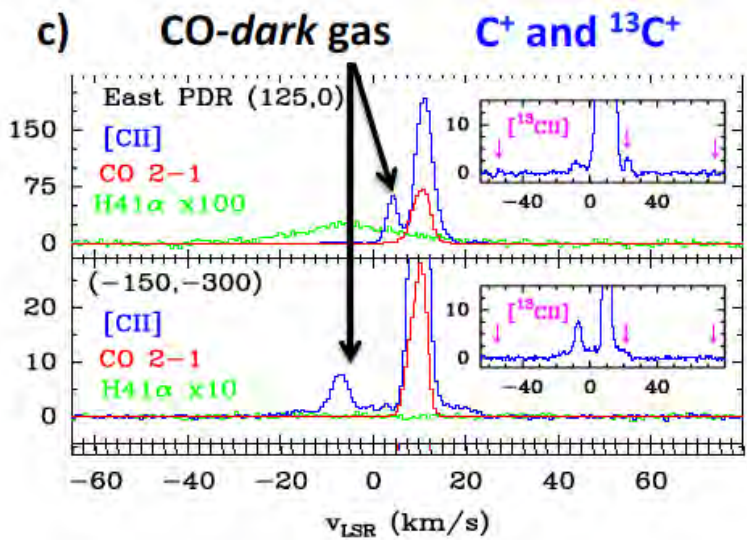

Figure 4-4 High spectral resolution observations (<1 km/s) allow resolving line profiles, studying the gas kinematics and characterising the C+ emission from the various ISM phases, including the CO-dark molecular clouds. Taken from Herschel/HIFI observations of the Orion cloud (Goicoechea et al. 2015a).

Herschel (pointed) observations along a very limited number of sightlines in the galactic plane suggest that $C^+$ is the best tracer of this CO-dark molecular gas, accounting for ≥1/3 of the [CII] emission in the Milky Way (e.g. Langer et al. 2014, Pineda et al. 2014). Note that only very high-spectral resolution line observations ($R>10^5$) can resolve the $C^+$ emission from the different gas phases (CNM, WNM, WIM). Unfortunately, while Herschel observations suggested the importance of CO-dark molecular gas, the degree-scale spatial resolution accessible to γ-ray observatories as well as the coarse spatial sampling of the Herschel survey, do not allow to quantitatively addressing the relative budget of *dark* gas in the molecular ISM component. The fine structure lines of neutral atomic carbon ([CI]370,610μm), although weaker than the [CII] line, are unique tracers of the C/CO transition in UV-illuminated molecular clouds. Hence, they add more constraints to the CO-dark gas characterization.

*Many questions still remain after Herschel discoveries; what is the mass, filling factor and physical properties of the CO-dark gas in a given Galaxy? What is the molecular gas mass one misses by observing the low-J CO lines alone? Obtaining large-scale, velocity-resolved maps of the FIR fine structure lines, combined with HI and CO observations, is clearly needed.*

## 4.4 The large-scale view of star-forming regions in the Milky Way

Spitzer, Herschel and Planck have revolutionized our understanding of the physical properties of molecular clouds and star forming regions in the Milky Way. In particular, Herschel FIR and submm photometric images of Giant Molecular Clouds (GMCs, Figure 4.5) have revealed spectacular networks of filamentary structures with chains of embedded cold cores where stars are born (e.g., André et al. 2010, Molinari et al. 2010).

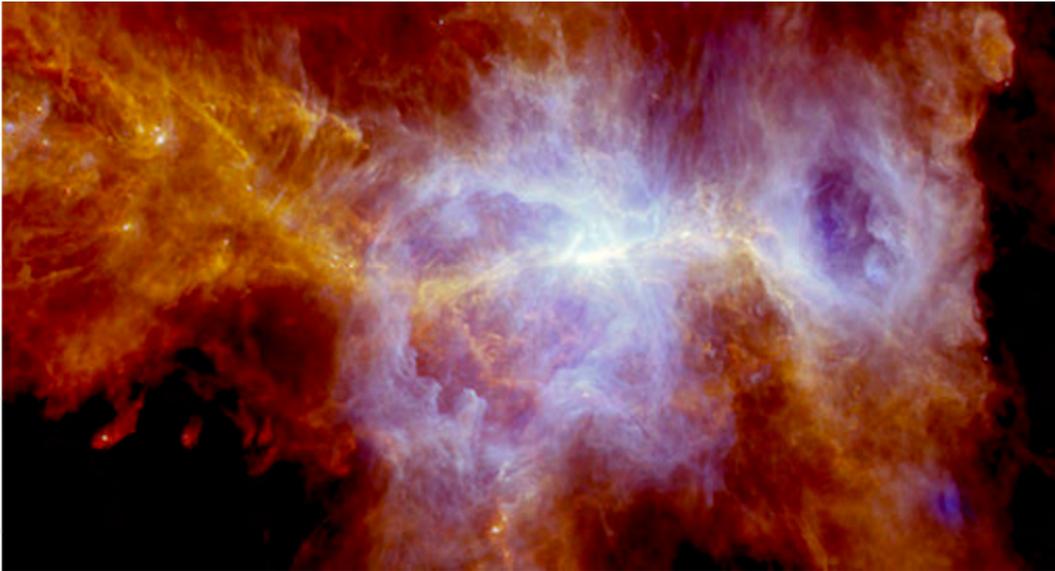

Figure 4.5 Filamentary network in the Orion A molecular cloud (ESA/Herschel Andre et al for the Gould Belt survey Key Programme.

Newborn OB massive stars release large quantities of energy as ionizing radiation and strong winds erode and disrupt their surroundings. Mediated by the poorly known effects of the magnetic field (e.g. Crutcher 2012), these processes induce a variety of thermal and hydrodynamic instabilities, powering turbulence, gas compression and chemical mixing (Hennebelle & Falgarone 2013 and references therein). These processes drastically modify and shape the parental molecular cloud: heating the gas and dust, creating elongated structures and pillars, excavating cavities and producing gravitationally unstable structures.

Nevertheless, while FIR photometric images of the dust thermal emission provide a static "snapshot" of the impact of the star formation process over entire molecular cloud complexes, *it is only by pursuing large scale maps of key spectrally resolved FIR lines that we can probe the extended cloud/filament/core dynamics and kinematics in detail*. Each of these physical stages leaves a particular signature in the prevailing physical conditions and chemistry, and vice-versa, the particular physical conditions and chemistry influences the cloud evolution (through gas cooling, ionization fraction, coupling with the magnetic field, etc.). WISE MIR- and Herschel FIR-photometric cameras have imaged a significant fraction of the Milky Way with angular resolutions down to ~35'' (that of Herschel/SPIRE at 500μm). Dust SED analysis has allowed the construction of $H_2$ integrated column density maps in many star-forming clouds. Herschel results support a "filamentary paradigm" for star-formation in two main steps (e.g. André et al. 2014): First, large-scale compression of interstellar material in supersonic MHD flows generates a co-web of ~0.1 pc-wide filaments in the ISM (this width seems to be surprisingly constant in most observed regions); second, the densest filaments fragment into pre-stellar cores (and subsequently proto-stars) by gravitational instability above the critical mass per unit length of nearly isothermal, cylinder-like filaments. Whether the exactly same scenario scales to high-mass star forming regions is still uncertain (Schneider et al. 2012). It has also been argued that filaments may help to regulate the star formation efficiency in dense molecular gas and may be responsible for a quasi-universal star formation law in the dense ISM of galaxies (cf. Lada et al. 2012).

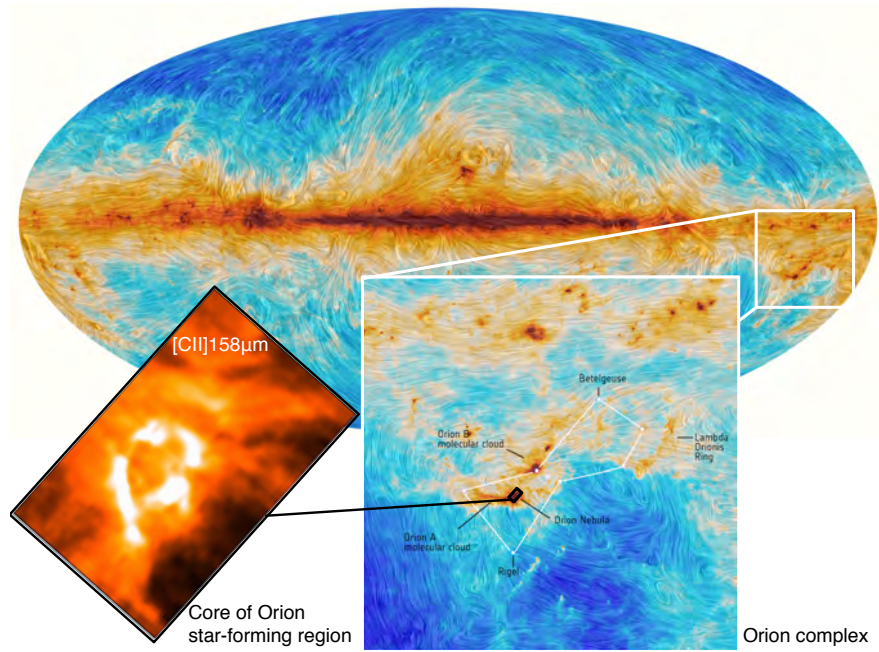

Figure 4-6 Interaction between interstellar dust in the Milky Way and the structure of the magnetic field. The color scale represents the dust thermal emission from interstellar clouds and star-forming regions. The texture is based on measurements of the direction of the polarised light emitted by the dust, which in turn indicates the orientation of the magnetic field (from Planck collaboration 2014, A&A, 586, 135). The inset shows an example of the limited size of velocity-resolved C+ maps currently available (a few hundred square arcmin).

Unfortunately, similar large-scale spectral-maps of the key FIR gas cooling lines at similar angular resolution do not exist (Figure 4-6). Velocity-resolved images of cold molecular gas tracers showing that some of the filaments observed in the dust formed by velocity-coherent sub-filaments or fibers (e.g. Hacar & Tafalla, 2013) complicate the interpretation of single-width filaments and reinforce the need for high-spectral resolution mapping of FIR tracers of the gas properties (physical conditions and kinematics) and energy sources at large scales (the environment).

In parallel, the Planck mission has led to major advances in our knowledge of the magnetic field geometry on large scales. The first all-sky-maps of dust polarized emission with 5' resolution (0.2 pc at the distance of the closest clouds) provided by Planck have revealed a very organized magnetic field on ~1-10 pc scales in Galactic interstellar clouds (See Figure 4.8 and Planck Collaboration Int XIX 2015). In particular low-density filamentary structures of the diffuse ISM tend to be aligned with the magnetic field (Planck Collaboration Int. XXXII 2016), whereas dense star-forming filaments tend to be perpendicular to the magnetic field (Planck Collaboration Int. XXXV 2016). This suggests that the magnetic field plays an important role in the dynamic formation of structures in the ISM. Unfortunately, the angular resolution achieved by Planck is far from that of Herschel photometric images and the comparison is not straight forwards. Hence, both velocity-resolved maps of the brightest FIR interstellar lines together with FIR dust polarization images at comparable angular resolution (≤35'') are needed to:

**1)** Resolve the kinematics and of the ISM gas (dynamics and turbulence),
**2)** Constrain the properties (and indirectly the energy and ionization sources) of the gas containing most of the mass in interstellar clouds.
**3)** Determine the topology and role of the magnetic field from diffuse to star-forming clouds.
**4)** Allow detailed scaling-laws of these bright FIR line diagnostics with the FIR dust, PAH, and CO luminosities that can latter be used to understand their emission from the distant universe (e.g. ALMA detections of [CII] and [NII] lines at very high red-shift).

*Large-scale, velocity-resolved maps of the Milky Way in the bright FIR fine structure lines will add the gas kinematics and turbulence piece of information needed to progress in our understanding of molecular cloud formation and evolution. Accessing the topology of the magnetic field will provide clues on how the ISM is structured and how the diffuse atomic clouds are converted into denser filamentary molecular clouds.* ALMA cannot access the FIR line emission from the Milky Way, and indeed ALMA is not designed to carry out all-sky maps of the extended interstellar emission. In addition, obtaining all-sky sensitive FIR polarimetric images at the dust SED peak of diffuse atomic and molecular clouds ($\lambda \sim 100$-$350 \mu m$) clearly requires dedicated space observations.

## 4.5  Massive stars and the impact on their surroundings

### 4.5.1  Understanding High Mass Star Formation

High-mass stars control most of the chemical and dynamical evolution of the ISM in galaxies, how they form, however, remains largely unknown. Do they form following a scaled-up version similar to that of low-mass star formation mechanisms involving turbulence and monolithic collapse (e.g. Zinnecker & Yorke 2007)? Do they form by competitive accretion (e.g. Tan & McKee 2002)? Massive stars are indeed rare and distant (a few kpc in the Milky Way) and evolve rapidly as they start nuclear fusion (thus radiating strong UV fields) while still accreting mass. Therefore, in order to understand their formation we need to observe sufficiently large samples. High-mass proto-stars and clusters are embedded in large quantities of gas and dust. Hence, they need to be observed at long-wavelengths where extinction is not an issue. FIR images show that most massive clusters lie at junctions of filaments (Schneider et al. 2012), but how these filaments are formed, and what controls the flow of material is debated. FIR spectroscopic observations are especially suited to understand the UV-irradiated gas dynamics in high-mass star-forming regions. Herschel was not able to provide large-scale FIR spectral-images, and only a few pointed spectroscopic studies were carried out.

The outflows and envelopes around individual massive protostars and clusters also emit copiously in the FIR. Together with the MIR $H_2$ lines (often obscured by extinction), the FIR rotationally excited lines of CO, $H_2O$ and OH are the most important coolants of the shocked gas, hence, they are an excellent filter to detect and characterize protostellar activity (Karska et al. 2014, Goicoechea et al 2015b). An example of the rich FIR spectrum from outflows around massive protostars is shown in Figure 4.7. These lines are also expected to arise from their circumstellar disks. A detailed spectroscopic characterization of individual protostars (disk and outflow cavitiy), however, would require reaching at least the challenging ~1" angular-resolution at FIR wavelengths (1"≈500AU at the distance to Orion, but 1"≈5,000 AU at the ~5 kpc typical distance of high-mass star-forming regions in the Milky Way).

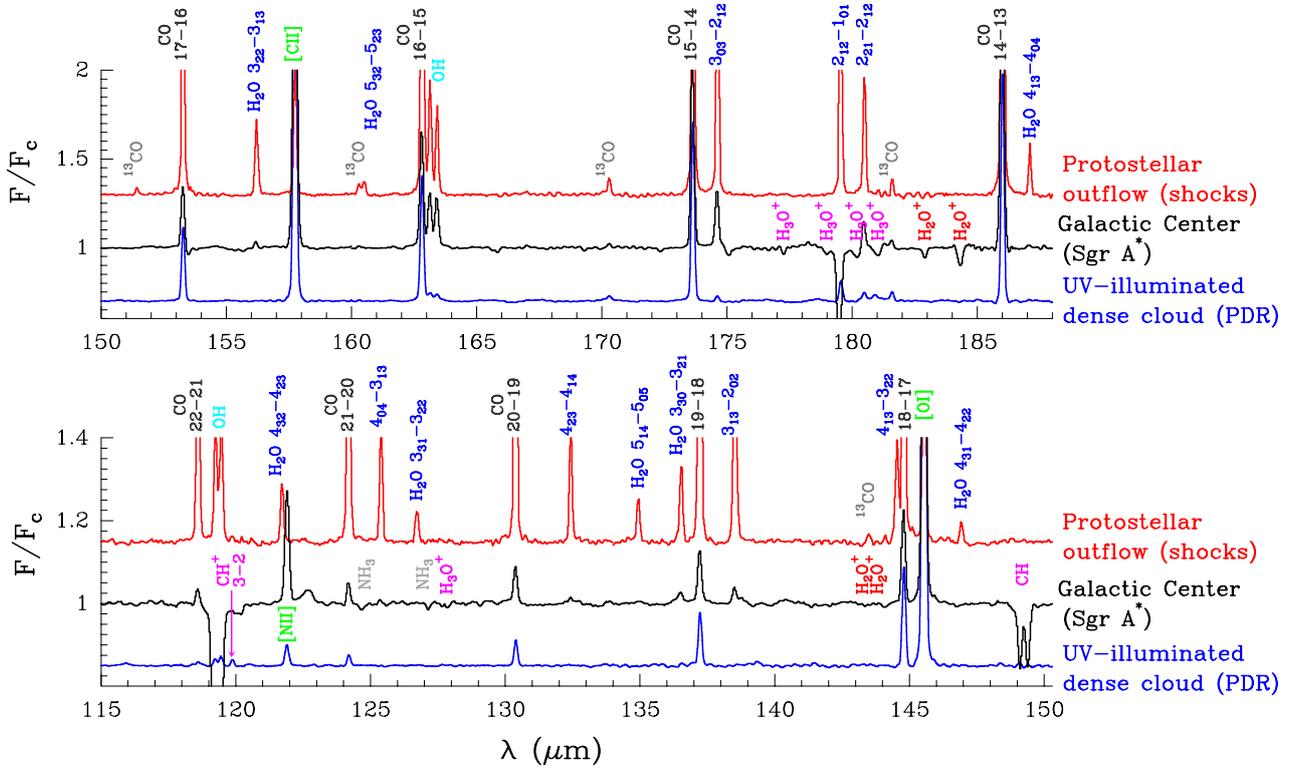

Figure 4.7 Comparative far-IR spectra of three template environments in the Milky Way: outflows around massive protostars (red), the Galactic Center (black), and a highly UV-irradiated PDR, the Orion Bar (blue). Observed by Herschel/PACS (Gerin, Neufeld, Goichoechea, ARAA 2016, and Joblin et al. in prep.).

### 4.5.2 Feedback from massive stars

One of the fundamental issues in understanding massive star formation is to quantify the magnitude and impact of the feedback processes induced by massive stars on their environment, and on the star formation rate in general. Previous continuum and CO surveys of the Galactic Plane have provided a complete map of the cold molecular gas in star forming regions. The filamentary structure testifies the strong dynamical effects, from the large-scale Galactic structure, the bar and spirals arms and the associated converging flows leading to the formation of molecular clouds. On smaller scales, the formation of massive stars provides an important *negative feedback* both radiatively, through the intense UV radiation, and dynamically, through the stellar winds, outflows and later the supernovae shocks. On the other hand, the expanding HII regions around young massive stars can also accelerate and shock the neutral material in adjacent molecular clouds, compressing the gas to high densities and triggering the formation of dense clumps that may ultimately form a new generation of stars (*positive feedback*, e.g. Hosokawa & Inutsuka 2006). Thanks to FIR photometric observations with Herschel, statistics shows that the column density probability distribution functions (PDF) around massive clusters and HII regions often show lognormal distributions (likely dominated by turbulence effects) and a high-density power-law tail at about $A_v$>8 mag (likely dominated by gravitational collapse). In some regions, the PDFs are consistent with the gas compression being induced by the effects of the expanding HII regions (Schneider et al. 2012, Tremblin et al. 2014). Determining the gas kinematics in the HII/cloud interfaces is a missing requisite to understand the true dynamics of these UV-irradiated regions.  High resolution spectral-maps of the bright FIR [CII], [OI], [CI] and rotationally excited CO lines, covering the same areas imaged by Herschel are needed to quantify the radiative and mechanical feedback of massive stars on their environment. The knowledge of the line profile is mandatory to isolate the different excitation processes, determine the role of turbulence and interaction of magnetic field.

[CII] and [OI] emission in star-forming regions also arises from dense PDRs at the interface between HII regions and their parent molecular cloud (e.g., Stacey et al. 1993). Combined with the FIR continuum, both lines can be used to trace the UV radiation field from massive stars and, indirectly, estimate the star formation rate (e.g., Stacey et al. 2010; Pineda et al. 2014; Rigopoulou et al. 2014, Magdis et al. 2014, Herrera-Camus et al. 2015). One of the too few examples in which the [CII] line has been mapped with Herschel/HIFI at high angular- and spectral-resolution at large scales (only about 10'x10') is the region surrounding the Trapezium stellar cluster in the Orion Nebula. *Velocity-resolved images of the [CII] and [$^{13}$CII] lines (0.2 kms$^{-1}$ resolution), combined with FIR dust thermal emission images, have provided an unprecedented view of the intricate small-scale kinematics of the ionized/PDR/molecular gas interfaces and of the radiative feedback from massive stars* (Goicoechea et al. 2015a, see their Figure 18).  For most massive star-forming regions, however, only small-scale mapping in the vicinity of very bright massive protostars and clusters have been carried out so far with HIFI and SOFIA/GREAT (e.g. Perez-Beaupuits et al. 2012; Beuther et al. 2014; Gerin et al. 2015). However, most of the [CII] luminosity is expected to arise from the extended cloud component.  Indeed, this UV-illuminated widespread gas in GMCs is likely to resemble the unresolved emission detected from distant Milky Way type galaxies. The physics is expected to be similar, but the properties of such regions have largely been unexplored mainly because of the lack of suitable instruments to measure large-scale [CII] emission with the required sensitivity, angular resolution  (tens of arc second) and spectral resolution (< 1 kms$^{-1}$) in the Milky Way.

Recent SOFIA observations of [CII] and the two [OI] lines in protostellar objects (identified as cold, deeply embedded submillimeter cores with luminous SPIRE emission) has made the surprising discovery that sources with comparable dust temperatures have [CII] emission strengths that differ by more than a factor of ten, and equivalent widths differing by more than a factor of twenty.  Similarly, the [CII]/[OI] ratio varies unpredictably between sources. Moreover, the results appear NOT to be correlated to the source luminosity, mass, temperature, dust column density, or other readily identifiable characteristic.  Instead, there are indications from submm spectroscopy that the [CII] emission is correlated to the chemical evolution of the embedded protostar.  A mission with high spectroscopic resolution and sensitivity together with a proper understanding of the excitation processes involved, enables us to use these abundant atomic species as a powerful new diagnostic of the earliest, deeply embedded phases of star formation in the Milky Way and beyond.

*The Galactic Center deserves special attention*, as it is the closest galactic nucleus we can study below the ~1pc scale in the FIR (e.g. Molinari et al. 2011). The properties of the ISM in the inner ~100pc of the Milky Way are markedly different from the galaxy disk. Widespread shocks, high-energy radiation, enhanced magnetic fields and strong tidal forces, all shape a very singular interstellar environment. Despite its uniqueness and relevance in a broader extragalactic context, the properties of the Galactic Center gas and dust that survive these harsh conditions and fuel star formation there are not fully understood. Owing to the lower extinction effects in the FIR compared to MIR observations, and because of the strong FIR emission from the interstellar component related to AGN and star formation (atomic fine structure lines and excited lines from CO, $H_2O$, OH, hydride ions...), the relevance of FIR spectroscopic observations to characterize extragalactic nuclei properties has greatly improved thanks to Herschel (see the low spectral resolution spectrum in Figure 4.7). Herschel/PACS has provided low-resolution spectral-maps of the brightest FIR lines around SgrA* (e.g. Goicoechea et al. 2013). Velocity-resolved line maps of the entire Galactic Center region in these FIR spectral line diagnostics (a 3D survey), will unveil the energy sources and true gas kinematic in this emblematic region and provide an unique template for extragalactic nuclei studies.

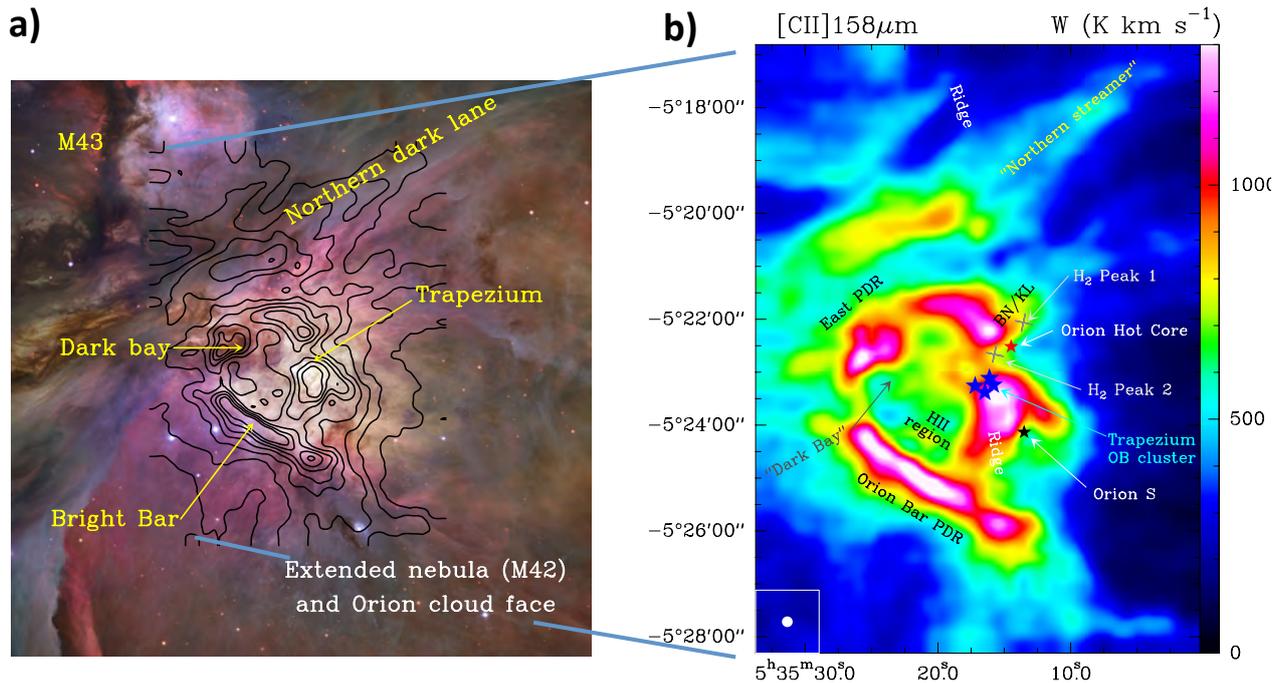

Figure 4.8 Spatially and velocity resolved observations of the [CII]158 μm line from ionized atomic carbon, of a ~7.5x11.5 arcmin field towards the Orion nebula observed by Herschel/HIFI at ~11'' resolution. The resulting data cube (or '3D view'), together with ancillary FIR dust emission and CO data provides a new view of the iconic Trapezium region, our closest high-mass star-forming region. a) Integrated [CII] emission over an image of the visible-light obtained by the Hubble Space Telescope. b) Herschel-HIFI map of the total [CII] 158 microns integrated line intensity in K kms$^{-1}$ from $v_{LSR}$ -30 to +30 kms$^{-1}$. (Goicoechea et al. 2015a)

## 4.6 Origin of dust grains from evolved stars and supernovae ("star dust")

Despite its importance in the physics of galaxy evolution, exactly how dust accumulates in their ISM is still not well understood. Asymptotic giant branch (AGB) stars, low- and intermediate-mass (1-8 $M_\odot$) evolved stars, have been confirmed to be the site of dust formation (e.g. Habing 1996; Blum et al. 2006; Cox et al. 2011; Ventura et al. 2012; Kervella et al. 2015). Dust grains formed via condensation within stellar winds of AGB stars are slowly expelled, and eventually incorporated into the ISM. Most recent studies of galaxies suggest that dust grains formed from AGBs constitute only a fraction of the dust in the ISM: no more than 50% in the Milky way (Dwek 1998; Tielens 2009), and only a few percent in the Magellanic Clouds (Matsuura et al. 2009; 2013; Boyer et al. 2012; Schneider et al. 2014).

The problem is even more severe at high-redshift, because these galaxies are too young for evolved AGB stars to be the dominant source of dust (Valiante et al. 2009; Michałowski et al.; 2010; Mattsson 2011; Michałowski 2015; Mancini et al. 2015). Hence, there is a crisis in the accounting of the total dust budget of the ISM, and at all redshifts other dust sources need to be identified to resolve this mass deficit.

In order to resolve this dust budget conundrum, two major alternative theories have been proposed. The first is core collapse supernovae (SNe) (Morgan & Edmunds 2003; Dunne et al. 2003; Maiolino et al. 2004; Sugarman et al. 2006; Dwek & Cherchneff 2011). The second is that AGB stars and SNe provide dust grain seeds for further grain growth in the ISM (Draine et al. 2009; Tielens 2009). Both scenarios are extremely challenging to test observationally. We focus on the dust mass generated from SNe, which is better quantifiable and in reach for a next FIR mission.

The quantity of dust formed in SNe is still controversial. So far, dust excess has been detected in over ten supernova remnants within a few years of the explosions, with a typical reported mass of only $10^{-6}$ to $10^{-4}$ $M_\odot$ (e.g. Wooden et al. 1993; Kotak et al. 2009; Gall et al. 2011). Supernova 1987A is a unique object whose dust mass has been measured over the past 25 years. Its initially reported mass was at least $\sim 10^{-4}$ $M_\odot$, 2 years after the explosion (Wooden et al. 1993), but 24 years later, the reported mass reached about half a solar mass (Matsuura et al. 2011). It is the only case where such a large amount of dust was measured in the late phase, so far. Nowadays, there is an alternative interpretation of the dust production in the early days of supernova evolution, a supernova can make half a solar mass of dust within a few years, but the large optical depth makes it difficult to measure the dust mass at mid-infrared wavelengths (Dwek & Arendt 2015). Whichever the case, observations at FIR wavelengths are crucial because the optical depth is much lower than those at MIR wavelengths.

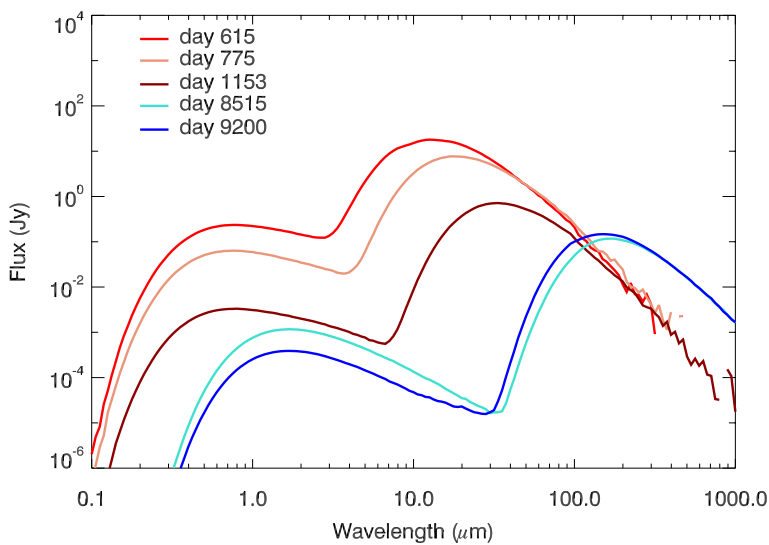

Figure 4.9 Time evolution of the dust in supernova 1987A fitted to observed SED (Wesson et al. 2015). In early days (day 615 and 775), the inferred dust mass was only a few x$10^{-3}$ $M_\odot$, while it increased to approximately half a solar mass in 20 years, although there is an alternative solution that dust mass was much larger but optically thick in mid-infrared (Dwek & Arendt 2015). So far, supernova 1987A is the only supernova where its SED and dust evolution has been monitored 25 years after the explosion. It is still unknown how the dust mass has increased in time, and whether such a large dust mass is unique to SN 1987A or commonly to supernova remnants. Future far-infrared space missions can answer these questions.

Simulated dust emission over time (Figure 4.9) demonstrates that dust becomes colder, so that its emission shifts from mid-infrared to far-infrared. Even at the time of JWST and the SPICA era, dust measurements in late phase of supernovae will be challenging. Except for SN 1987A, located at 50 kpc, the majority of nearby supernovae are found beyond 1 Mpc (Smart 2009; Otsuka et al. 2011). The JWST can detect dust from nearby newly exploded SNe, but it can cover only the mid-infrared range, being unable to detect cold dust. Future high-sensitivity FIR space missions will therefore be vital for answering the question of supernova dust production.

An alternative is to measure dust content in evolved supernova remnants (SNRs). So far, the dust masses has been measured in a handful of Galactic SNRs and the Magellanic Clouds (e.g. Williams et al. 2006; Rho et al. 2008; Barlow et al. 2010; Gomez et al. 2011). Again the dust mass discrepancy, found between MIR and FIR measurements (Rho et al. 2008; Barlow et al. 2010), could imply that strong emission from the hot dust component might hide emission from colder dust, while the cold dust tends to dominate the total dust mass. Herschel data of the SNRs in the Large Magellanic Cloud, located only at 50 kpc, suffer from severe ISM contamination problems (Lakicevic et al. 2015), hence high angular resolution and high sensitivity from future FIR space observatories are vital to resolve the origin of dust in galaxies.

## 4.7  Mass loss of AGB stars

As mentioned above, AGB stellar winds are thought to be a significant source of dust grains in galaxies. The mass loss of AGB stars is thought to originate from thermal pulsations in the stellar atmosphere and radiation pressure on dust grains formed in the coolest layers (e.g., Wood 1979; Caster 1981). As some material is levitated from the atmosphere by pulsations, radiation pressure on dust grains initiates outwards motion, the motion of dust grains carries with it surrounding gas, and mass loss starts. Although this is an attractive concept for mass loss triggering, modelling of the AGB contribution requires observational constraints out of reach and scope of currents instrumentation.

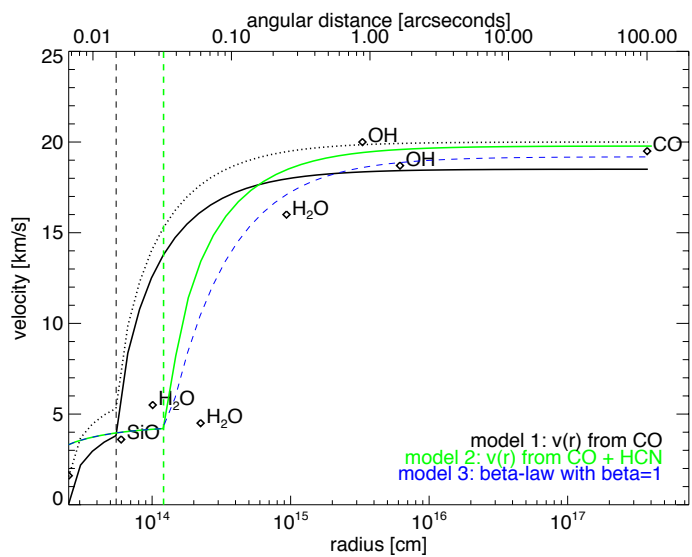

Figure 4.10 The radial velocity structure of the AGB star, IK Tau (Decin et al. 2010). The wind speed is accelerated near the dust forming region, where SiO is depleted to dust grains. Spatially resolved imaging of multiple molecular images at far-infrared will resolve this radial velocity structure precisely.

Early ISO observations of AGB stars helped our understanding of AGB mass loss through detections of silicates (in amorphous and crystalline form) and water (both gaseous and solid (ice) phases) (e.g., Barlow 1998; Sylvester et al. 1999; Molster et al. 2002a, b). However, the mass-loss behaviour of oxygen-rich stars differs from carbon-rich stars, and models are still not completely successful (e.g., Woitke 2006; Decin et al. 2010; Karovicova et al. 2013). The Herschel Space Observatory detected far-infrared rotational molecular lines including $H_2O$ vapour and other hydrides together with CO (Decin et al., 2010; Lombaert et al. 2013; Matsuura et al. 2014; Khouri et al. 2014; Danilovich et al. 2015, Maercker et al. 2016); these transitions originate from the gas at temperatures of about 1000 K. This temperature corresponds approximately to dust condensing, where wind acceleration starts (Figure 4.10). Herschel also provided observational constraints for dust formation mechanisms in the stellar wind with the 69µm forsterite feature (de Vries et al. 2014; Blommaert et al. 2014). Resolving these transitions on both image and velocity will provide a much clearer picture how dust is distributed, and thus how the gas is accelerated.

ALMA can resolve molecular emission (see Figure 4-11, Ramstedt et al. 2014, see also Decin et al. 2015), and free-free emission in AGBs has been identified by cm-VLA observations (e.g., Matthews & Karovska 2006), but transitions covered by ALMA tend to be from cold gas from the outer part of the circumstellar envelope. FIR transitions from 50 to 500µm are ideal for tracing hot/warm (100-1000 K) gas, making FIR the best regime for investigating the wind acceleration region. The required angular resolution is about a few arcsec to 1 arcsec to resolve the dust acceleration radius for nearby red-supergiants and AGB stars.

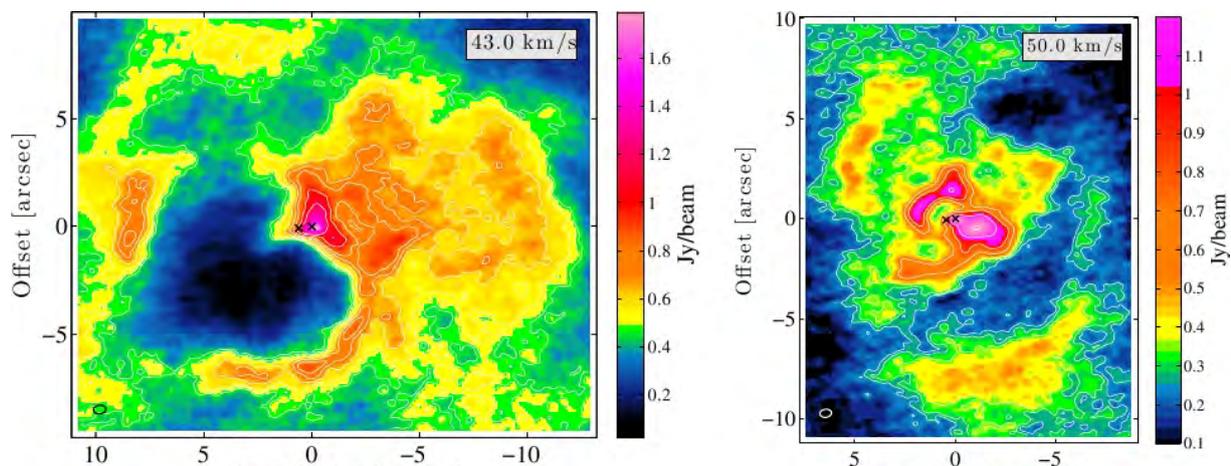

Figure 4-11 CO(3-2) ALMA map of the AGB binary Mira AB, velocity averaged around 43 km/s (left) and 50 km/s (right). The bubble (left) and spiral structure (right) are clearly visible (taken from Ramstedt et al. 2014).

## 4.8 Scientific Requirements

Here we assess the technical/observational requirements for each of the science cases described above. A summary of the requirements is given in the table at the end of the section. There are two key requirements that are common to most of the science cases we reviewed: high spectral resolution and mapping of large areas of the Galactic Plane and High Galactic latitude. The spectral resolution (0.5-1 km/sec) is crucial to disentangle the various components (ionized, neutral) of the ISM (Sections 4.2, 4.3 & 4.4) and in particular to measure the amount of CO-dark gas (4.3). Assessing feedback from massive stars would require resolution of about 1-10 km/sec albeit adequate sensitivity is also important.

| Science case | Area coverage | Spectral Resolution | Spatial Resolution | Sensitivity 5σ 1hr/FOV |
|---|---|---|---|---|
| Phases of the ISM | ~3600 Gal Plane | 0.5-1km/s | ~30" | $1 \times 10^{-19}$ Wm$^{-2}$ |
| CO-dark clouds | >3000 Gal Plane HGLS | 0.5-1km/s | ~30" | $1 \times 10^{-19}$ Wm$^{-2}$ |
| MW star formation | Large >5000sq. deg | 10km/s | ~30" | $1 \times 10^{-20}$ Wm$^{-2}$ |
| High Mass stars | >1000sq. deg | 1-10km/s | ~12" | $1 \times 10^{-20}$ Wm$^{-2}$ |
| Origin of dust | N/A | N/A | 10" | 3 mJy |

# 5 Feedback and dust in galaxies in the local universe

The proximity of galaxies in the Local Universe enables us to study star-forming regions with a spatially-resolved scrutiny that is not possible at high redshift. This is important because physical conditions of many nearby galaxies are more extreme than those found in the Milky Way itself. Even within the Local Group, R136, the massive star cluster powering the Tarantula Nebula (30 Doradus) in the Large Magellanic Cloud (LMC) is more than 10 times more massive than the most massive star cluster in the Milky Way (McLaughlin & van der Marel 2005). Beyond the Local Group, the Super Star Clusters (SSCs) in nearby (dwarf) galaxies NGC1569 and NGC1705 are more than 30 times brighter than R136, even after correcting for age differences (O'Connell et al. 1994); stellar densities are higher by similar amounts. This implies that to study the vast parameter space of star formation over cosmic times, we must go beyond the Milky Way to the study of nearby galaxies that offers access to a wealth of details that high-z galaxies do not. Here we discuss three key questions for nearby galaxies that are still outstanding and emphasized by recent IR space missions, and that can only be pursued in the FIR.

## 5.1 FIR fine-structure lines as probes of star-formation activity

With the advent of ALMA, the use of the FIR fine-structure (FS) lines such as [CII] 158µm and [NII] 205µm lines as measures of the star formation rate (SFR) is now routinely extended to the high-redshift universe (the lines shift from the FIR to submillimeter wavelengths). In fact, [CII] detections in redshift z∼6 star-forming galaxies with ALMA are more common than detection of molecular gas (e.g., Ota et al. 2014; Capak et al. 2015; Maiolino et al. 2015; Willott et al. 2015).

Yet, the validity of using FIR FS methods to estimate SFR still requires to be fully observationally and theoretically established. The key assumption is the dependence of the dust grain photo-electric heating rate on the specific physical conditions. Models have only been really tested through FIR observations for conditions relevant for dense and intense PDRs such as the Orion Bar (Tielens et al 1993) and for conditions relevant to diffuse clouds using UV absorption measurements (Pottasch et al 1979). Nevertheless, Herschel observations of [CII] in the Milky Way show that the [CII]-SFR relation extends over six orders of magnitude, but only by considering that [CII] emission originates from different ISM phases (e.g., dense PDRs, cold HI, CO-dark $H_2$, and ionized gas: Pineda et al. 2014). As shown in Figure 5-1 the [CII]-SFR relation for the combined phases in our Galaxy is consistent with that found for distant galaxies (e.g., Rigopoulou et al. 2014, de Looze et al. 2011, 2014; Herrera-Camus et al. 2015; Cigan et al. 2016). Recent models are also improving our theoretical understanding of this relationship (e.g., Vallini et al. 2015).

[NII] traces the ionized gas, coming both from the compact HII regions and from the more diffuse component. Herschel observations of our Galaxy (e.g. Persson et al. 2014, Kirk et al. 2010) show that the [NII]205µm line emission is extended and presents broad lines. In distant galaxies, because of its close relation with SFR as a tracer of ionized gas, [NII] is gaining popularity (Hughes et al. 2016; Zhao et al. 2016; Lu et al. 2015). [CII] 158µm and [NII]205µm, shifted to (sub)mm wavelengths, are observable with ALMA and provide important constraints on the SFR at high z. Other lines are also potentially valuable tracers of SFR as shown in Figure 5-2. Indeed, there is some indication that [CII] may not be the best probe, because of possible contributions from regions such as dense PDRs and diffuse gas

regions that are not directly associated with star formation (e.g., De Looze et al. 2014; Abdullah et al. 2016).

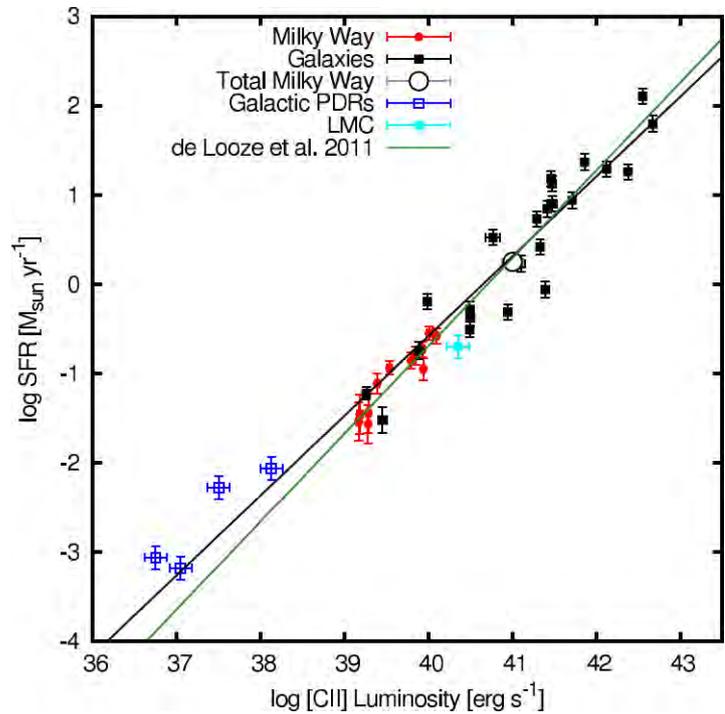

Figure 5-1 SFR vs. [CII] luminosity for different radii in the Milky Way, for individual PDRs, and for other galaxies (taken from Pineda et al. 2014).

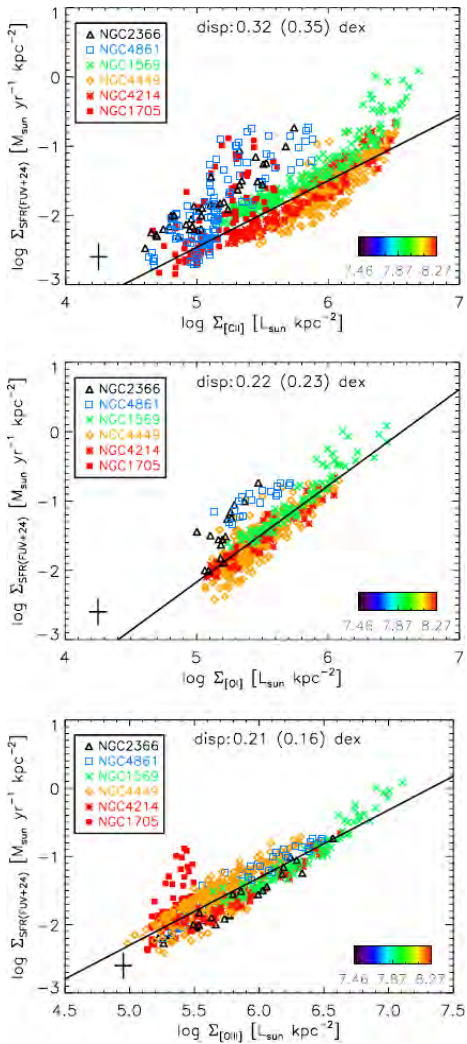

Figure 5-2 Spatially resolved relation between SFR and FIR FS line surface densities. Color coding corresponds to galaxy metallicity, 12+logO/H (taken from De Looze et al. 2014).

And there may be additional problems; at high FIR luminosities (i.e., high SFR), both [CII] and [NII] are deficient in emission with respect to what would be expected for a "perfect" (i.e., single power-law) SFR tracer (e.g., Ibar et al. 2015; De Looze et al. 2014; Diaz-Santos et al. 2013; Farrah et al. 2013; Sargsyan et al. 2012; Luhman et al. 1998). This means that in some regions, and/or over entire galaxies, physical conditions in the ISM are suppressing some of the emission of these lines; warm temperatures, high volume densities, and compact size are possible limiting conditions (e.g., Diaz-Santos et al. 2013).

Because [CII] and [NII] and other FIR FS lines are the best/brightest line diagnostics for high-redshift estimates of star-formation activity with ALMA, it is crucial to understand the origin of the deficit and develop a better theoretical understanding of the correlation. This can only be done by resolving ISM phases both spatially and kinematically in the Galaxy, the Local Group, and nearby galaxies, and can only be done with an FIR observatory.

## 5.2 Effects of feedback on the dust and gas in galaxies

Feedback from accreting super-massive black holes (SMBHs) to massive stars and SNe (see Sect. 4.5) plays a key role in galaxy evolution, as shown by the apparent success of semi-analytical models and hydro-dynamical simulations in modeling galaxy properties. While feedback from active galactic nuclei (AGN) shapes the galaxy stellar mass function at the high end (e.g., Vogelsberger et al. 2013, and references therein), the low-mass end is thought to be governed by stellar feedback via SNe and massive stellar winds (e.g., Oppenheimer & Davé 2008; Schaye et al. 2010; Hopkins et al. 2011). Even before the onset of SNe, massive stars provide significant feedback through deposition of momentum and energy into the ISM, thus also affecting metal enrichment and modifying the shape of the mass-metallicity relation (e.g., Davé et al. 2012; Hopkins et al. 2012).

### 5.2.1 Stellar feedback in low-mass galaxies

Despite their success at the high-mass end, a problem of the latest generation of models of galaxy evolution with feedback is that there are few observational constraints for stellar feedback, especially at the low-mass end (e.g., Weinmann et al. 2012). Massive star clusters are born embedded in dust and gas, and emerge from this cocoon only after 1 few Myr; however, the dominant mechanism responsible for structuring the ISM around massive stars and star clusters is still under debate (e.g., Rogers & Pittard 2013). The relative roles of radiation pressure from dust and different gas phases, mechanical energy and shocks are not clear, despite several recent studies (e.g., Lopez et al. 2011, 2014; Lebouteiller et al. 2012; Lopez et al. 2014; Sokal et al. 2015). A Local Group example of the observational complexity of stellar feedback is shown in Figure 5.3; the molecular gas coincides with dust emission while X-ray morphology shows cavities filled with hot gas.

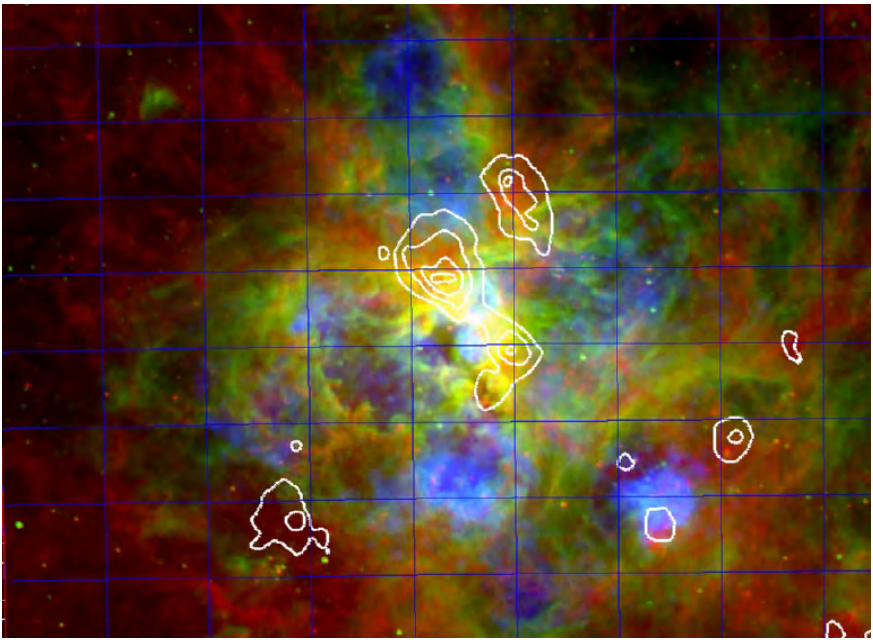

Figure 5.3: Color composite of 30 Doradus (LMC) with Spitzer/IRAC 8μm (red), Hα (green), 0.5-0.8 keV Xrays (blue). Image is ~18x24 arcmin$^2$; white contours correspond to CO(1-0) emission (figure taken from Lopez et al. 2011).

As discussed in Sect 4.2, FIR lines are important diagnostic tracers to determine physical conditions in the ISM. In particular, H$_2$O and OH trace shocks (e.g., Kaufman & Neufeld 1996; Wardle 1999), and the ISM energy budget as estimated using molecular cooling curves allows to distinguish between photo-dissociation and mechanical heating. Ionized gas tracers such [OIII] λ88 m are relatively insensitive to electron temperatures, but vary with density, while the two [NII] lines (122, 205 μm) can be used to measure electron densities (e.g. Croxall et al. 2013); thus this spectral region provides robust diagnostics for the ionized gas physical conditions including gas-phase metallicity even in the presence of dust. Such studies have already been conducted for specific regions in our Galaxy (see Sect. 4.4) and cosmological simulations at high spatial/spectral resolution at similar scales are fundamental understand the effects of feedback in low-mass galaxies.

Dwarf galaxies are the most numerous galaxy populations in the Local Universe, but they also suffer most from the effects of stellar feedback because of their weaker gravitational potential. The Dwarf Galaxy Survey (Madden et al. 2013, 2014) observed [CII], [OIII], and [OI] in ∼50 dwarf galaxies. The spectral and spatial resolution was insufficient for anything but global measurements beyond the Local Group. Like for our Galaxy (see Sect. 4.4), high spectral resolution, coupled with high spatial resolution is mandatory to assess dust content, heating and cooling, and the alteration of these processes through feedback at the various gas interfaces. A sensitive FIR space mission would be fundamental for dissecting stellar feedback in the Local Universe, and informing theoretical models for sub-grid recipes in theoretical models applicable to galaxy evolution at high redshift.

### 5.2.2  AGN feedback in massive galaxies: OH and $H_2O$ as tracers of energetic outflows

Early simulations suggested that AGN-driven outflows are main drivers of galaxy evolution in massive galaxies, now confirmed by observations. Indications of bi-polar ionized outflows were found in z ∼ 2 radio-loud AGN (Nesvadba et al. 2008) and ionized "ultra- fast outflows" in local radio-quiet AGN (e.g., Tombesi et al. 2010), but the first systematic observational evidence for AGN-driven molecular outflows was provided by Herschel.

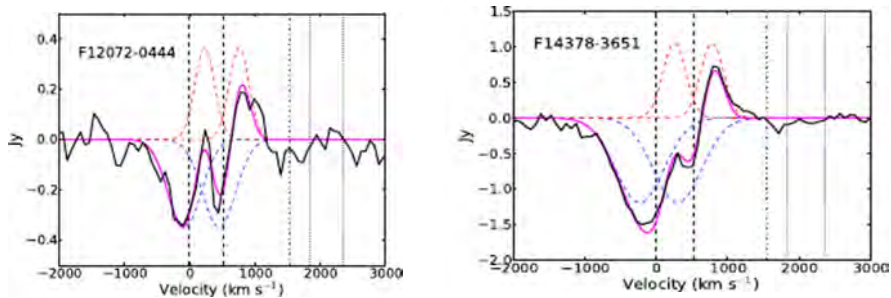

Figure 5.4: Spectral fits to the OHλ 119 μm profiles of selected objects from Veilleux et al. (2013) Blue- dashed lines represent the absorption component(s), and the red-dashed curves the emission. Vertical lines mark positions of the $^{16}$OH (dashed) and $^{18}$OH (dotted) doublets, and the CH$^+$ (dot-dashed) transition at 119.8μm. For more details, see Veilleux et al. (2013).

The direct signature of a massive galactic outflow discovered by Herschel/PACS observations was a P-Cygni-like profile in OH (the hydroxyl molecule, Fischer et al. 2010; Sturm et al. 2011; Veilleux et al. 2013). As shown in Figure 5.4, the maximum outflow velocities of these IR-luminous AGN can exceed 1000 kms$^{-1}$. In one object, Arp 220, a particularly well-studied ULIRG, P-Cygni profiles were also detected in other molecules including $H_2O$ (Rangwala et al. 2011; Gonzalez-Alfonso et al. 2012). Following up on these observations, the detailed spatially resolved kinematic maps in the molecular gas were obtained using CO (Feruglio et al. 2010; Cicone et al. 2014), showing outflow rates of several 100 $M_\odot yr^{-1}$, far exceeding the SFR in these galaxies. Further ground-based mm observations found evidence for similar outflows in the dense molecular gas traced by HCN (Aalto et al. 2012).

Inverse P-Cygni profiles in two luminous IR galaxies (LIRGs) NGC 4418 and Zw 049.057 showed evidence for molecular infall (Falstad et al. 2015; Gonzalez-Alfonso et al. 2012). Both these galaxies belong to the class of "compact obscured nuclei" and show a wealth of complex organic molecules (Costagliola et al. 2015).

The discovery of these molecular outflows/inflows was one of the key milestones of Herschel, but their study is currently limited to ~40 of the brightest nearby ULIRGs and quasars. A powerful FIR observatory is needed to capture the full demographics of local AGN and quantify how AGN feedback regulates star formation in high-mass galaxies.

## 5.3 Regulation of the dust content in galaxies

After ISO, Spitzer, and Herschel, it has become routine to measure dust masses in a variety of galaxies up to z≥2. Indeed, recent studies show that dust emission can be an important measure of total ISM mass both locally (e.g., Eales et al. 2012; Groves et al. 2015) and at high redshift (Scoville et al. 2014, 2015). However, the dust mass is apparently only a small fraction of the total ISM mass, so that dust-to-gas ratios and their variation with other properties such as metallicity give important insight into the limitations of such estimation techniques.

Dust-to-gas mass ratios (DGRs) in galaxies were thought to increase linearly with metallicity (as measured by nebular oxygen abundance: e.g., Draine et al. 2007), and such an assumption is incorporated in many, if not most, models of ISM physical conditions for star formation (e.g., Krumholz et al. 2009; Wolfire et al. 2010; Krumholz et al. 2012). However, recent results from Herschel have shown that below a threshold metallicity, of ~0.25$Z_\odot$, the DGR dependence appears much steeper than previously thought (Remy-Ruyer et al. 2014). To further complicate the matter, careful measurements of dust mass at extremely low metal abundance indicate that this may not be a general rule (Hunt et al. 2014), and that other parameters such as ISM density can influence the formation of dust grains and thus the total amount of dust mass (e.g., Schneider et al. 2016). DGRs plotted against metallicities are shown in Figure 5.5; it is evident (see the right panel) that only a handful of galaxies have measured dust masses below metallicities of 12+log(O/H)~7.5 and the Herschel Dwarf Galaxy Survey needs to be dramatically expanded.

However, even after the important Herschel Dwarf Galaxy Survey (Madden et al. 2013), as shown in Figure 5.5 there are still too few galaxies with dust masses measured at extremely low metallicities (i.e., 12+log(O/H) ≲7.7, ~0.1 $Z_\odot$). Such metallicities are important because they characterize the transition between pristine metal-free star formation in the early universe and the metal-rich chemically evolved galaxies common at the current epoch. Dust emission in low-metallicity galaxies tends to be quite warm (e.g., Hunt et al. 2005; Galametz et al. 2010; Remy-Ruyer et al. 2013), signifying different physical conditions in the ISM, similar to those of galaxies at higher redshift (see Sect. 6.2). A new sensitive FIR space observatory would help disentangle the current discrepancies in metal-poor local dwarf galaxies, and pave the way for a better understanding of large scale grain formation, dust evolution, and star formation in high-z galaxy populations.

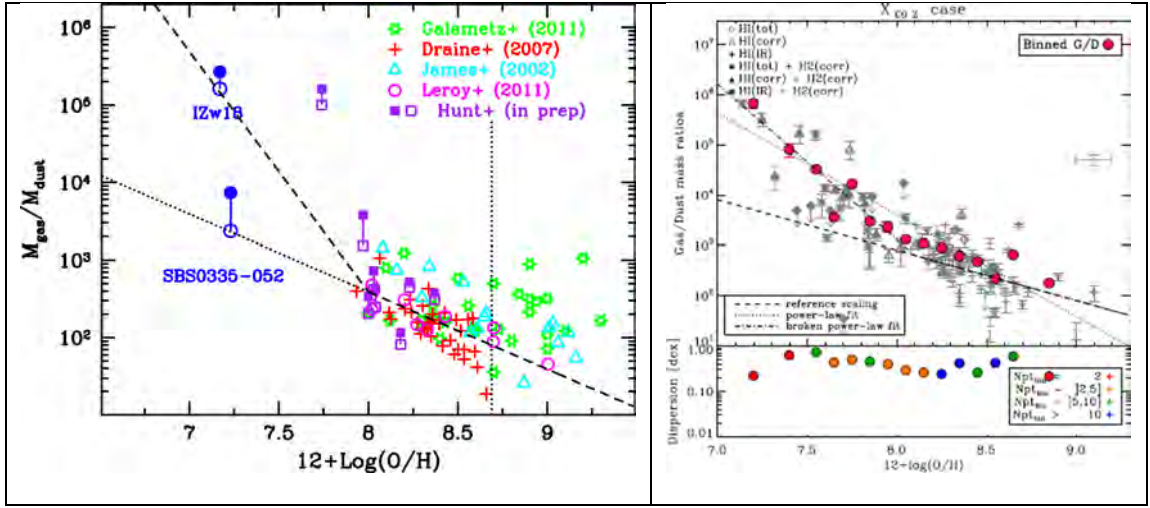

Figure 5.5: Gas-to-dust ratios as a function of nebular oxygen abundance. The left panel shows the compilation by Hunt et al. (2014) and the right panel by Remy-Ruyer et al. (2014). In the left panel, open blue circles show the DGR without a putative molecular gas component (CO is not detected in either of these galaxies), and filled ones with molecular gas as inferred from a Kennicutt-Schmidt relation (for more details, see Hunt et al. 2014). In the right panel, grey symbols are individual data points while filled magenta circles show binned results. Most galaxies follow a linear trend up to 12+log(O/H)~8.1 (0.25 $Z_{solar}$), but lower-metallicity galaxies show a steeper slope (~3) (see Remy-Ruyer et al. 2014, for more details).

## 5.4 Scientific requirements

| Science case | λ coverage | Sensitivity | Spatial resolution | Spectral resolution |
|---|---|---|---|---|
| FIR SFR tracers | ~50-250μm | $1 \times 10^{-20}$ Wm$^{-2}$ spectra | ~5'' (160μm) | 1500-2000 |
| z~0 AGN outflows | ~50-300μm | $3 \times 10^{-21}$ Wm$^{-2}$ spectra | ≤2'' (250μm) | 1500-2000 |
| z~0 stellar feedback | ~50-250 μm | $1 \times 10^{-20}$ Wm$^{-2}$ spectra | ≤2'' (250μm) | 1500-2000 |
| z~ 0 dust crisis | ~70-350μm | ~3 mJy continuum | ≤5'' (250μm) | 5 |

# 6 The far-infrared landscape of galaxy evolution

A key objective of previous IR missions was to resolve the cosmic infrared background (CIB). Thanks to ISO, Spitzer and Herschel, it is now generally accepted that the diffuse extragalactic background can be attributed to (resolved and still unresolved) galaxies (see Lutz 2014, and references therein). This is a excellent example of "mission accomplished" through several generations of FIR observatories.

Now we can concentrate on the key questions in extragalactic science that can only be answered with the next FIR mission: quantifying the star-formation history of the universe (Sect. 6.1); assessing the ISM physical conditions for galaxy assembly (Sect. 6.2); characterizing black hole and galaxy co-evolution (Sect. 6.3).

## 6.1 The history of galaxies

The epoch of galaxy formation and the average star-formation history (SFH) of galaxies is most commonly traced through the cosmic star-formation rate density of the Universe (SFRD). From $z \sim 10$ to $z \sim 6$, cosmic reionization occurs during which light from the first galaxies ionizes the neutral intergalactic medium, enhancing the SFRD. Toward lower redshifts, the SFRD peaks around $z \sim 1 - 3$, a period usually associated with the main epoch of galaxy assembly. During this epoch, when the universe was roughly 15-50% of its current age, almost half of the stars in present-day galaxies were formed (e.g., Reddy et al. 2008; Shapley 2011). The SFRD then declines dramatically, by roughly an order of magnitude, between $z \sim 1$ and today. The rise and fall of the SFRD as a function of redshift (and lookback time) is illustrated in Figure 6-1 (taken from Madau & Dickinson 2014).

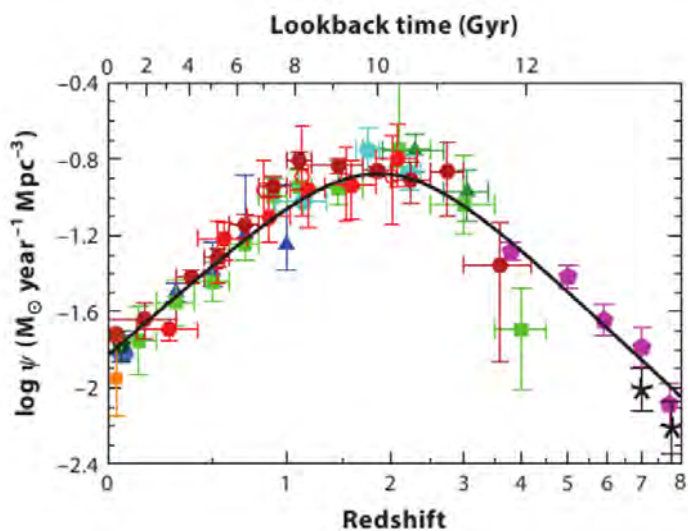

Figure 6-1 History of cosmic star formation from the combined FUV+IR rest-frame SFRD. The solid curve is the best-fit SFRD evolution in redshift derived by Madau & Dickinson (2014, more details given in their paper).

"Observed" (i.e., derived from integrating luminosity functions to account for completeness) UV, uncorrected for dust attenuation, and IR SFRDs are shown in Figure 6.2. There is a strong difference between SFRs derived from uncorrected UV and IR measurements. There is also some evidence that the UV correction is not universal, as it is luminosity dependent (e.g. Reddy & Steidel 2009), and it may also depend on ISM conditions and geometry of the obscuring dust (e.g., Oteo et al. 2013; Pannella et al. 2015). Figure 6.2 (left panel) demonstrates that IR wavelengths are intrinsically much better suited to trace SFRD to high redshifts. *The problem with recent IR facilities is that they did not obtain a robust census of galaxies potentially missed by UV surveys beyond z>~1, toward the cosmic peak of the star-* *formation activity at z>~2.*

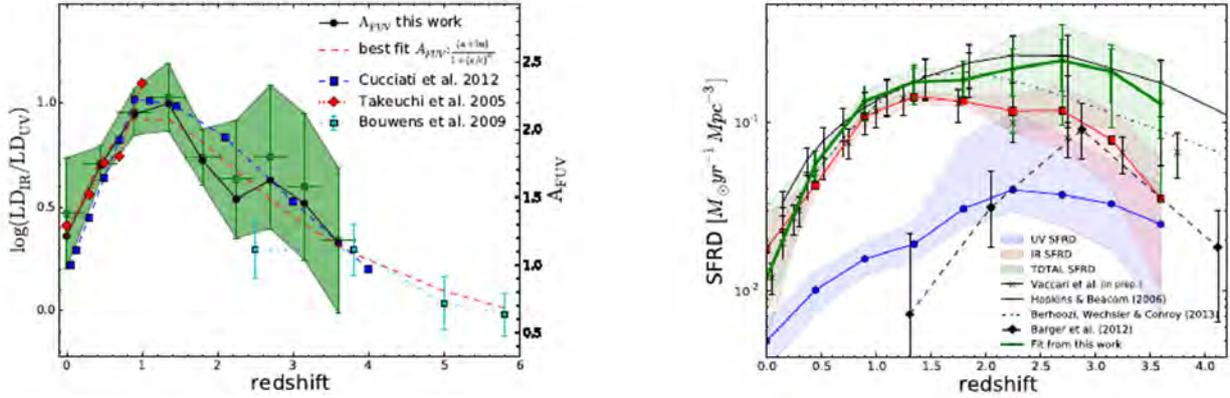

Figure 6.2: Right panel: a compilation from the literature of SFRDs in the FUV (uncorrected for dust attenuation) and in the far-IR; left panel: mean ratio of IR and UV luminosity densities as a function of redshift. The differences (integrated over galaxy luminosity functions) between observed FUV and FIR SFRDs can be almost an order of magnitude, and shows that most of the energy from star formation for z~<2 is reprocessed by dust. Figures are taken from Burgarella et al. (2013), more details given there.

Figure 6.3 shows two attempts to do this using IR-selected galaxies from the Herschel/PACS Extragalactic Probe survey (PEP: left panel, Gruppioni et al. 2013) and from the Herschel Multi-tiered Extragalactic Survey (HerMES) with SPIRE (right panel, Casey et al. 2012, 2014). The SFRDs from both IR-selected populations are consistent (peaking at z~1 with SFRD ~0.2 $M_\odot yr^{-1}$ $Mpc^{-3}$), and comparable to that from the optical/UV-selected galaxy populations up to the highest redshifts sampled (see Figure 6-1). However, the contribution of IR-selected dusty galaxies evolves strongly (e.g., Le Floc'h et al. 2005; Casey et al. 2014); the contribution of ultra-LIRGs (ULIRGs) is insignificant at z ~ 0, but increases up to ~50% at z>~2 as shown in Figure 6.3, but see also Rowan-Robinson et al. 2016. There is some indication that the IR selected sample contains different galaxy populations than in the UV selected sample (e.g., Heinis et al. 2014; Bernhard et al. 2014), and that current theories cannot yet explain the SFRDs obtained by combining UV- and IR-inferred SFRs (Heinis et al. 2014). *Ultimately, even by stacking the most sensitive existing Herschel observations, we are probably missing a significant fraction of faint dusty galaxies at the peak of the cosmic star-formation activity. To resolve this, and determine the importance of dust-obscured galaxies robustly and directly, more powerful IR facilities are needed.*

## 6.2   Galaxy assembly, star formation, and physical conditions in the ISM

The peak of star-formation activity at z~2-3 coincides with the peak of black-hole growth (see Sect. 6.1), but the complete answer to how galaxies are assembled, forming SMBHs in their nuclei and stars throughout their disks, remains elusive. The decline of the SFRD at z~<1 implies that star-formation activity in the Local Universe has faded. The Local Universe does not probe the dramatic differences in galaxy properties at earlier times. Indeed, observations have shown that high-z galaxies have different properties (size, mass, SFR, metal content, dust and gas temperature, density, etc.) than local ones. Nevertheless, the physical mechanisms by which the mass in galaxies is assembled are the same; available gas is converted into stars, and, in the process, dust grains are formed in the envelopes of evolving stars, in the ejecta of energetic supernovae (SNe) explosions, and inside dense clouds of molecular gas (see Sect. 4.6). Galaxies accrete gas from the surrounding intergalactic medium (IGM) and expel gas through galactic winds driven by SNe and AGN. Thus, the question becomes how these processes affect the physical conditions in high-z galaxies.

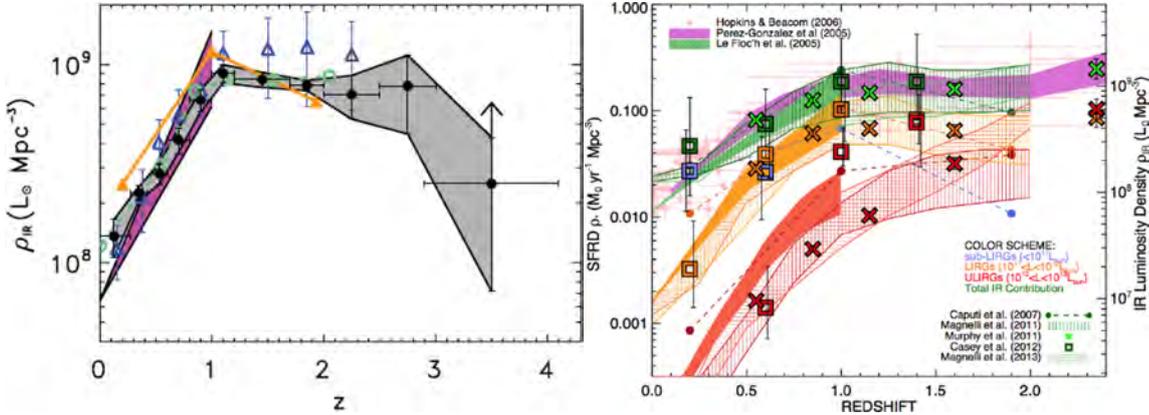

Figure 6.3: Left panel: co-moving IR luminosity density, $\rho_{IR}$ (equivalent to SFRD), obtained by integrating modified Schechter functions from PACS-selected galaxies at 100 μm and 160 μm (taken from Gruppioni et al. 2013): $\rho_{IR}$ =$10^9 L_\odot$ Mpc$^{-3}$ is equivalent to 0.17$M_\odot$yr$^{-1}$ Mpc$^{-3}$. The gray area represents the ±1σ uncertainty; for more details see Gruppioni et al. (2013). Right panel: estimates of dusty star-forming galaxy contributions to the SFRD (taken from Casey et al. 2014). As in the left panel, Schechter functions are used to extrapolate over luminosities not directly probed. Contributions to the SFRD are shown for total IR luminosities (green), IR luminosities $L_{IR}\sim 10^{11} L_\odot$ (blue), luminous IR galaxies ($10^{11}<L_{IR}<10^{12}L_\odot$, orange), ULIRGs ($L_{IR}>\sim 10^{12} L_\odot$, red). The right axes give the IR luminosity density that translates to the SFRD via usual scaling laws. For more details see Casey et al. (2012, 2014).

Quantifying physical conditions of star formation and understanding galaxy assembly is related to the ISM energy cycle, namely how mechanical and radiative energy within a galaxy interact with the IGM in order that gas forms stars. Because the main ISM cooling lines (e.g., [CII], [OI]) and molecular emission (e.g., high-J CO lines, $H_2$, HD, $H_2O$) are in rest-frame far-IR, and because of significant dust reprocessing of galaxy light (see Sect. 6.1), the FIR to submm spectral region provides the most definitive approach for redshifts beyond the Local Universe toward the peak of star formation and BH activity up to the epoch of reionization (z>6, where the lines shift into the ALMA wavelengths). Below, we explore three specific diagnostics of the ISM in galaxies, in order to illustrate the power of the FIR regime.

### 6.2.1 Dust as a tracer of gas content

Up to z>~2 (~10 Gyr lookback time), observations show a clear correlation between SFR and stellar mass, Mstar, commonly known as the "main sequence of star formation" (MS) (e.g., Brinchmann et al. 2004; Schiminovich et al. 2007; Noeske et al. 2007; Rodighiero et al. 2011; Elbaz et al. 2011; Karim et al. 2011; Whitaker et al. 2012). Although there is some evidence for curvature at the high-mass and high-redshift regimes (Lee et al. 2015; Schreiber et al. 2015), the slope of the correlation is roughly constant with redshift but the zero point shifts (e.g., Karim et al. 2011; Speagle et al. 2014). At a given $M_{star}$, high-z galaxies form stars at a higher rate than at z ~ 0 as illustrated in Figure 6.4. The shift of the SF MS position implies that the definition of "starburst" galaxies must change with redshift. In fact, this notion has been corroborated by Herschel observations of cosmic deep fields (e.g., Magdis et al. 2012; Schreiber et al. 2015; Pannella et al. 2015). At all cosmic epochs, galaxies on the MS have cooler dust than starbursts, but the typical dust temperature $T_{dust}$ of MS galaxies increases with increasing redshift (e.g., Magnelli et al. 2014), together with an increasing intensity of the interstellar radiation field (Magdis et al. 2012). Thus, the physical conditions in a local starburst would closely resemble a MS galaxy at z~1.5 (e.g., Pannella et al. 2015). Such changes almost certainly affect the molecular gas and alter the CO luminosity-$H_2$ mass conversion factor (Magnelli et al. 2012; Genzel et al. 2015). They also complicate the use of IR template schemes for high-z galaxies which rely on a correlation between $L_{IR}$ and $T_{dust}$, rather than on sSFR (e.g., Chary & Elbaz 2001; Dale & Helou 2002; Dale et al. 2014).

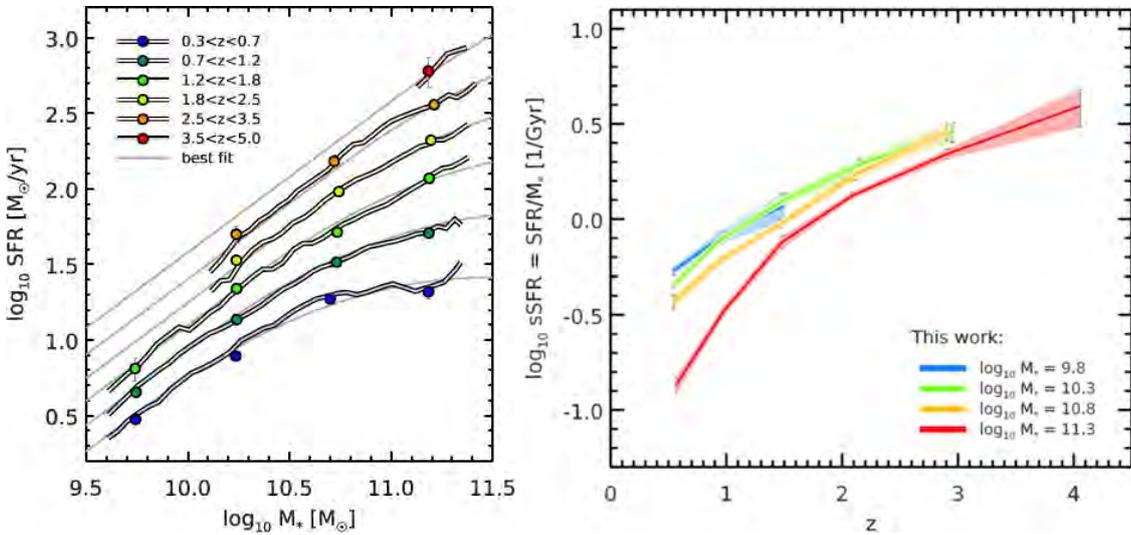

Figure 6.4: Left panel: Specific SFR vs. $M_{star}$ for different redshift bins. Right panel: Specific SFR as a function of redshift for different $M_{star}$ bins (taken from Schreiber et al. 2015, more details given in their paper). The shift in SFR with redshift is clearly evident in the left panel.

One of the main drivers of the behavior of the SF MS is thought to be the larger gas reservoirs available at high redshift (e.g., Lagos et al. 2015; Genzel et al. 2015; Bothwell et al. 2016). Because gas-to-dust ratios are well behaved, at least locally (e.g., Draine et al. 2007; Leroy et al. 2011), gas mass and its variation as function of redshift can be derived using dust mass measurements (Santini et al. 2014; Bethermin et al. 2015; Genzel et al. 2015); such measurements can be obtained using deep (mostly stacked) Herschel observations (Magnelli et al. 2012; Saintonge et al. 2013; Santini et al. 2014; Dessauges-Zavadsky et al. 2015; Genzel et al. 2015; Bethermin et al. 2015). Figure 6.5 shows gas fractions derived from dust masses obtained from stacked Herschel observations (Santini et al. 2014; Bethermin et al. 2015) compared to gas mass measured from CO observations (Magnelli et al. 2012; Saintonge et al. 2013; Dessauges-Zavadsky et al. 2015). In the absence of CO measurements, the gas mass is obtained by using the metallicity-dependent recipe for gas-to- dust ratios in the Local Universe determined by Leroy et al. (2011) coupled with metallicities derived using the scaling relations ($M_{star}$, SFR, and oxygen abundance) which are accurate to ~0.15 dex (e.g., Mannucci et al. 2010, 2011; Hunt et al. 2012, 2016). Despite several assumptions, so far the results are encouraging; as seen in Figure 6.5, the CO-based estimates of $H_2$ masses are within the same range of the dust-based estimates for total gas mass. Moreover, higher gas-mass fractions at a given redshift are associated with galaxies well above the SF MS (e.g., starbursts), and increase with decreasing stellar mass (see also Magdis et al. 2012). Currently only using stacking we can sample high stellar masses ($M_{star}>10^{11}M_\odot$) beyond z>~2, the peak of the cosmic SFRD. To go beyond the "tip of the iceberg" and measure dust masses that cover a range of stellar masses together with galaxy populations on and off the main sequence requires new IR facilities that probe the long-wavelength dust emission.

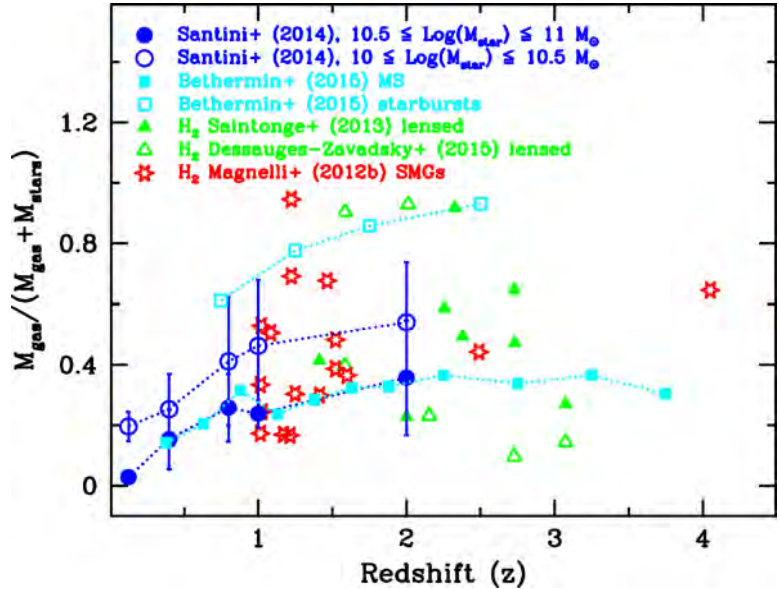

Figure 6.5 Gas fractions in different galaxy populations as a function of redshift. Samples from Santini et al. (2014) and Bethermin et al. (2015) rely on dust-mass measurements, converted to gas mass according to the dust metallicity function calibrated at z~0 (Leroy et al. 2011). The remaining samples include sub-millimeter galaxies (SMGs, Magnelli et al. 2012) and lensed galaxies (Saintonge et al. 2013; Dessauges-Zavadsky et al. 2015) with H2 mass estimated from CO observations converted to $H_2$ masses using a metallicity-dependent conversion factor (taken from Ginolfi et al. 2016, in prep.).

### 6.2.2  Molecules, ISM cooling, and star formation

Molecular gas ultimately provides the fuel for star formation, so molecular tracers are arguably the best diagnostics to assess heating and cooling of the ISM in galaxies and understand at a physical level how gas is converted into stars. After $H_2$, CO is the most abundant molecule in the ISM, followed by $H_2O$[1] (e.g. Omont et al. 2013; Carilli & Walter 2013; Yang et al. 2013).

Due to the lack of a permanent dipole moment and high excitation energies, $H_2$ emission is difficult to observe directly (lowest rotational transition lies at ~28 μm); thus, CO has been historically adopted as a proxy for tracing cool molecular gas. Although low-J CO transitions trace the bulk of the cool gas mass, they give virtually no insight into how the ISM cools. For this purpose, CO cooling curves (or spectral-line energy distributions, SLEDs, or excitation ladders) including higher-J transitions are necessary. While low-J CO transitions can effectively be studied from the ground, $H_2O$ and higher-J (J>6) CO transitions at z~<0.1 can only be studied from space. Combining different CO transitions is important because CO excitation distinguishes whether or not a high-redshift galaxy is on the SF MS (e.g., Carilli & Walter 2013, and references therein). The CO content in massive star-forming disks at z~1-2 is less excited than in turbulent starbursts (e.g., Narayanan & Krumholz 2014; Daddi et al. 2015; Liu et al. 2015), but it is difficult to distinguish between the two scenarios with only high-J or low-J transitions (e.g., Rigopoulou et al., 2013, Kamenetzky et al. 2014). Figure 6-6 illustrates the diagnostic power of high-J CO lines, and how temperature and density in the ISM change CO excitation.

---

[1] Here we refer to water vapour, rather than either water masers or water ice in grain mantles, the latter of which is more easily observed at near- and mid-infrared wavelengths (e.g. Boogert et al. 2015).

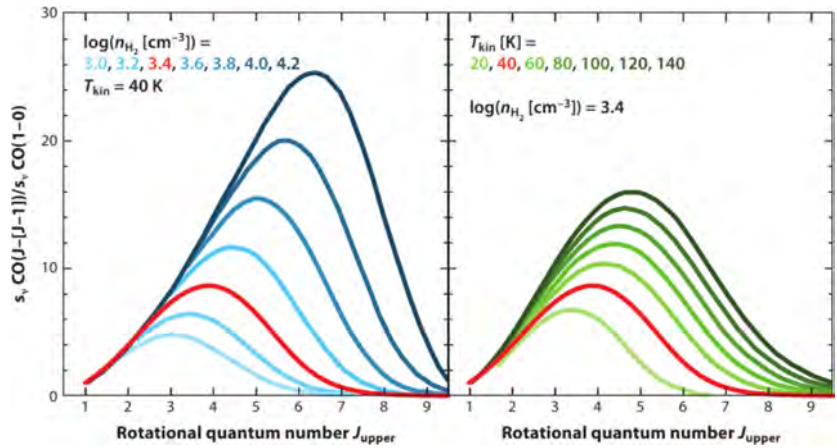

Figure 6-6 Normalized CO cooling curves as a function of temperature and density up to J=9 (taken from Carilli & Walter 2013).

In addition using models of single-component photon-dominated regions (PDRs) are usually insufficient to explain ISM energetics. In nearby starbursts, luminous IR galaxies (LIRGs), and galaxies hosting AGN, more PDR components are needed to explain the high-J emission, and frequently photon excitation is not enough; mechanical heating through shocks and/or turbulence is needed as well (e.g., Greve et al. 2014; Kazandjian et al. 2015) as shown in Figure 6-6 (Rosenberg et al. 2014, 2015). Understanding ISM heating and cooling through molecular excitation will constrain the process of star formation. Herschel made it possible to study the ISM cooling mechanisms through CO spectral line energy distributions (SLEDs) at low redshift, but to measure molecular excitation for galaxies with z>0.1 and toward the peak of the cosmic SFRD, more sensitive facilities are needed.

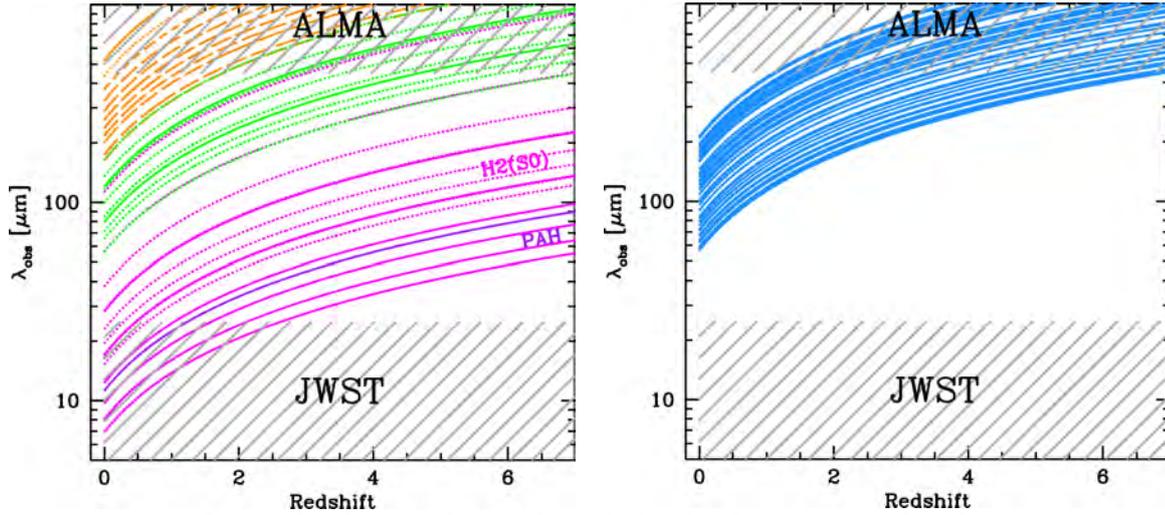

Figure 6.7: Left panel: Observed wavelength (μm) of various molecular transitions ($H_2$, HD, longest- wavelength PAH, OH, CO) plotted against redshift, z. Rotational $H_2$ transitions of $H_2$ [S(0) 28.2 μm to S(5) 6.9 μm] and HD [R(0) 112.1 μm to R(7) 15.3 μm] are shown as magenta curves (HD as dotted); the longest PAH (11.2 μm) as purple; OH as green (OH 119.4 μm and OH 79.2 μm as solid, remaining transitions as dotted); CO as orange. Right: Observed wavelengths of various $H_2O$ transitions plotted against z. Also shown in both panels are the wavelength limits of the two major complementary facilities, JWST and ALMA, showing the need for an observatory covering the wavelength gap.

$H_2O$ and other molecules (e.g., OH) are also important diagnostics to constrain the physics driving the ISM energy budget and its relation to star formation. The Herschel instruments PACS and SPIRE have made significant progress in our understanding of water vapor (non-maser) emission in galaxies; Herschel surveys of low-z LIRGs and ULIRGs reveal bright $H_2O$ lines that are comparable in strength to neighboring high-J CO lines (J = 8 – 7, J = 13 – 12) (e.g., van der Werf et al. 2010; Gonzalez-Alfonso et al. 2010; Rangwala et al. 2011; Kamenetzky et al. 2012; Gonzalez-Alfonso et al. 2012; Appleton et al. 2013; Omont et al. 2013; Meijerink et al. 2013; Pellegrini et al. 2013; Falstad et al. 2015). In SPIRE/FTS spectra, $H_2O$ is the strongest molecular emission line after CO (Yang et al. 2013). H2O emission is related to shocks or X-ray dominated regions (XDRs) (e.g., Meijerink et al. 2012, 2013; Fischer et al. 2014) and can be significantly enhanced in warm, dense environments. However, $H_2O$ vapor could be detected in only a handful of galaxies so far. Because $H_2O$ (and OH) is expected to dominate cooling in primordial galaxy formation, when oxygen has just appeared (e.g., Omukai et al. 2010; Schneider et al. 2012), it is of crucial importance to understand the mechanisms behind the emission of water vapor in galaxies.

Other simple molecular (e.g., $H_2$, HD, OH), hydride ions (e.g., $OH^+$ $H_2O^+$, $CH^+$), and macro-molecules such as Polycyclic Aromatic Hydrocarbons (PAHs) emissions are also important diagnostics for constraining star formation and ISM heating/cooling towards the peak of the SFRD at z ~ 2. Figure 6.7 shows the detectability of some of these as a function of redshift. The emission lines will only be detectable at z>0.5 with a sensitive FIR space mission; Herschel has just begun the story.

### 6.2.3 The ISM energy budget through FIR fine-structure lines

The energy budget of star-forming regions is regulated by balancing the heating and cooling in the atomic, ionized, and molecular gas phases. Stars form deep within molecular clouds, where dust and dense gas shield CO from photodissociation. On the other hand, at the surface of the clouds, the extinction is low, dust- and self-shielding are less effective, and UV radiation heats the gas, dissociating CO. Thus, there are chemically distinct transition regions within a star-forming region defined by relative abundances of various gas phases: from $H^+/H/H_2$ on the cloud inner surface (visual extinction $A_V\sim 0$) an intermediate depth ($A_V\sim 3-5$) where [CII], [CI], and CO coexist, to very deep within the clouds ($A_V>10$) where $H_2$ and CO are the dominant gas components (e.g., Hollenbach & Tielens 1999; Hollenbach et al. 2009). Since at low extinction (or low metallicity, see Bolatto et al. 2013), CO essentially disappears because of photo-dissociation, [CII] can be used as an effective tracer of "CO-dark" molecular gas (e.g., Wolfire et al. 2010; Leroy et al. 2011; Planck Collaboration et al. 2011; Lebouteiller et al. 2012; Pineda et al. 2014; see also Sect. 4.3).

FIR fine-structure (FS) lines are fundamental tools to understand the balance and energy budget of the different ISM gas phases and star formation; in the presence of dust extinction, they are virtually the only tools available. CO-dark gas can only be probed through FIR transitions. Neutral gas heating is thought to be governed mainly through photo-electric heating of the dust, namely electrons ejected from the surface of the grains by impinging UV radiation (e.g., Tielens & Hollenbach 1985). The cooling of the neutral gas is dominated by FIR FS lines with excitation potentials <13.6eV, mainly [OI] λ63, 146 μm and [CII] λ158 μm. Such lines, in particular [CII], can emit up to a few percent of the FUV energy, reprocessed by dust, from star formation. UV radiation from star formation also ionizes gas, which in the presence of dust is best traced by FIR FS lines including [OIII]λ52, 88μm and [NII]λ122, 205μm; [CII] also traces to some extent ionized gas, making its interpretation more complex than the other purely ionized tracers (e.g., Cormier et al. 2015; Croxall et al. 2015). First ISO and now Herschel observations have enabled us to make great progress in our understanding of [CII] emission in galaxies (e.g., Malhotra et al. 1997; Luhman et al. 1998, 2003; Gracia-Carpio et al. 2011; Sargsyan et al. 2012; Diaz-Santos et al. 2013; Farrah et al. 2013; De Looze et al. 2014; Rigopoulou et al. 2014; Magdis et al. 2014; Sargsyan et al. 2014; Ibar et al. 2015; Cormier et al. 2015; Herrera-Camus et al. 2015; Rosenberg et al. 2015; Spinoglio et al. 2015), and have paved the way to [CII] detections with ALMA in galaxies at very high redshift, close to the Epoch of Reionization (e.g., Maiolino et al. 2015; Capak et al. 2015; Willott et al. 2015).

[CII] emission, and to some extent [OI] and [NII], seem to correlate with other galaxy properties including $L_{IR}$, sSFR, $T_{dust}$, IR surface brightness $\Sigma_{IR}$, OH absorption depth, and other parameters (e.g., Luhman et al. 1998, 2003; Gracia-Carpio et al. 2011; Coppin et al. 2012; Sargsyan et al. 2012; Diaz-Santos et al. 2013; Farrah et al. 2013; Ibar et al. 2015; Gonzalez-Alfonso et al. 2015). Because of this, as discussed in Sect. 5.1, [CII] is also commonly used to trace the SFR (e.g., De Looze et al. 2014; Sargsyan et al. 2014; Herrera-Camus et al. 2015). However, these correlations break down (there is an emission-line deficit) at high $L_{IR}$ high specific SFR, warm dust temperature, high $\Sigma_{IR}$ and deep OH absorption. This causes a deficit that is attributed to strong far-UV radiation fields impeding the efficiency of gas heating by creating an excess of charged dust grains (e.g., Tielens et al. 1999; Croxall et al. 2012), although other mechanisms may play a role (e.g., Abel et al. 2009). Ultimately, because of the breakdown in the correlations, at some level the viability of using [CII] to trace SFR need to be fully calibrated. In this respect for metal-poor high-z galaxies, [OI] may be a better tracer of SFR at low metallicities in nearby dwarf galaxies (De Looze et al. 2014), and is in some cases brighter than the [CII] line (Cormier et al. 2015). [OI] is also brighter than [CII] in dense gas (Meijerink et al. 2007), and may ultimately prove to be a better tracer

of SFR at high redshift.

The FIR FS lines are potentially extremely powerful probes of the physical conditions of star formation in galaxies. However, despite the enormous advances made possible by Herschel (and previously ISO), the reasons for line deficits are still not understood. Currently, there is a huge gap in redshift between Herschel coverage of the FIR FS lines, and ALMA; this is illustrated in Figure 6-8 where the redshift variation of the rest wavelength of these lines is shown, together with the onset of the possibility of observation with ALMA. Because of the potential for ALMA observations of FIR FS lines at z>4, around the Epoch of Reionization, it is extremely important to understand the physics driving their emission; this will only be possible with a sensitive FIR space facility, capable of observing [CII] and [OI] up to z ~ 3, at the peak of the SFRD.

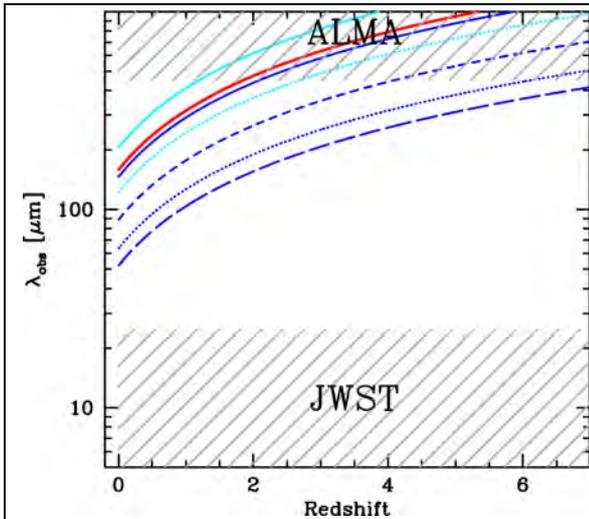

Figure 6-8 Observed wavelengths of various far-IR FS lines plotted against z. [OI] and [OIII] transitions are shown as blue curves ([OI] 145.5 μm is shown as a solid curve; [OI]63.2μm as dotted; [OIII]88.4μm as short-dashed; [OIII] 51.8 μm as long-dashed). Singly- ionized carbon [CII] is shown as a solid red curve; doubly-ionized nitrogen [NII] as cyan curves ([NII] 205μm as solid, [NII] 121.9μm as dotted). As in Figure 6.7, also shown are the wavelength limits of the two major complementary facilities, JWST and ALMA.

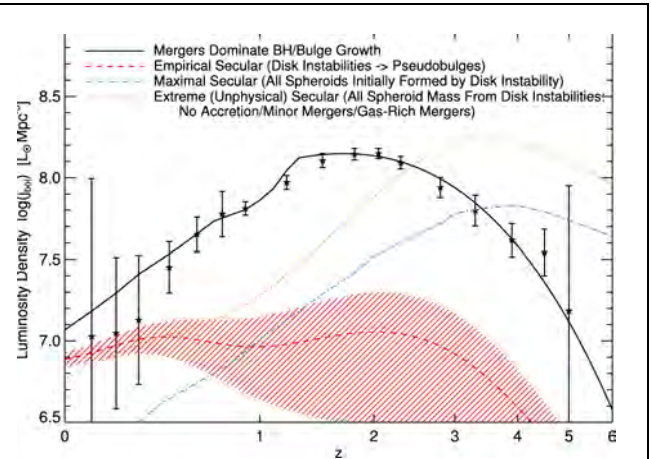

Figure 6-9 Bolometric quasar luminosity density as function of redshift (taken from Hopkins et al. 2008). Black stars show observations and various curves show estimates from different models (for more details, see Hopkins et al. 2008). The rise and fall of quasar activity mirrors that of the SFRD shown in Figure 6-1, Figure 6.2, Figure 6.3.

## 6.3 Co-evolution of supermassive black holes and their host galaxies

The observed correlations between SMBH masses and properties of the host galaxy (e.g., velocity dis- persion, bulge luminosity, mass: Ferrarese & Merritt 2000; Marconi & Hunt 2003; Gultekin et al. 2009) imply that BH accretion and galaxy growth are closely linked. This link is reflected in the form of co-moving SMBH bolometric luminosities as a function of redshift; as illustrated in Figure 6-9, it follows closely that of the co-moving SFRD, peaking at z ~ 2 (see Figure 6-1, Figure 6.2, Figure 6.3). Theoretical models suggest that the growth of black holes also regulates star formation, effectively quenching it in high-mass galaxies through powerful galaxy-scale outflows driven by AGN feedback (e.g., Croton et al. 2006; Hopkins et al. 2006, 2008; Somerville et al. 2008).

### 6.3.1 Identifying AGN feedback at high redshift

Molecular outflows are thought to be the "smoking gun" of AGN feedback because they clear the circumnuclear environment of gas that could have fueled subsequent episodes star formation. Herschel identified significant numbers of molecular outflows in LIRGs and ULIRGs in the Local Universe (see Sect. 5.2), but because of limited sensitivity, it was impossible to carry out similar observations at cosmological redshifts. Because of the co-dependencies between the outflow rates, the SMBH accretion rates (e.g., Cicone et al. 2014), and AGN gas accretion rates, the frequency of outflows and their properties are expected to evolve. Indeed, according to Veilleux et al. (2013), "there is some indication that molecular outflows subside once the quasar has cleared a path through the obscuring material".

Thus, powerful AGN-driven molecular outflows are probably only a phase of the AGN duty cycle, and their occurrence almost certainly changes with redshift, and may be related to the growth of SMBHs and to the evolution of the SFRD (see Sect. 6.1). OH outflows at 119µm can be traced up to z~3 with an FIR observatory that covers wavelengths to 500µm, and to z~5 using the 79µm OH transition (e.g., Sturm et al. 2011). *In order to characterize the rise and fall of the effects of AGN activity directly through systematic observations of outflow tracers, a sensitive space-borne FIR facility is needed.*

### 6.3.2 FIR fine-structure lines as tracers of black-hole accretion

AGN activity can also be traced using MIR/FIR FS lines that are excited by the hard UV continuum, i.e., at high-ionization potential; the brightest of these lines is [OIV]λ25.9µm (Sturm et al. 2002; Schweitzer et al. 2006; Armus et al. 2007; Diamond-Stanic & Rieke 2012; Shipley et al. 2013). Thus the [OIV] line is considered an important diagnostic to identify AGNs, and separate its contribution to the IR luminosity from that of star formation. Moreover, black-hole accretion rates are found to be proportional to [OIV] luminosities (e.g., Melendez et al. 2008), making [OIV] an important tool for measuring accretion rates onto SMBH in the presence of dust. The [NeV]14.3/24.3µm lines are also efficient AGN tracers, as their high ionization potential (~100eV) can only be excited by an extremely hard UV radiation field. These lines are generally weaker than [OIV], but can be extremely effective probes of AGN in FIR surveys (e.g., Bonato et al. 2015). Blind surveys with planned FIR space-borne missions in the [OIV] and [NeV] lines will prove to be an important diagnostic for the coevolution of star formation and AGN out to the peak of SFRD at z ~ 2 (e.g., Spinoglio et al. 2012a, 2014; Bonato et al. 2014b).

### 6.4 Scientific requirements

Here we examine the technical requirements, evaluated as a function of the science drivers described above. A summary of the requirements is given in the table at the end of the section.

In order to sample the peak (50–160µm) of dust emission in galaxies up to redshift z≲ 6, 500 µm is the minimum long wavelength required; such wavelength coverage will enable the verification of the trend of increasing dust temperatures with increasing redshift and constrain dust (and gas) mass and ISM physical conditions in galaxies within ~1 Gyr of the birth of the Universe. The required sensitivity of 5σ ~ 3mJy at 250μm is equivalent to the current most sensitive stacked Herschel measurements (Bethermin et al. 2015), and would enable direct detection of the SEDs of massive main-sequence galaxies up to z ~ 4; Milky-Way-like galaxies would be directly detected to z ~ 1.5. A resolution of ~12" at 250μm would give a confusion level ~10 times lower than with Herschel. A diffraction-limited 5m single-dish diameter would give ~12" resolution at 250 μm, sufficient to significantly improve sensitivity and ameliorate confusion constraints.

For the spectroscopic observations we require wavelength coverage λ~50-500 μm in order to trace the molecular and FIR FS line transitions in galaxies up to z ~ 2. Using model estimates from Bonato et al. (2014a) and published Herschel measurements a conservative estimate for the required line sensitivities would be of the order of a few × $10^{-21}$ $Wm^{-2}$. A spectral resolution of λ/Δλ~1500–2000 would be required for the science cases outlined above.

| Science case | λ coverage | Sensitivity | Spatial resolution | Spectral resolution |
|---|---|---|---|---|
| History of galaxies | ~100 to 500 μm | ~3 mJy cont | <12"(250μm) | 5 |
| Dust as tracer of gas content | ~100 to 500μm | ~3 mJy cont | <12"(250μm) | 5 |
| Molecules, ISM | ~50 -500μm | a few x$10^{-21}$$Wm^{-2}$ spectra | 15 to 20" (250μm) | 1000 - 2000 |
| FIR FS lines | ~50-500μm | a few x$10^{-21}$$Wm^{-2}$ spectra | ~15-20" (250μm) | 1000-1500 |
| AGN outflows | ~50-300μm | 3x$10^{-21}$$Wm^{-2}$ spectra | 12" (250μm) | 1500-2000 |
| SMBH accretion | ~25-150μm | 3x$10^{-21}$$Wm^{-2}$ spectra | ~15-20" (250μm) | 1000-1500 |

# 7 Technology and techniques[2,3]

## 7.1 Capability and summary of science driver requirements

In order to achieve the challenging science goals described in the previous chapters, substantial technological developments will be necessary. In many cases ultra-high sensitivity is needed, often in combination with medium to high spectral resolution and fast mapping speed. In the far-infrared this requires deep cooling (20-50 K for Schottky mixers, 4 K for HEB mixers, 100 mK or less for TES/KID or QCD detectors). Most science applications will benefit from large arrays, which put heavy demands on all system aspects.

To avoid photon noise from self-emission, telescopes also need to be cooled. For concepts such as CALISTO and SPICA, active cooling to 4-6 K is needed, challenging cryo-cooler technology and system design. Last but not least, high angular resolution can only be achieved through light-weigthed large single dishes, or by combining apertures to form an interferometer.

We should not underestimate the importance of early development. It took nearly 15 years before 1 THz mixers for Herschel/HIFI could be made. For the TES detectors of Athena/XIFU, almost two decades were needed. But even more important is that selection of science concepts by the Space Agencies ensures that the development is accelerated and well funded.

### 7.1.1 Sensitivity

To achieve sensitivity limited by astrophysical backgrounds rather than the thermal emission of the telescope itself, far-infrared telescopes need to be cooled to temperatures of a few K (Figure 7.1), and in addition their direct detection instruments must involve sub-kelvin cooling of the detectors. To achieve this, Herschel was equipped with a large cryostat for the instruments and the telescope itself was passively (radiatively) cooled to a temperature of ~ 85 K, with recyclable $^3$He refrigerators cooling bolometer arrays to 0.3 K in the PACS and SPIRE instruments. The Planck telescope used a design with several V-grooves that radiate away the heat and maintain each shield at a stable temperature. By stacking the V-grooves, the Planck telescope was passively cooled down to 40 K. Because cooling the telescope to a few K reduces the background photon noise by orders of magnitude, huge gains in sensitivity can be achieved. Studies show that 4 K can be reached for approximately a 2–3-m class telescope within an M-class budget (ESA CDF study Next Generation far-IR Telescope 2014/2015 http://sci.esa.int/trs/56108-next-generation-cryogenic-cooled-infrared-telescope/). Larger apertures will need a combination of active and passive cooling. The requirements of heterodyne systems on telescope temperature are much less stringent

---

[2] This section is based on inputs from Brian Ellison, Jian-rong Gao, SPACEKIDS consortium, Juan Bueno, Colin Cunningham, Matt Griffin, Matt Bradford, Marc Sauvage, Dimitra Rigopoulou, Bruce Swinyard, Peter Roelfsema, Fabian Thome and Gilles Durand
[3] A THz roadmap also exists (Dhillon et al. 2017), but this only has a small section on space applications.

than for direct detection instruments, and passive cooling of the aperture is all that is required.

Herschel was possibly the last satellite to carry a large liquid helium cryostat into space. In the future cryocoolers, which are now space qualified, will be the obvious choice. Since it is impossible to cool down from ambient (either room temperature or 70 K for passively cooled systems) to 4 K with a single cooler, chains of coolers need to be used, each providing enough heat lift for the level below. In both SPICA and FIRSPEX the problem is solved by using a chain involving 20-K class two-stage Stirling coolers, and additional Joule-Thomson coolers providing 4-K and 1-K stages (ESA CDF study).

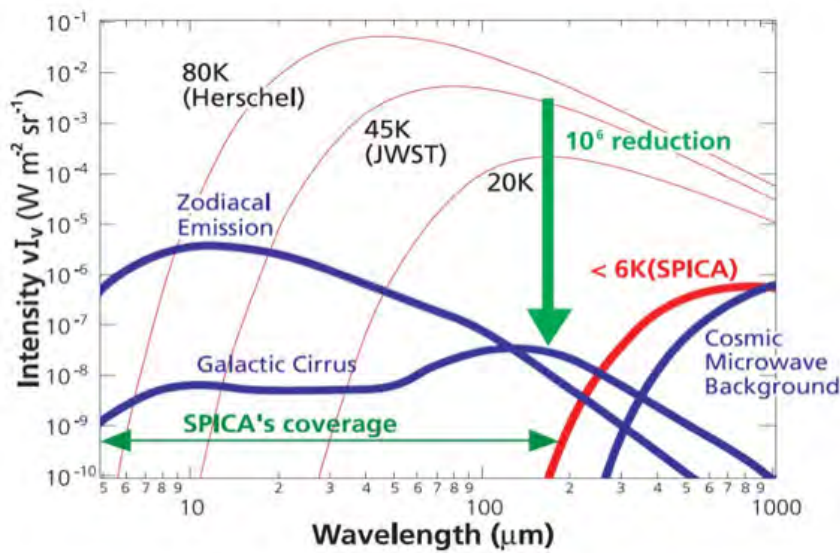

Figure 7.1: Comparison of natural backgrounds (zodiacal emission, galactic cirrus, and cosmic microwave background radiation) with those of thermal radiation from telescopes as a function of temperature. Credit: SPICA consortium.

### 7.1.2 Spectrometers

In the far-infrared one can use both coherent and incoherent systems as spectrometers. The choice of system should be determined by the science objectives; that is, it must possess the necessary sensitivity, spatial and temporal resolution, and spectral resolving power. When moderate spectral resolution ($R <$ a few 1000) is adequate (for instance, to measure line intensities) then the best sensitivity can be achieved using a direct detection system (in which the phase of the incident radiation is not measured), which, with suitably sensitive detectors, is subject only to photon noise from the incident radiation. If high resolution ($R >$ around $10^4$) is needed, for instance to resolve narrow line profiles, then heterodyne detection (in which both phase and amplitude are measured) is necessary. Simultaneous measurement of amplitude and phase introduces, via the uncertainty principle, a component of quantum noise that imposes an ultimate limit on sensitivity. Heterodyne systems thus prioritise spectral resolution over sensitivity whilst direct detection systems do the opposite.

**Incoherent (direct detection) systems**

Classical direct-detection spectrometer concepts, which have been implemented in space astronomy missions, include the Fourier Transform Spectrometer (e.g. Herschel/SPIRE) and the grating spectrometer (e.g. Herschel/PACS; SPICA/SAFARI) and Fabry-Perot spectrometer (e.g. ISO/LWS). All of these have their own advantages and disadvantages when it comes to spectral resolution, mapping speed and instantaneous wavelength coverage, which can be optimised for the science goal. An FTS is compact and suitable for imaging, and provides instantaneous coverage of its complete wavelength range, which makes it very suitable for survey spectroscopy. A disadvantage is that the detectors see the background from sky and/or the telescope over the full wavelength range and so the photon noise contribution is higher than for a monochromator system such as a grating. For the European instrument on SPICA the original choice was first to have an imaging FTS, but higher sensitivity for point sources pushed the design to a grating spectrometer with a Martin-Puplett interferometer added for higher spectral resolution for bright sources. A price paid for this additional sensitivity is that the imaging capability of the system will be very limited.

Alternative schemes have been studied as well. In the BLISS concept (Bradford et al. 2010) the classic grating spectrometer is morphed into a set of waveguide/grating slices, together forming a low-to-medium resolution spectrometer. Concepts designed to provide spectral sensitivity at pixel level in a detector array using superconducting transmission line technology are also being developed. μ-Spec (Cataldo et al. 2014) proposes the use of superconducting microstrip transmission lines on a single 4-inch silicon wafer. μ-Spec is equivalent to a grating spectrometer, in which the phase changes introduced by grating grooves are instead produced by propagation along transmission lines of different lengths. Superspec (Hailey-Dunsheath et al. JLTP) uses half-wave resonators to implement a moderate-resolution ($R \sim 500$) filter-bank coupled to lumped element kinetic inductance detectors (KIDs), all integrated on a single silicon chip. A prototype has been operated in the 180–280 GHz range. DESHIMA (Endo et al. 2012) is based on a similar concept using microwave KIDS (MKIDs).

### Heterodyne (coherent) systems

FIR direct detection instrumentation, whilst offering broad spectral coverage and high sensitivity, is typically limited to a resolving power $< 10^4$, which is insufficient to resolve narrow spectral lines. Many applications will need high spectral resolving power ($\lambda/\Delta\lambda$) of order $10^6$ in the FIR. This requirement is necessary to avoid spectral line confusion, to discriminate between emission and absorption, and to probe the kinematics of interstellar gas. For highest spectral resolving power the heterodyning technique must be used, in which the phase of the incident radiation is preserved. The sky signal is mixed with an extremely pure local oscillator (LO) signal. The resultant output is a down-converted intermediate frequency (IF) signal that contains the same spectral information as the original, but typically in the GHz frequency range where it can be electrically amplified and more easily processed.

Although possessing limited spectral bandwidth, the heterodyne technique provides a spectral resolution capability that is limited by the spectral purity, of the LO, and resolution orders of $10^7$ are achievable. Additionally, in preserving signal phase, the heterodyne technique provides another important advantage of interferometric observation. With the latter, multiple individual telescopes (antennas) can be united to form phased arrays, e.g. NOEMA and ALMA, that synthesize a large single dish aperture and provide ultra-high spatial resolution.

Heterodyne radiometry has been very successfully used in a variety of spaceborne missions in support of both Earth observation and astronomy remote sounding experiments (Gaidis *et al.*, 2000, Waters *et al.,* 2006, and de Graauw *et al*, 2010). The sensitivity of a THz radiometer is usually dominated by the mixer which, in combination with the LO, performs the necessary frequency translation. Efficient coupling of the free-space signal to the mixer is important. It important that the resistive losses in the fore-optics components are minimised in order to avoid excessive increase in system noise. Coupling of the focused energy to the mixer diode is achieved through use of a miniature antenna structure or feedhorn. Fabrication of feedhorns suitable for operation in the THz region is highly demanding. Corrugated feeds produce excellent antenna patterns and are relatively wide-band, but at the higher frequencies they are difficult to manufacture due to the small feature sizes involved. Smooth walled feedhorns, whilst being mechanically easier to construct, have a smaller bandwidth, but their antenna patterns can approach the quality of corrugated feeds with careful design. Examples of both feedhorn types are shown in Figure 7.2.

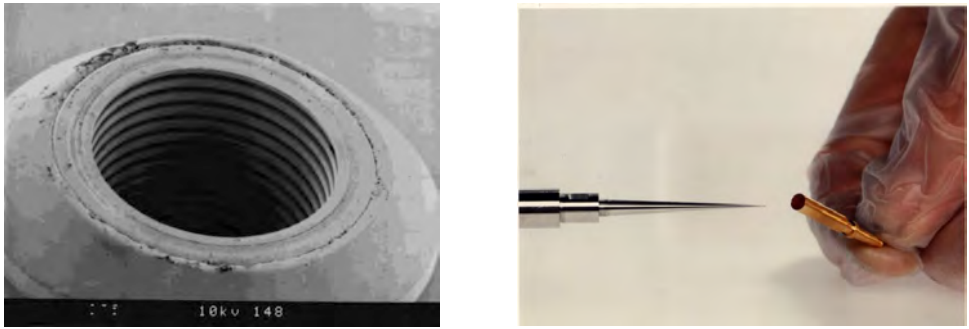

Figure 7.2: THz scale corrugated and smooth wall feedhorn examples.

The THz noise performance of the non-linear mixing element has been substantially enhanced in recent years due, primarily, to a transition from the use of semiconductor (Schottky diode) technology to that of superconducting thin film technology. Superconducting mixers require cooling to temperatures of 4 K or less. Although Schottky mixers can operate ~80 K, offering considerable advantages in terms of mission viability, cost and lifetime, this is at the expense of significantly reduced sensitivity.

### 7.1.3    Angular resolution

Angular resolution is proportional to telescope size and inversely proportional to wavelength. The long wavelengths of the far infrared waveband therefore require large telescope size. Sauvage et al. (2013) write: "Science exploiting the FIR domain is thus relatively young, yet already demonstrates an impressive track record: a succession of facilities (KAO, IRAS, COBE, ISO, Spitzer, AKARI, Planck and Herschel) allowed us to gaze into the obscured Universe, advancing our understanding of cosmology, star and galaxy formation, and the origin of planetary systems. Despite these developments, FIR observational capabilities remain primitive in comparison with the optical/NIR region. Our most advanced facility, Herschel, delivered an angular resolution no better than Galileo's telescope and was operated against a blinding thermal background." While the Herschel telescope represented highly advanced technology at the time (~300 kg mass for almost 10 m$^2$ area - a factor of five lighter than a classical glass mirror), larger area mirrors are still prohibitively heavy and alternatives are needed, either by light-weighting the dishes, stepping away from filled apertures or a combination of the two. We note that there is only one exception to the need for larger apertures - mapping the ISM of the Milky Way and beyond in the far-IR fine-structure lines, of which the 1.9 THz [CII] line is the most prominent. In this case a telescope diameter of 1 - 2 m is sufficient, because the ISM is considered as a whole rather than as individual sources and because this allows for direct comparison between [CII] and Herschel/SPIRE 500-μm photometric images.

In 2007 Sauvage and colleagues proposed TALC, a segmented annular mirror of 20 m outer diameter and 3 m width. This would provide 20 times the collecting area of the Herschel telescope. The ring configuration is maintained with cables attached to a deployable mast, which provide structural rigidity through compression. Within ESA, large-aperture telescope technology has been studied (e.g. Gambicorti et al. 2012 and references therein), with a concept involving surface density of less than 18 kg m$^{-2}$ including backplane and actuators for getting the right shape. Alternatives were also investigated over a decade ago. The Dual Anamorphic Reflector Telescope, which was studied for the far-IR by NASA (Morgan et al. 2004), is an architecture for large aperture space telescopes that uses membranes. A membrane can be shaped in one direction of curvature using a combination of boundary control and tensioning, yielding a cylindrical reflector. Two orthogonal and confocal cylindrical reflectors constitute the unobstructed 'primary mirror' of the system.

Keeping membranes in shape with electrostatic pressure was also studied (e.g. Errico et al. 2002). More recently Ball Aerospace and DARPA studied large Fresnel lens-like etched membranes in the MOIRE program (Membrane Optical Images for Real Time Exploitation, press release Ball May 2014) yielding very large mirrors suitable for spying from geostationary orbit in the optical. In the same area of spy satellites, photon sieves have also been studied (Andersen 2005). The same technique could be scaled to far-infrared wavelengths, but large focal length and wavelength-dependent focus may be difficult to avoid.

In the USA, the ATLAST program studied segmented deployable mirrors up to a limit of 16 m as successors to JWST in the UV/optical. The expected capabilities of the Ares V launch system are such that heritage of JWST is preferred over completely new developments. ATLAST is technology for a High Definition Space Telescope. . Because of the start of the study for a far-infrared observatory for the next decadal plan, new impetus is given to finding new materials for lightweight mirrors. P. Stahl (private communication) showed test blanks of laser-sintered Aluminium which can be polished to the accuracy needed for the far-IR. The Active Structures Laboratory has, on its website, information on lightweight telescopes: http://scmero.ulb.ac.be/project.php?id=8&page=index.html

The far-infrared is less affected by wavefront errors than the optical. However, if mirrors get bigger and thinner, the incoming wavefronts will need to be actively controlled. One possibility is correction using free-form optics along the optical train (e.g. http://www.nasa.gov/feature/goddard/out-with-the-old-in-with-the-new-telescope-mirrors-get-new-shape or http://www.astro-opticon.org/fp7-2/jra/wp5_active_freeform_mirrors.html), but large-aperture telescopes will also need an active secondary or tertiary to compensate for static and dynamic (e.g. thermal drift) errors. If flatness or shape of large membranes is a problem, actuation will also be needed on the primary. In general, large deployable mirror structures inherently require active adjustment, alignment, and control in space; and, because of their long focal lengths, they also need other deployable structures like masts and booms in a deploy-and-lock situation. This is recognized by ESA who write in ESA ITT 8248: "Large apertures (antennas and telescopes) and long baselines and focal lengths in space have applications for telecommunications, Earth observation and scientific missions. Astrophysics missions need deployable structures mostly for the creation of long base lines of interferometers and large focal lengths, e.g. for X-ray telescopes. Missions currently in early phases may require for instance deployable booms of up to 100 m deployed length and positioning accuracy within a sphere of 1 mm radius or better and rotations lower than 0.005 degree."

Structureless telescopes have also been considered, inspired by Labeyrie's hyper-telescope (see also Bekey 2003). These concepts are based on free-flying, a technique discussed under interferometry, although a hyper-telescope has extra challenges for station keeping, since large distances and large numbers of elements are involved.

### 7.1.4 Polarimetry

To date, most dust emission polarisation measurements have been done from Earth-based observatories. An exception was the Planck satellite, which has revealed the large scale magnetic fields in our Milky Way, based on its polarimetric capabilities. However, now that Herschel has mapped the intensity of dust emission from large star-forming regions in the Galaxy, it is important to investigate polarimetric imaging concepts enabling the interplay between magnetic fields, turbulence and kinematics to be studied as an important aspect of ISM and star-formation research. Numerous ground-based, balloon- and aircraft-borne polarimeter instruments will be available in the coming decade for such studies. While these facilities will have powerful capabilities over a range of angular resolutions, a spaceborne imaging polarimeter on a cold-aperture telescope would have far superior sensitivity and could achieve high-fidelity polarimetric imaging over a range of angular scales as Herschel was able to do in intensity.

### 7.1.5 Interferometry

The TALC concept shows that a large telescope can be built with relatively simple building blocks and in a scalable fashion. However, there will be a point when interferometers will be easier to handle than large dishes, because of the sun-shield/V-grooves scaling with telescope size.

In the last decade two interferometer concepts have been proposed, based on different detection schemes: coherent, or heterodyne, detection and incoherent, or direct, detection. The first technique is similar to that used by ALMA, the second to VLTI. In interferometry a few main concepts/parameters drive the design: collecting area, baseline length, uv-plane filling and image quality.

A comprehensive study of a direct detection interferometer, using the double Fourier technique to achieve simultaneous spectral and spatial interferometry, was undertaken by Savini and collaborators in the EU FP7 project FISICA (http://www.fp7-fisica.eu/). The European concept was very much like the SPIRIT concept in the NASA studies (Leisawitz et al. 2007a) and the earlier SPECS study (Leisawitz et al. 2007b). NRAO (Condon and collaborators) has done a detailed study also of the heterodyne interferometer in 2014, but the report is not freely available.

The sensitivity of an interferometer depends on the total collecting area of the apertures. The maximum baseline determines the angular resolution on the sky, so large baselines are needed for high angular resolution science: to reach an angular resolution of 0.1" at 100 μm requires a baseline of ~200 m. Combining two elements in interferometric mode does not provide an image but a so-called visibility function, which can be compared with the visibility function of the model of the astronomical source. For many applications, imaging is much preferred over fitting the visibility function. This can be achieved by moving the telescopes in such a way that the Fourier plane of the baselines (also called the uv-plane) is sampled as homogeneously as possible. In space, the telescopes could be moved through electric propulsion in free-flying configurations, by means of tethers, which are shortened and lengthened, or by movement along a spinning rail to achieve the uv-plane filling. It is important to note that none of these methods provides instantaneous imaging and that precise metrology is needed between the optical elements and the location at which photons are combined for the incoherent system and that complete phase knowledge is needed for the coherent system. Full restoration of an image with structure larger than the primary beam (e.g. nearby galaxies, outflows, evolved stars etc.) also requires access to the zero and short spacings.

Free-flyers have been considered by several space agencies as risky, but in the last five years NASA's Earth gravity GRACE mission and in the near future ESA's PROBA-3 mission will have shown that precise free-flying is achievable, and future distributed space systems have standard collision avoidance. American studies have shown that tethers are a viable alternative for free-flying, but repointing is likely more difficult in a tethered system. For shorter baselines, up to at least 50 m, deployable booms with rails for the telescopes can also be used. Studies for the gravitational wave interferometer mission eLISA show that station keeping can be done with high precision metrology. Far-infrared interferometric imaging is conceptually much easier than operation of eLISA to detect gravitational waves, since distances involved are of order hundreds of meter instead of millions of km.

## 7.2 Ultrasensitive detectors

### 7.2.1 Incoherent detectors

Future space science and Earth observation missions will rely on the availability of imaging detector array technology for the mm-FIR wavelength range (3 mm to 30 μm). For the wavelength range 28-45 μm there is currently no high performance technology with space heritage. For longer wavelengths, 45-2000 μm, there is European expertise in a range of detector technologies, including photoconductors, semiconductor and transition edge superconducting (TES) bolometers, and kinetic inductance detectors (KIDs). Except for KIDs, these technologies present significant fabrication difficulties, and lead to a high degree of complexity of system integration and readout electronics for the large format arrays demanded by the next generation of astronomical missions such as SPICA, FIRSPEX, CoRE+, LiteBIRD, TALC or FIRI. Large format arrays are also needed for low-to-medium resolution spectroscopy, where confusion is less of a problem because of the spectral dimension.

There are a number of approaches to far-IR and sub-mm photodetection using large-format superconducting detector arrays which are being explored as possible technologies to obtain the very low noise equivalent powers (NEPs), as low as $10^{-20}$ W/Hz$^{1/2}$, required for optimal exploitation of the cold-aperture telescopes in space. One such technology, the kinetic inductance detector (KID), relies on the sensitivity of the surface inductance of a superconducting film to the absorbed electromagnetic power through the phenomenon of Cooper pair breaking. Two others, the transition edge sensor (TES) and the nano-hot electron bolometer (nano-HEB), make use of the sharpness of the superconducting transition. The approach of the quantum capacitance detector (QCD) (e.g. Shaw et al. 2009, Bueno et al., 2010, 2011) is based on the extreme susceptibility of the single Cooper-pair box, a mesoscopic superconducting device, to pair-breaking radiation.

Of these detectors the TES are most mature because of their use in submm camera's for measuring the Cosmic Microwave Background and general astrophysics from dry sites on Earth (PolarBear (Hattori et al. 2016), BICEP-2 (Ade et al. 2015), etc., etc.). For space, without the atmosphere as photon noise source, the required sensitivity is at least two orders of magnitude better posing new problems, just as multiplexing thousands of pixels in space: problems which currently are being solved under the SPICA/SAFARI umbrella (see e.g. Van der Kuur et al 2015 and references therein).

**Kinetic Inductance Detectors**

Microwave Kinetic Inductance Detectors (MKIDs) are pair breaking detectors, in which radiation is absorbed in a small superconducting film at very low temperatures $T \ll T_c$ ($T_c$ being the critical temperature of the film). Photons at an energy $h\nu > 2\Delta$ ($\nu > 80$ GHz for aluminium) can break Cooper pairs, paired electrons that form the ground state of the superconducting film, into single particle excitations (quasiparticles). The result is a change in the complex surface impedance of the superconducting film. In a KID this superconducting film is placed inside a high quality factor microwave resonance circuit that is coupled to a microwave feed-line. Absorption of a photon in the film modifies the resonant frequency $f_0$ and quality factor, $Q$, of the circuit due to the changes in the complex impedance of the superconductor. As a consequence photon absorption alters the transmitted phase and amplitude of a microwave readout signal at the resonance frequency. By coupling many KIDs, each with a slightly different resonance frequency to a single feed-line, one can read-out a large number (several thousand) of KIDs in a bandwidth of a few GHz using a standard HEMT amplifier. KID arrays are now being deployed on a number of balloon-borne and ground-based telescopes. Figure 7.3 shows KID arrays manufactured for the APEX AMKID instrument.

In the FP7 SPACEKIDS project, an extensive study was done of the different kind of KIDs (antenna or absorber coupled devices), new materials, optimization of the radiation coupling, methods of reducing susceptibility to cosmic rays, KID multiplexing, minimizing crosstalk, and the development of the necessary readout electronics. Other KID studies currently being carried out include spectrometers-on-a-chip and multi-object spectrographs.

The SPACEKIDS project has made significant strides in many areas of KID technology and has also shown that a CMB satellite experiment has very different requirements than astrophysics mission. As a follow-up to SPACEKIDS, further work remains be done on reaching high sensitivity, especially in wavelength bands other than the submillimeter. For interferometry the speed of response needs be substantially reduced.

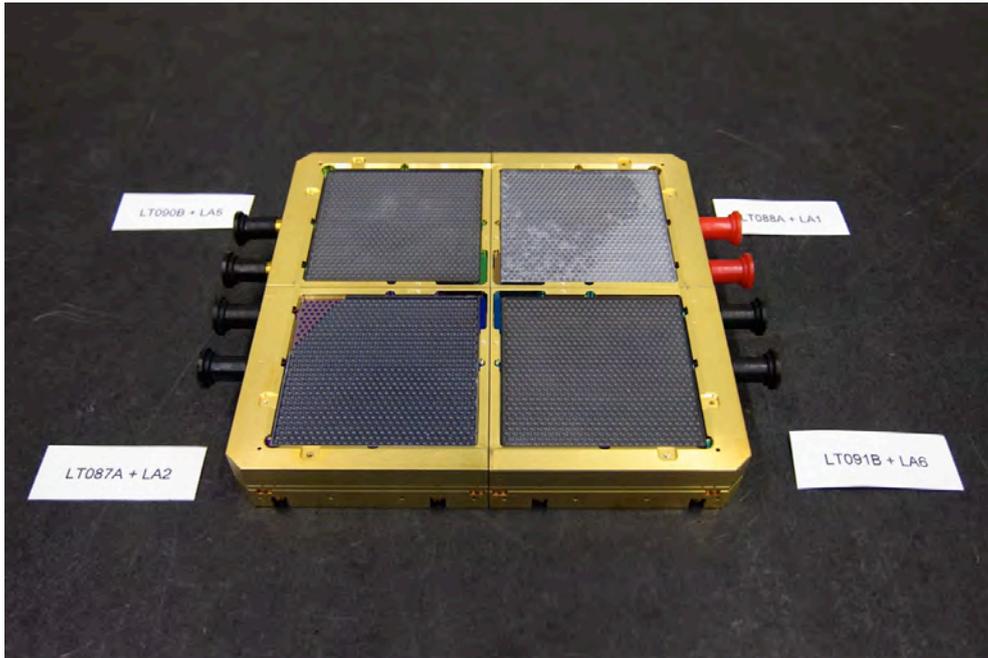

Figure 7.3 MKID arrays for the AMKID low band, ready for shipment to the APEX telescope

**Quantum Capacitance Detectors (QCDs)**

In a QCD, radiation coupling via the antenna breaks Cooper pairs in the absorber, generating quasiparticles that can tunnel in and out of a so-called "island". The rate of tunnelling in is proportional to the quasiparticle population in the absorber, but the rate of tunnelling out is largely independent of the quasiparticle population in the absorber. When the gate voltage is swept, one observes a series of peaks in the capacitance of the island (Duty et al. 2005) with amplitude proportional to the optical signal power. The change in capacitance is measured by the phase shift of an RF signal transmitted through the feedline caused by the change in the resonant frequency of the resonator. One can think of the QCD as a KID with a qubit capacitively coupled to it. Therefore, QCDs are naturally as easy to multiplex as KIDs. QCDs have demonstrated photon shot noise-limited performance with respect to the absorbed power (Echternach et al. 2013) for optical loadings between $10^{-20}$ and $10^{-18}$ W, corresponding to an NEP below $10^{-20}$ W Hz$^{-1/2}$ at 1.5 THz.

QCD fabrication is more complicated than for KIDs since a Josephson junction is required. So far, small arrays (25 pixels) have been fabricated at JPL/NASA. Scalability to larger arrays may be an issue because it is difficult to control the quality of hundreds or thousands of junctions over a whole four-inch wafer. The optical loading range in which QCDs are photon noise limited is only 0.01 – 1 aW, although it is possible to increase it at the expense of reducing the detector sensitivity by making larger absorbers.

### 7.2.2 Coherent systems and components

**SIS mixers**

During the last 20 years most astronomical heterodyne receivers have used Superconductor-Insulator-Superconductor (SIS) devices in which quantum tunnelling is the mechanism for mixing. SIS devices provide large bandwidths and high sensitivity (close to the quantum limit). They can only operate up to twice the superconducting gap width of the materials used, so the most commonly used niobium-based mixers are only usable below 1.5 THz. SIS devices are still the prime choice for frequencies below 1 THz.

### HEB mixers

A hot electron bolometer mixer consists of a thin film superconducting bridge that is contacted with an Au antenna structure. The most suitable material is a niobium nitride film, which requires a typical operating temperature of 4-5 K. Low-noise HEB mixers (Zhang et al. 2010) have been demonstrated up to 5.3 THz. For example, a DSB noise temperature of 800 K was measured at 4.7 THz (Kloosterman et al. 2013). HEB mixers, depending on their volume and critical temperature ($T_c$), require LO power as low as ~ 100 nW and, unlike SIS mixers, don't need a magnetic field, making them excellent candidates for large focal plane unit arrays. A small array of four pixels has been demonstrated at 1.4 THz in the lab using a Fourier Phase reflective grating to distribute the single LO into multiple beams for the array, which consists of quasi-optical twin slot antenna coupled HEB mixers. Arrays with up to 16 pixels at frequencies up to 4.7 THz are currently in construction for SOFIA. The development of array receivers was stimulated by programs within the EU 6$^{th}$ and 7$^{th}$ Framework Program under the umbrella of RadioNet.

Typical IF bandwidths are < 4 GHz, and an outstanding question in the field of HEB mixers is how to increase the IF bandwidth without compromising sensitivity to serve for extra-galactic astrophysics where larger bandwidths are needed.

### The Schottky Barrier Diode Mixer

The Schottky barrier diode (see Figure 7.4 for an example picture), can achieve good sensitivity over a wide input signal and IF bandwidth, and wide operational temperature range. For instance, previous development work at both the University of Virginia, RAL and the Jet Propulsion Laboratory demonstrated, between over 15 years ago, a room temperature system noise performance at 2.5 THz suitable for detection of atmospheric constituents (Crowe 1996, Ellison 1996, Gaidis 2000).

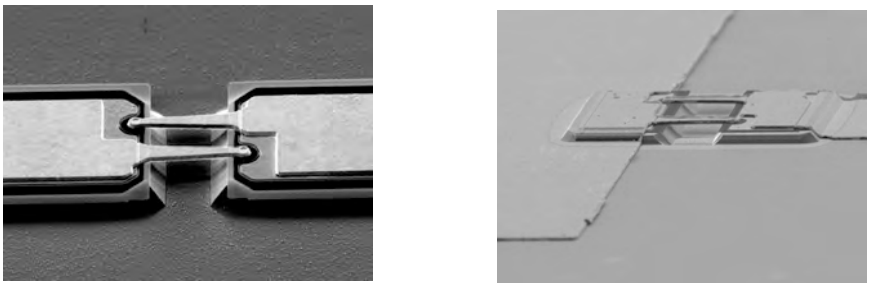

Figure 7.4: Example pictures of RAL fabricated air-bridge planar Schottky diodes – balanced diode (left) and fully integrated structure, i.e. including filtering, (right). Anode sizes are in the region of 1 to 2μm diameter.

Subsequent developments in diode fabrication, mixer embedding circuit design and post mixing low noise amplifier performance, have produced increasingly impressive results. From this it is clear that although there is significant improvement still to be achieved in the 3 – 5 THz range, between 0.8 THz and 1.1 THz heterodyne systems offer a reasonable level of technical maturity and with mixer noise performance typically about 2,000K in DSB operation and for room temperature operation.

### Other detectors for heterodyne systems

Another development is to step away from heterodyne mixing and do direct detection of radiation in microstructures (Microwave Monolithic Integrated Circuits or MMIC) or even in CMOS technology driven by telecom needs. If the frequencies can be scaled up (and results at JPL indicate this is possible) and sensitivity improved at the same time this may turn out to be the technology that can replace SIS. http://publica.fraunhofer.de/dokumente/N-145908.html A good overview of developments in the world of solid-state integrated circuit amplifiers is Samoska 2011.

### Local Oscillators

The provision of local oscillator power is a demanding requirement in the THz region, especially at higher frequencies, until lasers take over around 10 micrometer. Tunability, purity and stability are also important requirements.

Quantum cascade lasers (QCLs) have been demonstrated as local oscillator (LO) sources at THz frequencies (Dean 2011, Valavanis 2009 and see summary by Williams, 2007, Kloosterman et al. 2013). In a QCL the active region of the heterostructure consists of a stack of repeated identical quantum well modules (typically 200), which enables a single electron to cascade down and emit a photon in each module. The precise operating frequency is determined by a distributed feedback (DFB) grating structure of the laser. Currently a selectivity of around 5 GHz has been achieved around the QCLs natural output frequency, with 1 GHz being typical.

Despite the requirement for cryogenic operation, QCLs offer very considerable promise of achieving the required power levels and in a highly integrated form that is compliant with the spacecraft available resources. An example QCL device, fabricated by the Institute of Microwaves and Photonics at the University of Leeds is shown in Figure 7.5.

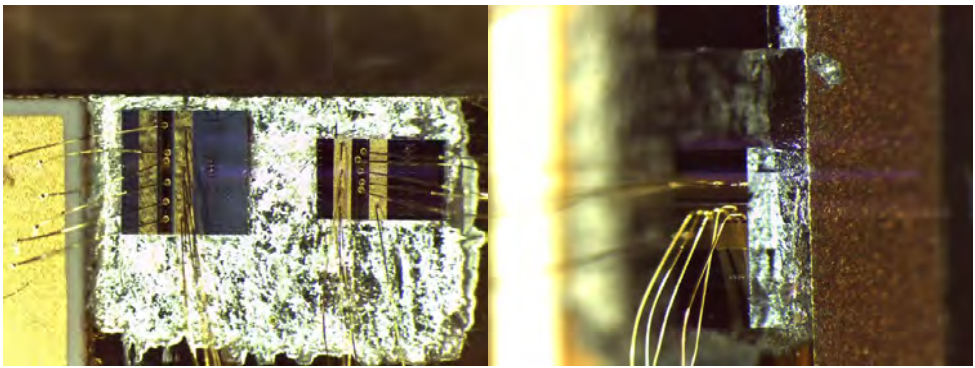

Figure 7.5 Quantum Cascade Lasers developed by the University of Leeds (left top view, right end view).

The QCL possesses an inherently narrow linewidth, limited ultimately by quantum fluctuations, but it is affected by the properties and quality of the resonant cavity in which it is embedded, thermal variations, and electrical noise. A spectral resolution approaching 1 MHz requires the introduction of some form of active frequency stabilisation circuit, i.e. a frequency or phase lock loop (e.g. Sirtori et al. 2013, Hayton et al. 2013). Such a scheme will be tested in a balloon borne telescope (STO-2) from NASA's Long Duration Balloon facility in Antarctica at the end of 2016.

Future development work on QCLs should focus on: a) increasing the frequency tuning range from currently ~ 1 GHz to ~ 5 GHz; b) further increasing the maximum operating temperature from ~ 70 K to ~ 200 K; c) increasing output power up to 1 mW, enabling operation of a large array of 64 pixels; d) development of superlattice harmonic mixers for frequency locking. need to be better understood and thus become more reliable to operate in a real instrument.

**Real-time spectrometer backends**

Spectrometer backends in heterodyne systems operate at a relatively low frequency (the intermediate frequency) where electronic signal processing methods can be applied. This allows in principle to achieve arbitrary high spectral resolution. Since the down-converted FIR signals to be analysed are generally very weak in relation to the receiver temperature, usually millions of spectra are averaged before satisfying signal-to-noise ratios are achieved. This means that the signal has to be converted instantaneously, so that real-time signal processing techniques have to be applied (in order to achieve a duty cycle of 100 %). Classical spectrometer backends are filterbanks, Autocorrelator Spectrometers (ACS), Acousto-Optical Spectrum Analysers (AOS), Chirp Transform Spectrometers (CTS) and meanwhile also the Fast Fourier Transform Spectrometer (FFTS). While filterbanks, ACS, AOS and CTS have space heritage, large bandwidths FFTS are relatively new and were not flown yet. Nevertheless the FFTS technique (see Figure 7.6) is very promising, since once developed for space application (e.g. low power consumption and mass, radiation tolerant), it is straightforward and cost efficient to produce FFTS' in masses. The latter is mandatory in order to cover all the many IF-bands required for the spectral coverage between 400 and 500 GHz at once. For a future FIR space borne observatory it is desirable to achieve an instantaneous bandwidth of 8 GHz with a spectral resolution of 50 kHz at the same time and a power consumption of < 1 W/GHz bandwidths.

A schematic layout is shown in Figure 7.6, in which the IF signal from the front-end is sampled at a GHz rate and with multi-bit analogue to digital conversion precision. The data stream is accumulated and then processed via a digital signal processor (DSP), which performs a Fast Fourier Transform (FFT) and produces a power spectrum. The spectral data is, after compression, either transmitted to Earth or stored for transmission at a later time.

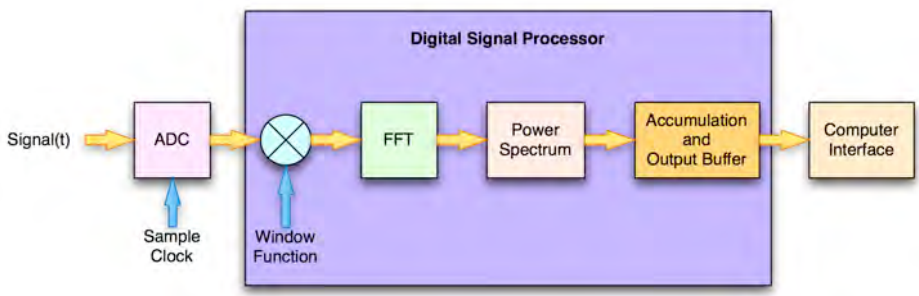

Figure 7.6: Top-level schematic concept of the DSP base back-end spectrometer unit.

# 8 Mission Descriptions

## 8.1 FIRSPEX (proposed for M5 Cosmic Vision)

The Far-Infrared Spectroscopic Explorer (FIRSPEX) is a European-led mission developed to enable sensitive large area, high spectral resolution (>$10^6$) surveys of the IR sky in the THz regime. FIRSPEX uses superconducting mixers configured as tunnel junctions and hot electron bolometers in conjunction with frequency stable local oscillators (LOs) and advanced digital sampling and analysis techniques.

FIRSPEX comprises a heterodyne payload cooled to 4K and a 1.2m primary antenna to scan large areas of sky from L2 (Figure 7.7). There are four parallel receiver channels that can operate simultaneously and therefore independently sample neighbouring regions of the far-IR sky. Each receiver channel is located within the focal plane of the 1.2m primary and offers angular resolution of the order of 1 arcmin. The instrument is passively cooled to 50K (L2 orbit) and, with active cooler technology providing sub-stages with necessary heat lift at 4K and 15K. The 4K stage cools the sensitive mixers and the first stages of low noise amplification. Each receiver operates in a double sideband configuration. In total there are 7 sampling pixels on the sky. The frequency band allocations are described in the Table along with estimated system sensitivities. Bands 1 through 3 use two independent mixers per frequency band giving multiple pixel sampling to compensate for the smaller beams. Coupling to the primary quasi-optical focal plane is accomplished via relay optics comprising a series of re-imaging mirrors. For Band 1, 2 and 3, conventional harmonic frequency up-convertors provide suitable sources of LO power injected into the mixer using simple beam splitter. For band 4, the LO source is provided by a quantum cascade laser (QCL) cooled to ~50K. Each mixer is followed by a cooled LNA and a further stage of ambient temperature amplification, while the IF final output is processed by a dedicated fast Fourier Transform spectrometer (FFTS).

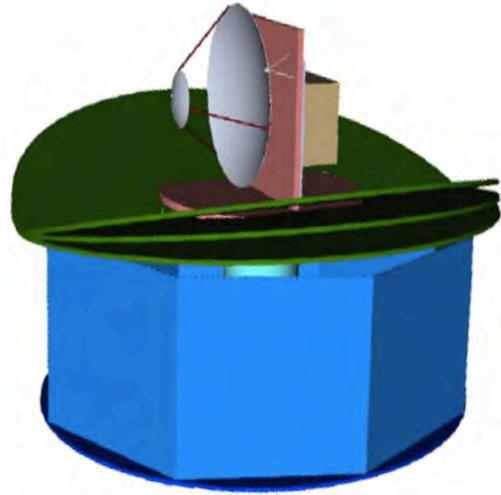

Figure 8.1: FIRSPEX mission concept

| Designation | Frequency (THz) | Primary Species | Secondary Species | No of Pixels | System Noise (K) |
|---|---|---|---|---|---|
| Band 1 | 0.81 | CI | CO(7-6) | 1 | 180 |
| Band 2 | 1.45 | NII | $SH^+$,SO,$CF^+$,$H_2O^+$ | 2 | 350 |
| Band 3 | 1.9 | CII | $^{13}C+$ | 2 | 500 |
| Band 4 | 4.7 | OI | - | 2 | 800 |

FIRSPEX opens up a previously unexplored spectral and spatial parameter space that will produce an enormously significant scientific legacy by focusing on the properties of the multi-phase ISM, the assembly of molecular clouds in our Galaxy and the onset of star formation; topics which are fundamental to our understanding of galaxy evolution.

## 8.2 SPICA (proposed for M5 Cosmic Vision)

The SPICA satellite will have a 2.5 meter class Ritchey-Chrétien telescope, cooled to a temperature in the 6-8 K range. The telescope is mounted on the service module with its axis perpendicular to the satellite axis. The payload will be cooled using mechanical coolers in combination with V-groove radiators as was done for the PLANCK satellite. The mission is to operate nominally for 3 years, with a goal of 5 years. This mission configuration is based on the results of the recent ESA *Next Generation Cryogenic Infra-Red* telescope study and subsequent further analysis by JAXA/ISAS – the resulting spacecraft design is shown in Figure 8-2.

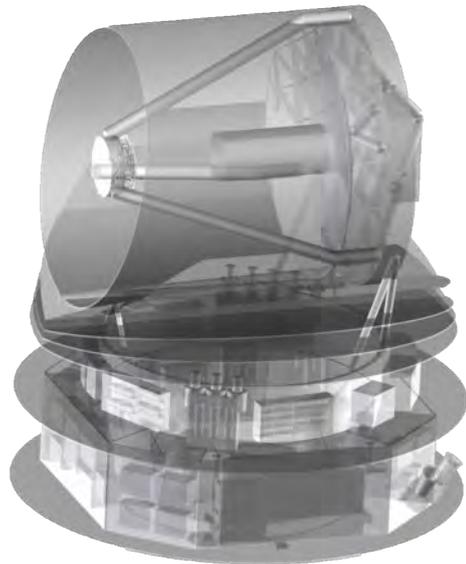

Figure 8-2 SPICA mission concept

For the payload two core instruments are foreseen; a mid-infrared imager/spectrometer (SMI) and a far-infrared grating spectrometer/ polarimeter (SAFARI). SAFARI will be provided by a European consortium of space research institutes, and the SMI will be implemented by a Japanese consortium. The two instruments together provide continuous spectroscopic coverage over the full 20 to 230 μm range, with a resolving power ($R = \lambda/\Delta\lambda$) between a few hundred and a thousand. With a Martin-Puplett interferometer in the SAFARI signal path the resolving power can be further increased up to ~11000 allowing more detailed line profile studies between 35 and 230 μm. Additionally SMI will provide a high resolving power (R ~ 25,000) capability in the 12 - 18 μm window. The mid-infrared camera will allow high sensitivity mapping of large areas in the 20 - 34 μm range. Both instruments utilise state of the art detector technologies, providing the high sensitivity required by the main science goals (see also Figure 8-3).

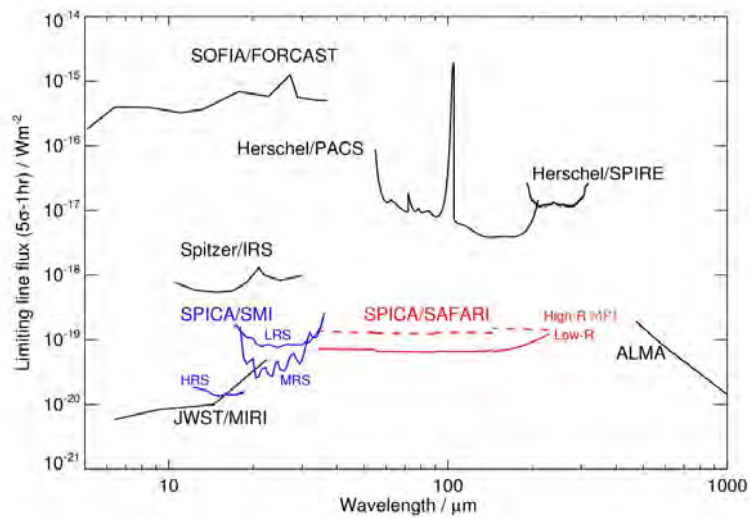

Figure 8-3 projected spectroscopic sensitivity of the SPICA instruments as compared to other facilities. SAFARI sensitivity is based on 2×10$^{-19}$W/√Hz NEP detectors.

## 8.3 TALC (to be proposed for L4)

Fairings put a strict limit to the size of a single dish aperture that can be launched in space. TALC (Thinned Aperture Light Collector) is a 20-m diameter deployable concept that explores some unconventional optical solutions (between the single dish and the interferometer) to achieve a very large aperture. Its collecting area is 20 times larger than Herschel's, giving access to very faint and/or distant sources. With an unconventional optical design comes the necessity to combine data acquisition with unconventional data processing techniques, which are being developed today, based on the notion of sparsity in astronomical signals (e.g. Starck et al. 2010).

The deployable mirror structure exploits the concept of *tensegrity*, i.e. when structural rigidity is achieve through compression. The TALC mirror (see Figure 8.4) is a segmented ring of 20 m diameter and 3 m width. For launch the identical mirror segments are stored on top of each other and a deployable mast pulls a series of cables that deploy the stack of mirror into the required shape. Tension on the cables applied by the central mast provides stiffness to the inner diameter of the deployed ring. On the outer diameter, a degree of freedom persists that allows optimization of the mirror geometric shape by adjusting each of the segments with respect to its neighbors. We foresee an active system using reference stars in the NIR to optimize the segments' position for FIR operations.

Because the aperture is not filled, TALC exhibits a main beam size that is narrower than that of a 20-m single dish, and reaches 0.9" at 100 µm. While this main beam contains only 30% of the total energy, simulations of typical observations campaigns demonstrate that we have the numerical tools at hand to restore a clean map at nominal resolution.

The mirror surface is passively cooled, with a concept that borrows from the JWST sunshield. Sensitivity estimations have been performed with a mirror temperature of 80 K. Simulations with an 80-K mirror show that a sensitivity of 0.1 mJy 5 sigma 1 hr is reached.

The available field of view is 2" and the instrument bay can find ample room just below the secondary, allowing for a suite of instruments to be implemented on the telescope. TALC is currently foreseen to accommodate imaging instruments (to exploit the field of view and the sky accessibility), with polarimetric and medium spectroscopic in-pixel capacities (such as those studied in the context of the FOCUS collaboration (http://ipag.osug.fr/Focus-Labex/). Preliminary investigations show that implementation of very-high resolution heterodyne spectroscopy can be envisioned as well.

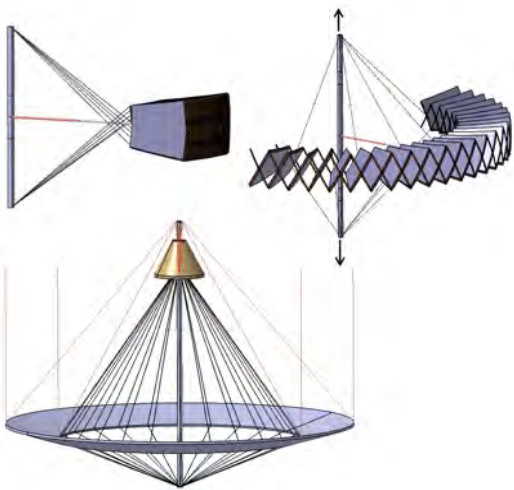

Figure 8.4 TALC deployment (clockwise). At top left the mirror segments are still stowed, but have been pushed away from the central mast, which extends, deploying the whole structure. Red lines indicate the optical path to the instrument platform (Figure from A. Bonnet).

## 8.4 Origins Space Telescope (candidate mission for next decadal plan, after WFIRST)

The Origins Space Telescope is under study right now and the exact configuration is not yet known. The largest concept is a 9.4m telescope (Figure 8.5) cooled down to approximately 4 K and operating between 30 (or lower) and 600 µm. It has been chosen by the American science community as the main candidate for the FIR surveyor study that has started early 2016.

In order to bring down the telescope and instruments to 4 K, a system with sun-shields, stacked V-grooves and closed cycle coolers. The primary instrumentation for CALISTO is a suite of 5–8 moderate resolution ($R \sim 500$) wideband spectrometers, which combine to span the full 30 to 600 µm range instantaneously with no tuning. The detailed arrangement of the modules in the focal plane and the degree to which multiple modules can couple to the same sky position simultaneously is a subject for the detailed study, but any given frequency channel will couple at least tens and up to 200 spatial pixels on the sky. Fourier transform modules will provide higher resolution in the grating orders.

Broadband imagers (cameras) are also studied, and this could be particularly powerful for the short wavelengths where the beam is small and the confusion limit is thus deep.

Finally, we note the potential for heterodyne spectrometer arrays (studied in Europe). While not benefitting from the cryogenic aperture, phase-preserving spectrometers offer the only means of obtaining velocity information and detailed line profiles for Galactic ISM studies as well as protostars and protoplanetary disks. As a guide to the sensitivity, the magenta curve increasing with frequency in Figure 8.6 shows the sensitivity of a quantum-limited receiver to at 10-km/s wide line. If the line profile itself is not of interest, and line confusion and line-to-continuum concerns are not a concern, then the direct detection system is more sensitive even at very narrow line widths.

Figure 8.5 OST will look very similar to JWST with large, actively cooled segmented mirror and shields

Figure 8.6 Mapping speed for different facilities

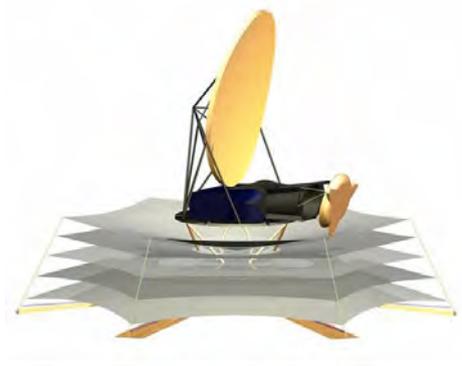
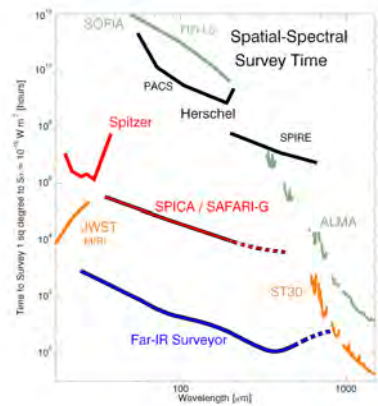

## 8.5 Vision Interferometry Missions

In addition to the previously described missions which have a definite collocation in a near- to mid- future timeframe, there have been more than a few studies relating to what is perceived to be the inevitable post-single dish Era for Far-Infrared astrophysics (Blandford et al. 2010).
Interferometry is performed preferentially in two ways which define distinctly the nature of the interferometer and the technology it makes use of (already addressed in most of the technology section): heterodyne and direct detection. Mission concepts have been proposed for both cases and are reported here. The common advantage in pursuing interferometry in space is the huge increase in angular resolution which results from the very large baselines which these concepts propose and the only way to guarantee the sought after sub-arcsecond angular resolution. The main disadvantage which then ensues is the relatively small collection aperture when compared to large single dishes in both cases. To this date we have no record of combined large baseline space interferometers proposed which present both of the above technologies combined.

### 8.5.1 Heterodyne Space Interferometers

The **ESPRIT** concept was proposed in 2008 (Wild et al. 2008) featuring free-flying formation of satellites presenting a number of dishes (4 or more) of 3.5m diameter with baselines of the order of (but with a potential of an order of magnitude larger) 50m with the need for precise positional metrology and timing (but relaxed with regards to positional control). The latter is the main advantage of such a concept in addition to the inherent high-resolution spectroscopy available to the heterodyne back-end spectrometers. A consequence of the lower sensitivity is the additional advantage on reduced requirements for a cooled telescope.

### 8.5.2 Direct detection (VLBI) Space Interferometers

The **SPECS** concept (and its pathfinder **SPIRIT**) proposed in 2002 (Leisawitz et al. 2001) were studied thoroughly between 2004 and 2007 (papers in Leisawitz et al. 2006) and proposed a double Fourier modulation technique which allows Michelson interferometry to achieve spectroscopy modulated dynamically with the change of baselines to yield spectral datacubes. These concepts presented a two-dish single variable connected baseline (100-class m tether and 2x18m extendable boom respectively). This technique was also further explored in an ESA CDF study (FIRI, Lyngvi et al. 2007) and later with the FISICA-FP7 activity (www.fp7-fisica.eu) which produced a publicly available open source instrument simulator PyFIInS (Rinehart et al. 2007) to explore the potential of this concept. While the sensitivity limitation from the relatively small disc aperture (2m and 1m respectively) is compensated by high sensitivity detectors, disadvantages include the limitation in spectral resolution ( < $10^4$ ) and requirement for cryogenic mirrors (<4K). In addition, the VLBI nature of the interferometer implies a connected structure with strong emphasis on the fractional wavelength level positioning of the two antennas.

### 8.5.3 Existing proposals for Space Interferometers at the time of submission

At the time of submission, the only known existing proposal of a space interferometer is the **SHARP-IR** concept (submitted to the NASA-Probe Class call and intended as a smaller incarnation of the SPIRIT concept, Rinehart et al. 2007). With two 0.8m class cooled telescopes and a max baseline of 12m, SHARP-IR is the current vision for an all connected deployable interferometer.

www.fp7-fisica.eu and http://cordis.europa.eu/project/rcn/106557_en.html - (2015)

# 9 Acknowledgements

The technology section is based on inputs from Brian Ellison, Jian-rong Gao, SPACEKIDS consortium, Juan Bueno, Colin Cunningham, Giorgio Savini, Matt Griffin, Matt Bradford, Marc Sauvage, Dimitra Rigopoulou, Bruce Swinyard, Peter Roelfsema, Fabian Thome and Gilles Durand

Special thanks go to Vanessa Doublier for a very careful reading of the document.

# 10 List of Acronyms

AGB - asymptotic giant branch
AGN - active galactic nucleus
ALMA - Atacama Large Millimeter Array
A-MKID camera – APEX Microwave Kid Inductance Detector camera
APEX - Atacama Pathfinder Experiment
ATLAST - Advanced Technology Large Aperture Space Telescope
ACS - Autocorrelator Spectrometer
AOS - Acousto-Optical Spectrometer
BH - black hole
BICEP - Background Imaging of Cosmic Extragalactic Polarization
BLISS - Background-Limited Infrared-Submillimeter Spectroscopy
CALISTO - Cryogenic Aperture Large Infrared Space Telescope Observatory
CIB - Cosmic Infrared Background
CIRS - Composite Infrared Spectrometer
CMB - Cosmic Microwave Background
CNM - cold neutral medium
COBE - Cosmic Background Explorer
COrE+ - Cosmic Origins Explorer
CTS - Chirp Transform Spectrometers
D/H - deuterium/ hydrogen
DARPA - Defense Advanced Research Projects Agency
DGR - Dust-to-gas mass ratio
DART - Dual Anamorphic Reflector Telescope
DFB - distributed feedback
HEB - Hot Electron Bolometer
IR - infrared
ISM - interstellar medium
E-ELT - European Extremely Large Telescope
FFTS - Fast Fourier Transform Spectrometer
FIR - far-infrared
FIRAS - Far Infrared Absolute Spectrophotometer
FIRSPEX - Far-Infrared Spectroscopic Explorer
FISICA - Far Infrared Space Interferometer Critical Assessment
FS - fine-structure
FTS - Fourier Transform Spectrometer
FUV - far-ultraviolet

GMC - giant molecular cloud
GRACE - Gravity Recovery and Climate Experiment
GREAT - German Receiver for Astronomy at Terahertz Frequencies
HD – Deuterated Hydrogen
HDO – Deuterared water
HEB - Hot Electron Bolometer
HerMES - Herschel Multi-tiered Extragalactic Survey
HEMT - High-electron-mobility transistor
HIFI - Heterodyne Instrument for the Far-Infrared (Herschel)
HIM - hot ionized medium
HST - Hubble Space Telescope
IDP - interplanetary dust particles
IGM - intergalactic medium
IF - intermediate frequency
IRAS - Infrared Astronomical Satellite
ISAS - Institute of Space and Astronautical Science (Japan)
ISM - interstellar medium
ISO - Infrared Space Observatory
JUICE - JUpiter ICy moons Explorer
JFC - Jupiter family comet
JLTP - Journal of Low Temperature Physics
JPL - Jet Propulsion Laboratory
JWST - James Webb Space Telescope
KAO - Kuiper Airborne Observatory
KBO - Kuiper belt object
KID - Kinetic Inductance Detector
LIRG - luminous infrared galaxies
LiteBIRD - Light satellite for the studies of B-mode polarization and Inflation from cosmic background Radiation Detection
LMC – Large Magellanic Cloud
LNA - low-noise amplifier
LO - local oscillator
LWS - Longwave Spectrograph
MBC - main belt comet
METIS - Mid-infrared E-ELT Imager and Spectrograph
MHD - Magnetohydrodynamics
MIPS – Multiband Imaging Photometer
MIR - mid-infrared
MIRI - Mid Infrared Instrument
MKID - Microwave Kinetic Inductance Detector
MMIC - Microwave Monolithic Integrated Circuits
MOIRE - Membrane Optical Images for Real Time Exploitation
MS - main sequence
NEP - noise equivalent power
NIR - near-infrared
NIRSPEC - Near Infrared Spectrograph
NOEMA - Northern Extended Millimeter Array
NRAO - National Radio Astronomy Observatory
OCC - Oort cloud comet

OPR – ortho-to-para ratio  
OST - Origins Space Telescope  
PACS – Photodetector Array Camera and Spectrometer  
PAH - Polycyclic Aromatic Hydrocarbon  
PDF - probably distribution function  
PDR - photo dissociation region  
PDR - photon-dominated region  
PEP - PACS Extragalactic Probe  
PROBA - Project for On-Board Autonomy  
QCD - Quantum Capacitance Devices  
QCL - Quantum cascade laser  
RAL - Rutherford Appleton Laboratory  
RF - radio frequency  
ROSINA - Rosetta Orbiter Spectrometer for Ion and Neutral Analysis  
SAFARI - SpicA FAR-infrared Instrument  
SED - Spectral energy distribution  
SFH - star-formation history  
SFR - star formation rate  
SFRD - star-formation rate density  
SIS - Superconductor-Insulator-Superconductor  
SLED - spectral-line energy distribution  
SMBH's - super-massive Black Holes  
SNe - supernovae  
SNR - Signal to noise ratio  
SNR - supernova remnant  
SOFIA - Stratospheric Observatory for Infrared Astronomy  
SMI - Spica Mid-infrared Instrument  
SPICA - Space Infra-Red Telescope for Cosmology and Astrophysics  
SPIRE - Spectral and Photometric Imaging Receiver  
STO - Stratospheric Terahertz Observatory  
SWI - Submm Wave Instrument  
TALC - Thin Aperture Light Collector  
TES - Transition Edge Sensor  
ULIRG - ultraluminous infrared galaxy  
UV - ultraviolet  
VLTI - Very Large Telescope Interferometer  
VSMOW - Vienna Standard Mean Ocean Water  
WFIRST - Wide Field Infrared Survey Telescope  
WIM - warm ionized medium  
WISE - Wide-field Infrared Survey Explorer  
WNM - warm neutral medium  
XIFU - X-ray Integral Field Unit